\documentclass[10pt]{iopart}
\usepackage{graphicx} 
\usepackage{epstopdf}
\usepackage{amsfonts}
\usepackage{harvard}	
\usepackage{color} 
\usepackage{amssymb}

\graphicspath{{./fig/}}

\definecolor{olive}{rgb}{0.0, 0.6, 0.0}
\definecolor{CH2}{rgb}{0,0,1}

\begin{document}
\topical[Helical Liquids in Semiconductors]{Helical Liquids in Semiconductors}

\author{Chen-Hsuan Hsu$^{1}$, Peter Stano$^{1,2}$, Jelena Klinovaja$^{3}$, and Daniel Loss$^{1,3}$}

\address{$^{1}$RIKEN Center for Emergent Matter Science (CEMS), Wako, Saitama 351-0198, Japan}
\address{$^{2}$Institute of Physics, Slovak Academy of Sciences, 845 11 Bratislava, Slovakia}
\address{$^{3}$Department of Physics, University of Basel, Klingelbergstrasse 82, CH-4056 Basel, Switzerland}

\vspace{10pt}
\begin{indented}
\item[] \today 
\end{indented}

\begin{abstract}
One-dimensional helical liquids can appear at boundaries of certain condensed matter systems. Two prime examples are the edge of a quantum spin Hall insulator and the hinge of a three-dimensional second-order topological insulator. For these materials, the presence of a helical state at the boundary serves as a signature of their nontrivial electronic bulk topology. Additionally, these boundary states are of interest themselves, as a novel class of strongly correlated low-dimensional systems with interesting potential applications. Here, we review existing results on such helical liquids in semiconductors. Our focus is on the theory, though we confront it with existing experiments. We discuss various aspects of the helical liquids, such as their realization, topological protection and stability, or possible experimental characterization. We lay emphasis on the hallmark of these states, being the prediction of a quantized electrical conductance. Since so far reaching a well-quantized conductance has remained challenging experimentally, a large part of the review is a discussion of various backscattering mechanisms which have been invoked to explain this discrepancy. Finally, we include topics related to proximity-induced topological superconductivity in helical states, as an exciting application towards topological quantum computation with the resulting Majorana bound states. 

~ \\ 
{\it [Note]  This is a revised version accepted for publication in Semicond. Sci. Technol. 36, 123003 (2021). The final version of this article can be found at} 
https://doi.org/10.1088/1361-6641/ac2c27. 

\end{abstract}

\vspace{2pc}

\submitto{\SST}

\maketitle
 
\ioptwocol

\section{Introduction \label{Sec:Introduction}}
\begin{figure}[!bht]
\centering
\includegraphics[width=0.52\linewidth]{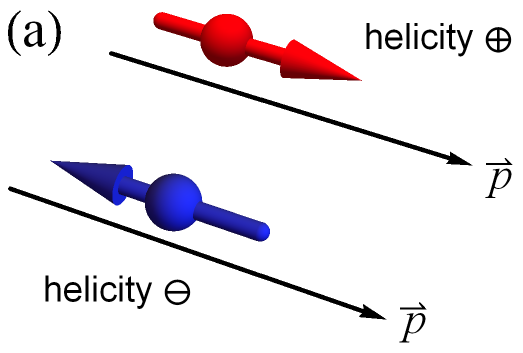}
\hspace{0.02\linewidth}
\includegraphics[width=0.42\linewidth]{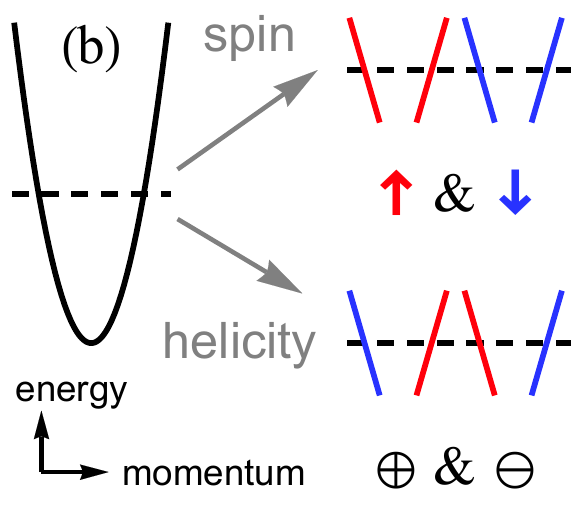}
\caption{Notion of helicity. (a) In particle physics, the helicity of a particle is defined through the relative orientation between its spin and momentum $\vec{p}$. 
(b) In a condensed matter system hosting spin-$1/2$ fermions with quadratic dispersion, we can label the degenerate states near the Fermi level either by their spins or by their helicities. Unlike the spin, the states with opposite helicities can be split in a time-reversal-invariant system. 
}
 \label{Fig:helicity}
\end{figure}

In relativistic quantum field theory, spin-$1/2$ particles are governed by the Dirac equation. The Dirac Hamiltonian commutes with the helicity operator, the projection of a particle spin on the direction of its momentum. Therefore, helicity, defined as the sign of the eigenvalue of the helicity operator, is an invariant of motion.\footnote{Strictly speaking, the helicity of a massive particle is not an intrinsic property, as the sign of momentum might change upon a Lorentz boost, thus depending on the reference frame. However, being an invariant of motion, the helicity can serve as a good quantum number in a given reference frame.}
It allows one to assign a definite---negative or positive---helicity to eigenstates, as illustrated in \Fref{Fig:helicity}(a).

\begin{figure}[t]
\centering
 \includegraphics[width=0.4\linewidth]{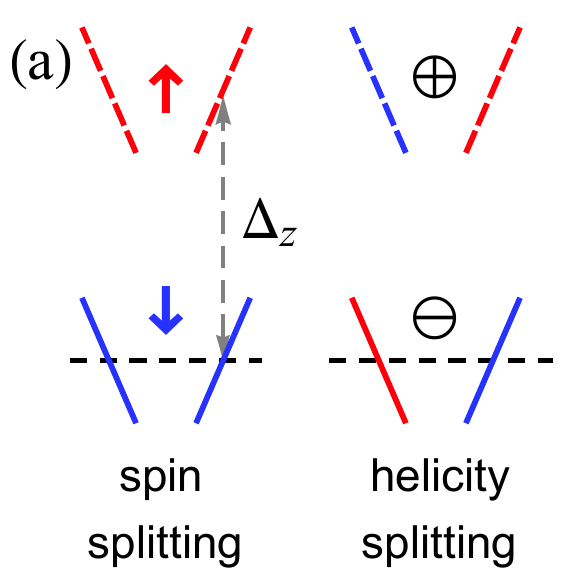}
	\hspace{0.0\linewidth}
 \includegraphics[width=0.57\linewidth]{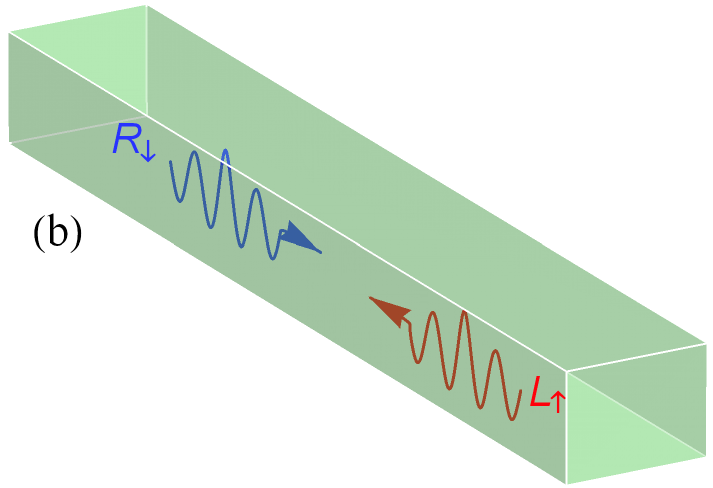}
\caption{(a) The degeneracy for spin-$1/2$ fermions can be lifted either by the Zeeman splitting $\Delta_z$ or by helicity-dependent interactions. When the Fermi energy crosses only one of the branches, the system is spin-polarized or has a definite helicity. 
(b) Assuming that the chemical potential lies within an energy window where only states with a definite helicity are populated, the conduction modes are made of helical states whose spin orientation is fixed by the propagation direction.
}
 \label{Fig:helicity3}
\end{figure}

In a condensed matter system consisting of spin-$1/2$ fermions, the low-energy Hamiltonian can mimic Dirac fermions and one can define the helicity of a fermionic state according to its momentum and spin in a similar manner. Assuming that the Fermi momentum is nonzero, one can label the states at the Fermi level (which have intrinsic spin degeneracy) according to their helicity instead of spin.\footnote{In analogy to particle physics, we define the helicity through the spin projection onto its quantization axis, even though the latter does not need to be in parallel to the momentum. In more general terms, states of the same helicity are defined such that they form a time-reversal (Kramers) pair. } 
We illustrate the two sets of labeling in \Fref{Fig:helicity}(b) for particles with quadratic dispersion.
The first set of labeling can be straightforwardly examined in experiments. Namely, upon applying a magnetic field, which has a Zeeman coupling to the spin, the degeneracy of the opposite spin states can be lifted, as illustrated in \Fref{Fig:helicity3}(a). 

Alternatively, one can image that, if the two states with the opposite helicities can be split in energy, one can create a {\it helical liquid}, with one of the helicity states being occupied [see \Fref{Fig:helicity3}(a)]. 
In real space, such a system hosts conduction modes made of helical states with spin orientation fixed by the propagation direction; an example with the negative helicity is shown in \Fref{Fig:helicity3}(b).
Furthermore, in contrast to lifting the spin degeneracy, since the helicity is invariant under time reversal, the helical liquid can be generated in a time-reversal-invariant setting.

While the fermion doubling theorem has proven that a helical liquid with an odd number of components (thus including a single pair of helical states) cannot be formed in a purely one-dimensional system~\cite{Wu:2006}, the theorem can be circumvented by having a helical liquid as a part of two- or three-dimensional systems. 
Indeed, recent progress in condensed matter physics has demonstrated that it is possible to stabilize such a helical liquid at boundaries of a higher-dimensional bulk.

In this review, we focus on such gapless helical states flowing along the edges of two-dimensional or hinges of three-dimensional bulk materials. 
The helicity degeneracy is lifted due to the electronic topology of the bulk, which results in one-dimensional helical channels appearing on the boundaries.
The research on helical channels has both fundamental and practical motivations. 
First, as mentioned above, they appear on surfaces of certain materials in a way analogous to the chiral edge states in the quantum Hall effect under external magnetic fields.
Since their existence is related to the electronic bulk topology~\cite{Kane:2005b}, their presence serves as a signature for the topologically nontrivial phase (quantum spin Hall effect) that goes beyond the notion of Landau's spontaneous symmetry breaking.
Second, being spatially confined in a narrow channel, the role of electron-electron interactions increases, offering a possible realization of a (quasi-)one-dimensional strongly correlated fermion system~\cite{Wu:2006,Xu:2006}. As we will see, while interactions can potentially destabilize the helical liquid, in other scenarios they can drive the helical liquid into various phases ranging from magnetic orders to topological superconductivity.
In other words, combining the electron-electron interactions with other ingredients such as magnetic impurities, spin-orbit coupling and superconducting pairing, helical states provide a platform for unconventional states of matter.

Apart from academic motivations, the helical channels are also candidates for potential applications, for example in spintronics or topological quantum computation~\cite{Moore:2009}. 
First, in contrast to ordinary one-dimensional channels, the topological origin of helical channels protects them from Anderson localization due to weak disorder, possibly offering low-dissipation charge and spin transport at nanoscales~\cite{Sheng:2005,Sheng:2006}.
Second, as their spin degeneracy is lifted, they can be used to produce Majorana or parafermion modes for quantum computation~\cite{Fu:2009,Mi:2013,Klinovaja:2014}. 

As we discuss in depth below, many of these expectations turn out to be much more involved in reality. Nevertheless, these prospects initiated extensive research on topological and strongly correlated systems over the past decade. 
Especially the investigations of the quantized charge conductance, as the paramount property predicted for helical channels, continued unabated in both theory and experiments. 
In particular, the possible mechanisms for the unexpected deviation from the quantized conductance have been the subject of numerous studies on helical channels. 
On the other hand, the theoretical investigations, including those relying on the helical Tomonaga-Luttinger liquid (hTLL) model, are rather scattered in the literature. In addition to being not easy to track, they include several sets of mutually contradicting results.
This situation was among our motivations to undertake a comprehensive review on this topic.

We organize the review as follows. 
In section~\ref{Sec:Realization}, we discuss how the helical states arise on the boundaries of topologically nontrivial systems.
In section~\ref{Sec:hTLL}, we discuss the hTLL realized in the edge or hinge channels of the topological materials, setting the foundations for the next sections. We also discuss how to detect and characterize the helical channels.
In section~\ref{Sec:Transport}, we review experimental progress (section~\ref{Sec:Exp}) and discuss mechanisms which can lead to backscatterings and therefore affect the electrical conductance of a helical channel. We divide the backscattering mechanisms into two types--perturbations which break the time-reversal symmetry (section~\ref{Sec:Scattering_noTRS}) and those which preserve it (section~\ref{Sec:Scattering_TRS}).
\Tref{Tab:nomenclature} summarizes the diverse nomenclature for backscattering mechanisms induced by spin-orbit interactions.
Since the resistance mechanisms are distinguishable through their temperature dependence, we summarize the latter in \Tref{Tab:mechanism} and \Tref{Tab:mechanism2} for time-reversal symmetry breaking and time-reversal-invariant mechanisms, respectively.
In section~\ref{Sec:TSU}, we discuss how topological superconductivity arises upon adding proximity-induced superconducting pairing, and how Majorana bound states arise in various setups.
We give an outlook in section~\ref{Sec:Conclusion}.

We point out review articles on related topics, both recent~\cite{Sato:2017,Rachel:2018,Haim:2019,Gusev:2019,Beenakker:2020,Culcer:2020} and less recent ones~\cite{Hasan:2010,Qi:2011,Maciejko:2011,Alicea:2012,Beenakker:2013,DasSarma:2015,Chiu:2016,Dolcetto:2016}. 
Compared to those, we focus on the charge transport properties of the one-dimensional helical channels themselves.
Also, we cover more recent developments, such as the possibility of helical hinge states in higher-order topological insulators or realizations of Majorana bound states using them.

\section{Realization of helical edge states \label{Sec:Realization}}
\subsection{Quantum spin Hall effect}
We start with how the helical states are realized in solid-state systems. 
As mentioned in section~\ref{Sec:Introduction}, even though a single pair of helical states cannot arise in a purely one-dimensional system, it can appear on the one-dimensional edge of a two-dimensional system hosting the quantum spin Hall state, a time-reversal-invariant analog of the quantum Hall state~\cite{Kane:2005a,Kane:2005b,Bernevig:2006b,Wu:2006,Xu:2006}.
This mechanism is closely related to the chiral edge channels in a quantum Hall system. Here, the up- and down-spin states feel effective magnetic fields with opposite signs, leading to two copies of quantum Hall liquids with the opposite Hall conductance.
Viewed separately, each spin subsystem realizes a quantum Hall liquid hosting a chiral edge state with the opposite chirality for the opposite spins.
When combined, the two spin subsystems host helical edge states, thus preserving the time-reversal symmetry of the entire system. 

As an initial prediction, Kane and Mele proposed that the quantum spin Hall effect can be realized in graphene~\cite{Kane:2005a}, a state-of-art material at that time~\cite{Novoselov:2004,CastroNeto:2009}. The key ingredient driving this effect is spin-orbit coupling: entering as an imaginary spin-dependent hopping term in a tight-binding model of graphene, it results in the spin-dependent magnetic field required for the quantum spin Hall state. 
Similar to chiral edge channels of a quantum Hall liquid, which led to the notion of topological order, the helical edge channels of a quantum spin Hall state are protected by the energy gap of the bulk. 
The quantum spin Hall state in the Kane-Mele model was subsequently identified as a $\mathbb{Z}_2$ topological order in a time-reversal-invariant system~\cite{Kane:2005b}, a novel state distinct from an ordinary insulator or a quantum Hall liquid with broken time-reversal symmetry.
The new classification was characterized by a $\mathbb{Z}_2$ topological invariant constructed from the bulk Hamiltonian~\cite{Sheng:2006,Fu:2006,Moore:2007,Roy:2009b} and inspired the naming for the ``topological insulator'' phase~\cite{Moore:2007}. 
In a bulk-boundary correspondence, the invariant is related to the number of the Kramers pairs of helical states on the boundary~\cite{Fu:2006,Moore:2007,Roy:2009b,Hasan:2010}.

Even though later it became clear that the spin-orbit coupling, and the resulting gap, in graphene is too small to provide a quantum spin Hall phase under realistic conditions~\cite{Min:2006,Yao:2007}, the works by Kane and Mele were seminal for subsequent investigations for more realistic setups and for broader research on topological materials. 
In particular, the identification of the $\mathbb{Z}_2$ topological order further motivated works on topological classification of gapped systems based on their symmetry classes. 
Specifically, one can characterize gapped systems, including insulators and superconductors, based on the time-reversal and particle-hole symmetries described by the relations~\cite{Ryu:2010,Hasan:2010},
\begin{eqnarray}
\mathcal{T} H ({\bf p}) \mathcal{T}^{-1} &=& H (-{\bf p}), \label{Eq:TRS} \\
\mathcal{C} H({\bf p}) \mathcal{C}^{-1} &=& -H(-{\bf p}), \label{Eq:PHS}
\end{eqnarray}
where $H$ is the Bloch Hamiltonian in momentum (${\bf p}$) space, and the time-reversal (particle-hole) symmetry is represented by an antiunitary operator $\mathcal{T}$($\mathcal{C}$). 
The gapped systems are then categorized according to the values of $\mathcal{T}^2$ and $\mathcal{C}^2$ (as well as the product of $\mathcal{T}$ and $ \mathcal{C} $, if both $\mathcal{T}^2$ and $\mathcal{C}^2$ are zero). 
The quantum spin Hall insulator phase in the Kane-Mele model, characterized by $\mathcal{T}^2=-1$ and $\mathcal{C}^2=0$, is in fact an entry in the periodic table for topological insulators and superconductors~\cite{Kitaev:2009,Ryu:2010}.

As reviewed here, while the term ``topological insulator'' was originally coined for the nontrivial phase in the Kane-Mele model and its three-dimensional generalization~\cite{Moore:2007}, it was later also used to cover a broader class of topologically nontrivial insulating systems in the periodic table~\cite{Kitaev:2009,Ryu:2010,Hasan:2010}. The latter includes, for instance, quantum Hall states in the absence of time-reversal symmetry.
Nonetheless, since we restrict ourselves to insulating systems hosting helical states, we adopt the original terminology and use the terms ``quantum spin Hall insulator'' and ``two-dimensional topological insulators'' (2DTI) interchangeably.\footnote{In addition, since the term 2DTI does not imply that helical states can be labeled by a spin index, it also covers more generic settings that fulfill \eref{Eq:TRS} but do not conserve spin.  }
Below we review their realizations in heterostructures based on semiconductors.

\subsection{Quantum spin Hall effect in a semiconductor quantum well}
Among other theoretical proposals~\cite{Sheng:2005,Bernevig:2006b,Murakami:2006,Qi:2006}, a key contribution was made by Bernevig~et~al., who predicted the quantum spin Hall state in a composite quantum well made of HgTe and CdTe~\cite{Bernevig:2006}. Owing to its significance, here we review the BHZ model, named after the authors of \cite{Bernevig:2006}.
It is based on the $k \cdot p$ theory, a standard perturbation theory for semiconductors, based on a restriction onto a few energy bands around the Fermi level. 
The BHZ model is constructed from symmetry considerations for a quasi-two-dimensional quantum well grown along $z$ direction with the in-plane momentum $\hbar {\bf k}=(\hbar k_x, \hbar k_y)$ measured from the $\Gamma$ point (${\bf k} = 0$).

\begin{figure}[t]
\centering
 \includegraphics[width=\linewidth]{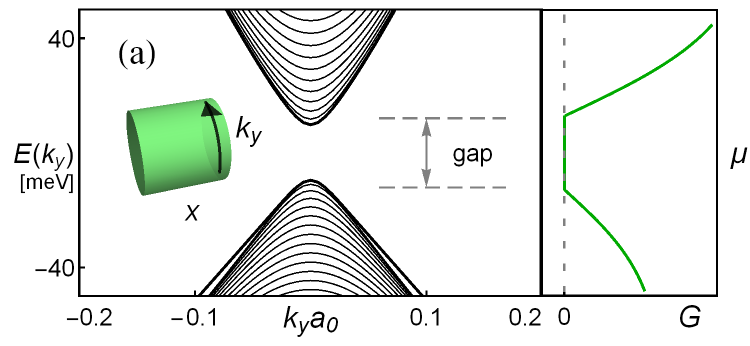}\\
 \includegraphics[width=\linewidth]{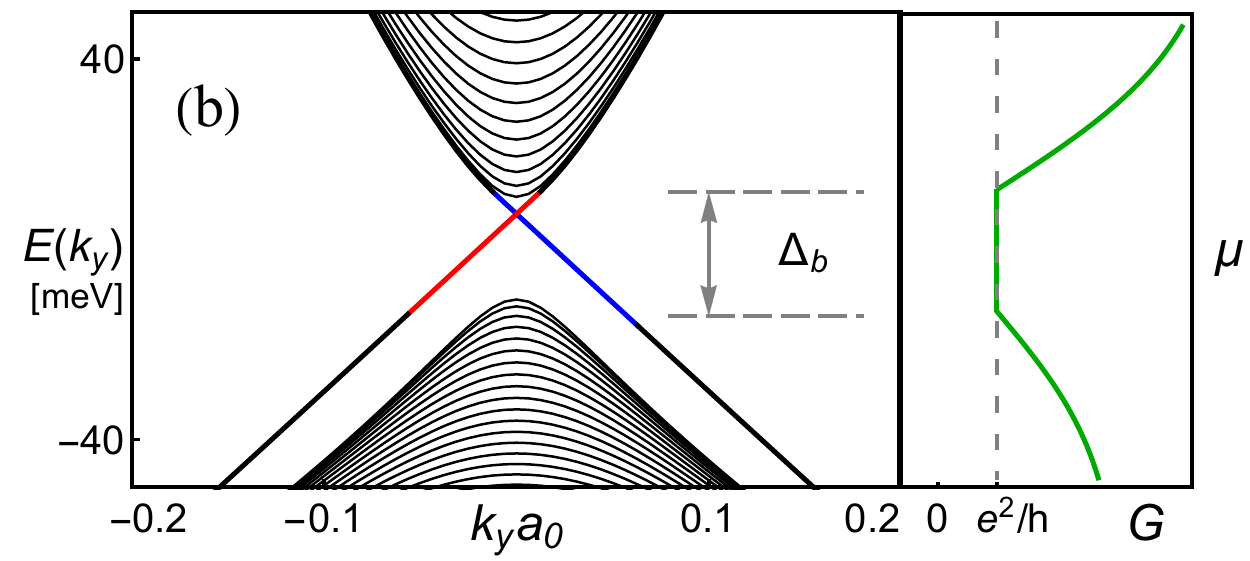} 
 \caption{Energy spectra of the BHZ model in a lattice of 500 sites and the edge conductance $G$ as a function of the chemical potential $\mu$.
As indicated in the inset of Panel (a), we take zero boundary conditions along $x$ and periodic along $y$. 
(a) $M_{\rm bhz}/B_{\rm bhz}<0$. There is no state within the gap, leading to a trivial insulator with zero conductance when $\mu$ lies in the gap.
Here we adopt the parameter values: $A_{\rm bhz} = 3.87$~eV~\AA, $B_{\rm bhz} = -48.0$~eV~\AA$^2$, $C_{\rm bhz} = 0$, $D_{\rm bhz} = -30.6$~eV~\AA$^2$ and $M_{\rm bhz} = 0.009$~eV, corresponding to the quantum well width of $55$~\AA~\cite{Qi:2011}.
(b) $M_{\rm bhz}/B_{\rm bhz}>0$. The bands are inverted, accompanied by gapless states propagating along the edges; see \Fref{Fig:BHZ_edge}. 
When the chemical potential lies within the bulk gap $\Delta_{\rm b}$, the edge conductance is quantized. 
The adopted parameter values are $A_{\rm bhz} = 3.65$~eV~\AA, $B_{\rm bhz} = -68.6$~eV~\AA$^2$, $C_{\rm bhz} = 0$, $D_{\rm bhz} = -51.2$~eV~\AA$^2$ and $M_{\rm bhz} = -0.01$~eV, corresponding to the width of $70$~\AA~\cite{Konig:2008,Qi:2011}.  
}
 \label{Fig:BHZ_spectrum}
\end{figure}

The BHZ model Hamiltonian takes the following form,
\begin{eqnarray}
H_{\rm BHZ} &=& \left( 
\begin{array}{cc}
h_{\rm bhz} ({\bf k})  & 0 \\
0 & h_{\rm bhz}^{*} (-{\bf k}) 
\end{array}
\right).
\label{Eq:BHZ1}
\end{eqnarray}
Here, the upper block is $h_{\rm bhz} ({\bf k}) = h_0({\bf k}) \tau_0 + h_{\mu} ({\bf k}) \tau_{\mu} $, with the Pauli matrices $\tau_{\mu}$ for $\mu \in \{ 1, 2, 3 \}$, and $\tau_0$ is the identity matrix. Finally,  the lower block is related to the upper one through the time-reversal symmetry.
The basis for the above Hamiltonian comprises $|E,+ \rangle $, $|H,+ \rangle $, $|E, - \rangle $, and $|H, - \rangle $ with $E,H$ denoting the electron- and (heavy-)hole-like bands in HgTe/CdTe and $\pm$ denoting the time-reversal indexes. 
In addition to the time-reversal symmetry imposed in~\eref{Eq:BHZ1}, the form of $h_{\rm bhz} ({\bf k})$ is further restricted by parity, the eigenvalue under the operation of spatial inversion.  
Using $\uparrow$ or $\downarrow$ to label the spin, and $s$ or $p$ the orbitals, the electron-like band $|E,\pm \rangle$ consists of states  $|s,  \uparrow\hspace{-3pt}/\hspace{-3pt}\downarrow  \rangle$ while the hole-like band $|H,\pm \rangle$  of $| \pm (p_x \pm i p_y), \uparrow\hspace{-3pt}/\hspace{-3pt}\downarrow \rangle$. Since these two sets have opposite parity, the matrix elements $ h_0 $ and $h_3$ must be even and the elements $ h_1 $ and $h_2$ must be odd under inversion. 
Taken together, the time-reversal, inversion and crystal symmetries impose the following functional form for the matrix elements Taylor-expanded in momentum components around ${\bf k}=0$~\cite{Bernevig:2006,Konig:2007},
\begin{equation}
\eqalign{
h_0({\bf k}) &= C_{\rm bhz} - D_{\rm bhz} (k_x^2 + k_y^2), \\
h_1({\bf k}) &= A_{\rm bhz} k_x, \\
h_2({\bf k}) &= -A_{\rm bhz} k_y, \\
h_3({\bf k}) &= M_{\rm bhz} - B_{\rm bhz} (k_x^2 + k_y^2),}
\label{Eq:BHZ2}
\end{equation}
with the material- and structure-dependent parameters $A_{\rm bhz}$, $B_{\rm bhz}$, $C_{\rm bhz}$, $D_{\rm bhz}$ and $M_{\rm bhz}$. The values of these parameters cannot be obtained from symmetry analysis. 
Nevertheless, one sees that there is a band inversion when the ratio $M_{\rm bhz}/B_{\rm bhz}$ changes its sign. Crucially, this ratio is experimentally controllable through the width of the HgTe layer, sandwiched by CdTe layers in the quantum well.
As a remark, the bulk-inversion asymmetry in the zinc-blend lattice induces an additional term not included in the BHZ model.  
However, detailed studies~\cite{Dai:2008,Konig:2008,Rothe:2010}
demonstrated that while adding such a symmetry-breaking term can affect the energy spectrum, it does not destroy the topological phase transition that emerges in the simplified model described by \eref{Eq:BHZ1} and \eref{Eq:BHZ2}.

\begin{figure}[t]
\centering
 \includegraphics[width=0.49\linewidth]{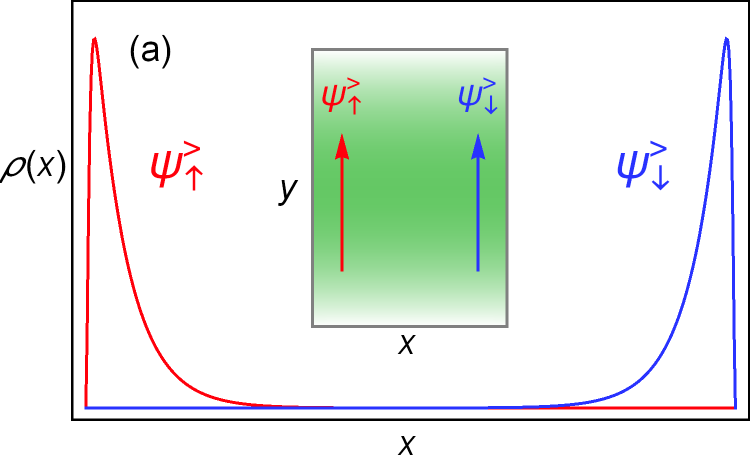} 
 \includegraphics[width=0.49\linewidth]{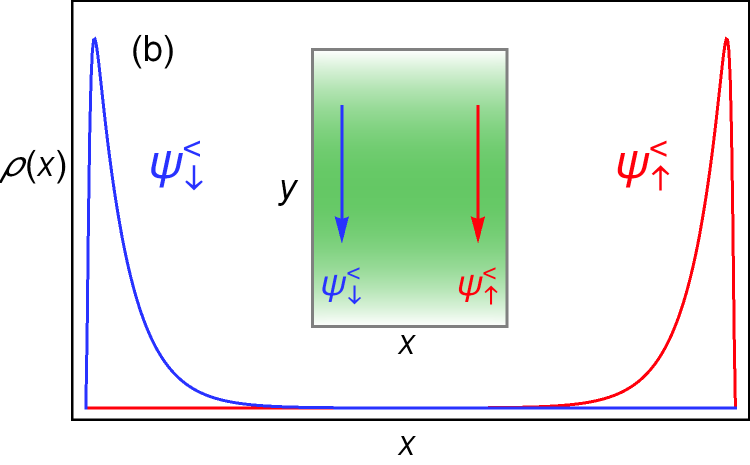}
\caption{Spatial density profile, $\rho(x)$, of the in-gap states in the band-inverted regime of the BHZ model in a lattice of 500 sites. The states localized at the edge along $x$ correspond to ones from \Fref{Fig:BHZ_spectrum}(b) for the energy within the bulk gap and  for  the propagation directions along $y$ as indicated in the insets. (a) Profile of $|\psi_{\sigma}^{>}|^2$ for the eigenstates with positive velocity along $y$ and spin $\sigma$. Here we set the chemical potential to $3.7$~meV; the other parameters are the same as those adopted in \Fref{Fig:BHZ_spectrum}. (b) Similar to (a) but for negative velocity denoted as $|\psi_{\sigma}^{<}|^2$. The gapless states with the opposite spins propagate in the opposite directions. 
}
 \label{Fig:BHZ_edge}
\end{figure}

For illustration, we solve numerically a tight-binding version of the BHZ model, using a two-dimensional rectangular grid with the lattice constant $a_0=6.5~$\AA. The Hamiltonian $H_{\rm BHZ}$ keeps the form given in \eref{Eq:BHZ1} with $h_{\mu}$ for $\mu \in \{ 0, 1, 2, 3 \}$ replaced by~\cite{Konig:2008,Qi:2011}
\begin{equation}
\eqalign{
h_0({\bf k}) &= C_{\rm bhz} - 2 \frac{D_{\rm bhz}}{a_0^2} [2- \cos (k_x a_0) -\cos (k_y a_0) ], \\
h_1({\bf k}) &= \frac{A_{\rm bhz}}{a_0} \sin (k_x a_0) , \\
h_2({\bf k}) &= -\frac{A_{\rm bhz}}{a_0} \sin (k_y a_0) , \\
h_3({\bf k}) &= M_{\rm bhz} - 2 \frac{B_{\rm bhz}}{a_0^2} [2- \cos (k_x a_0) -\cos (k_y a_0) ].}
\label{Eq:BHZ3}
\end{equation}
These expressions reduce to \eref{Eq:BHZ2} at small ${\bf k}$. To calculate the energy spectrum, we consider a cylindrical geometry with zero boundary conditions along $x$ and periodic boundary conditions along $y$. To reflect this geometry, we perform inverse Fourier transform in the $x$ coordinate and plot the eigenvalues as a function of $k_y$, which remains a good quantum number.
In \Fref{Fig:BHZ_spectrum}(a) where $M_{\rm bhz}/B_{\rm bhz}<0$, the system is fully gapped, indicating a trivial insulator with zero conductance.
In contrast, in \Fref{Fig:BHZ_spectrum}(b) with $M_{\rm bhz}/B_{\rm bhz}>0$, the bands are inverted and gapless states emerge within the bulk gap. Looking at the corresponding eigenfunctions plotted in \Fref{Fig:BHZ_edge} reveals the following two properties of these gapless states. First, they are localized at the edges. Second, they are spin polarized, and the spin polarization swaps on inverting the velocity. In other words, these states are helical. 

\begin{figure}[t]
\centering
 \includegraphics[width=\linewidth]{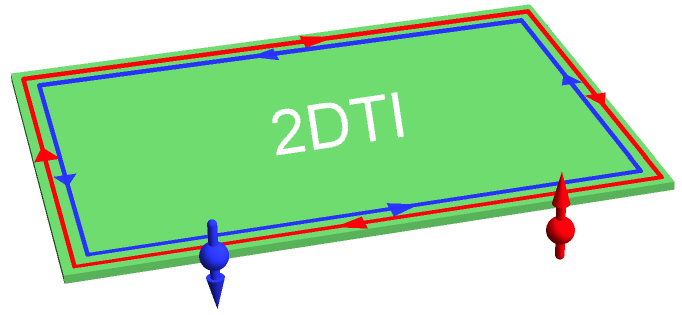}
\caption{Schematic of a two-dimensional topological insulator (2DTI). 
While the interior (bulk states) of the system is gapped, there are gapless states located in the exterior, forming a helical liquid circulating the edge.}
 \label{Fig:QSHI}
\end{figure}

A realistic sample is finite in both $x$ and $y$ directions. This case is illustrated in \Fref{Fig:QSHI}. There are gapless edge states circulating around the sample, whereas the interior of the system is gapped.
Along a specific edge,  gapless states with opposite spins flow in the opposite directions; hence they are helical. 
To distinguish the normal and the band-inverted regimes, a straightforward probe is through the edge conductance. As illustrated in \Fref{Fig:BHZ_spectrum}, in the normal regime, the conductance is zero when the chemical potential is in the gap. In contrast, when the band is inverted, we expect a finite---and in an idealized case, quantized---edge conductance when the chemical potential lies within the gap.

Having demonstrated the presence of the helical edge states in the bulk gap, we now discuss how they are related to the bulk topology by examining the Berry phase of the bulk eigenstates. Since the upper and lower blocks in \eref{Eq:BHZ1} are decoupled (which can be viewed as the up- and down-spin components), the Berry phase for the two blocks can be computed separately. 
To this end, we define the Berry curvature as~\cite{Griffiths:1995} 
\begin{equation}
 {\bf \Omega}_{\sigma,\pm} \equiv i {\bf \bigtriangledown}_k \times \langle \Phi_{\sigma,\pm}({\bf k})| {\bf \bigtriangledown}_k | \Phi_{\sigma,\pm}({\bf k}) \rangle
\end{equation}
for the eigenstate $|\Phi_{\sigma,\pm} \rangle$ with the periodic boundary conditions along both $x$ and $y$ directions. Here, for each spin $\sigma$, the $\pm$ sign corresponds to the upper-/lower-band of the BHZ Hamiltonian with the (spin-degenerate) eigenvalues 
\begin{equation}
h_0 ({\bf k} ) \pm \sqrt{h_1^2 ({\bf k} )+ h_2^2 ({\bf k} )+ h_3^2 ({\bf k} )},
\end{equation}
where $h_{\mu}$ are given in \eref{Eq:BHZ3}.
As long as there exists a finite bulk gap, we can define the unit vector $\hat{h}_{\sigma} \equiv \vec{h}_{\sigma}/|\vec{h}_{\sigma}|$ with $\vec{h}_{\sigma} = (\sigma h_1, h_2, h_3)$, and express the nonzero ($z$) component of the Berry curvature as
\begin{eqnarray}
 \Omega_{\sigma,\pm} &\equiv& {\bf \Omega}_{\sigma,\pm} \cdot {\bf e}_z =  \mp {\frac {1}{2}} \hat{h}_{\sigma} \cdot ( {\frac {\partial{\hat{h}_{\sigma}}}{\partial k_x}} \times {\frac {\partial{\hat{h}_{\sigma}}}{\partial k_y}}).
\end{eqnarray}
The Berry phase is obtained upon integrating the curvature over the momentum space.
For convenience, we define the following quantity as the Berry phase divided by 2$\pi$,
\begin{eqnarray}
N_{\sigma,\pm} \equiv \int_{\rm BZ} {\frac{d^2\it{k}}{2 \pi }} \Omega_{\sigma,\pm},
\label{Eq:winding}
\end{eqnarray}
with the integral over the Brillouin zone. 
It can be shown that the above expression is the skyrmion number, which measures how many times the unit vector $\hat{h}_{\sigma}$ covers the unit sphere around the origin while $(k_x,k_y)$ spanning the Brillouin zone~\cite{Hsu:2011}. 
Therefore, it is quantized and cannot be continuously varied unless the vector $\vec{h}_{\sigma}$ shrinks to zero, which would require closing the bulk gap. 
Thus, the integer $N_{\sigma,\pm}$ is a {\it topological invariant} protected by the bulk gap, in analogy to Thouless-Kohmoto-Nightingale-den Nijs (TKNN) invariant in the integer quantum Hall states~\cite{Thouless:1982}.

\begin{figure}[t]
\centering
 \includegraphics[width=0.49\linewidth]{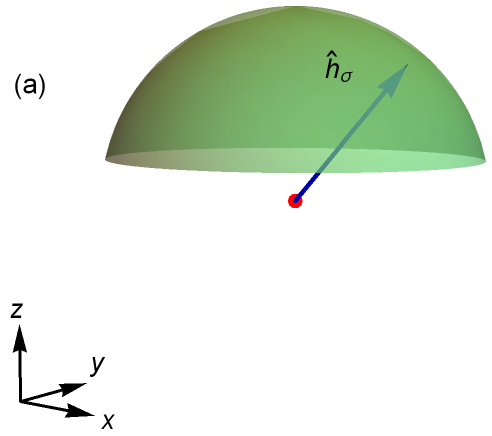} 
 \includegraphics[width=0.49\linewidth]{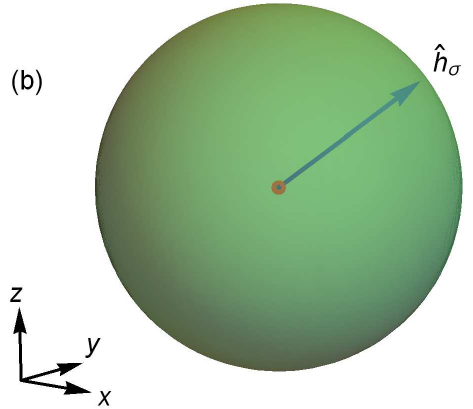}
\caption{Parametric surface formed by the  trajectory of the unit vector $\hat{h}_{\sigma}$ (blue arrow) when ${\bf k}=(k_x,k_y)$ spans over the Brillouin zone. (a) In the trivial regime, the surface does not enclose the origin (red dot), making the skyrmion number in \eref{Eq:winding} zero. (b) In the topological regime, the origin is enclosed by the surface once, leading to a skyrmion number of unity (up to a spin-dependent sign). 
For the integer to change through varying the Hamiltonian parameters, the surface has to pass through the origin, corresponding to the vector $\vec{h}_{\sigma}$ vanishing at certain momentum and, thus, to a gap closing point. }
 \label{Fig:winding}
\end{figure}

As illustrated in \Fref{Fig:winding}, in the normal regime the trajectory of $\hat{h}_{\sigma}$ does not enclose the origin and we have $N_{\sigma,\pm}=0$, whereas in the inverted regime we get a nontrivial value $N_{\sigma,\pm} = \mp \sigma$. Assuming that the chemical potential lies within the gap so that the lower band is occupied and the upper band is empty, we evaluate the total Chern number $N$ and the spin Chern number $N_{\rm s} $,
\begin{eqnarray}
N \equiv \sum_{\sigma} N_{\sigma,-}  = 0 , \; N_{\rm s} \equiv \sum_{\sigma} \sigma N_{\sigma,-} = 2.
\end{eqnarray}
As a result, unlike the quantum Hall states labeled by the total Chern number, here the bulk topology is characterized by the 
spin Chern number, 
which is a $\mathbb{Z}_2$  invariant.
Similar to the relation between the Chern number and the quantized Hall conductance \cite{Thouless:1982}, here a nontrivial spin Chern number indicates a quantized spin Hall conductance.

From the low-energy effective Hamiltonian of a quantum spin Hall insulator, we can find a correspondence between the topology of the system and the sign of the mass of a Dirac fermion. Namely, the topological band-inverted regime is analogous to a massive Dirac fermion with a negative mass while the trivial regime is characterized by a positive mass. 
At a boundary where two regions with opposite mass meet, the mass has to pass through zero, thus forming a domain wall, which can trap gapless states~\cite{Jackiw:1976}.  For a 2DTI, the band inversion leads to a bulk gap with negative mass and the vacuum surrounding it corresponds to a trivial insulator with a positive mass.  Therefore, we have the bulk-boundary correspondence--at the boundary separating the topologically distinct regions, gapless edge states are stabilized~\cite{Fu:2006,Moore:2007,Roy:2009b,Hasan:2010}.

\subsection{Helical channels in various materials}
In addition to HgTe composite quantum wells, the quantum spin Hall effect was predicted in InAs/GaSb heterostructures~\cite{Liu:2008}. Here, it can be described by an extended BHZ model including additional terms induced by the bulk inversion asymmetry and the surface inversion asymmetry. 
These additional terms modify the location of the phase transition between the quantum spin Hall and the trivial insulating phases (phase boundary in the parameter space). However, they do not alter the character of the phases that transit to each other, so the helical edge states are protected by the $\mathbb{Z}_2$ invariant as in HgTe. 
In addition, the heterostructure consisting of an electron layer and a hole layer allows inducing topologically nontrivial phase electrically, by a gate. 
Soon after the theoretical proposals, quantum spin Hall states were reported in experiments in HgTe~\cite{Konig:2007} and InAs/GaSb~\cite{Knez:2011}.
For HgTe, a finite edge conductance is observed when the quantum-well width exceeds a critical value, whereas a narrower well remains insulating. 
For both HgTe and InAs/GaSb, a conductance close to the quantized value was observed for sufficiently short edges.
The nonlocal conductance expected for edge transport was also demonstrated in \cite{Roth:2009} and \cite{Suzuki:2013} for these materials.
Concerning the InAs/GaSb heterostructures, it is known that their material properties strongly depend on the details of fabrication processes. This fact leads to conflicting experimental results, which may cast doubts on their topological nature, as we will discuss in section~\ref{Sec:Exp}. 

There are other quantum spin Hall systems potentially hosting helical edge channels.  
Accompanied by the rapid progress on novel van der Waals heterostructures~\cite{Geim:2013}, the quantum spin Hall effect was predicted in two-dimensional transition-metal dichalcogenides~\cite{Cazalilla:2014,Qian:2014}, including 1T'-WTe$_2$ monolayer~\cite{Tang:2017}, further boosting the community's interest in topological phases of monolayer materials.  
The experimental indication of edge channels was reported in 1T'-WTe$_2$ monolayers~\cite{Tang:2017,Wu:2018}.
On the one hand, spectroscopic observations of the gapless edge channels accompanied by a bulk gap at the Fermi level through the scanning tunneling microscope (STM) and the scanning tunneling spectroscopy (STS) were reported~\cite{Jia:2017}. On the other hand, a contradicting STM study of \cite{Song:2018} concluded instead a semimetal-like gapless bulk band structure.  
Furthermore, topological edge channels were observed in spectroscopic measurements on bismuthene on SiC~\cite{Reis:2017} and ultrathin Na$_3$Bi films~\cite{Collins:2018}, albeit these materials so far lack transport measurements.
Remarkably, there was an experimental indication of a hTLL along the edge channels of bismuthene on SiC~\cite{Stuhler:2019}. 
In comparison to earlier semiconductor-based materials HgTe and InAs/GaSb, the more recent van der Waals heterostructures tend to have a larger bulk gap and thus better topological protection for the edge states.
 Whereas currently there are only a handful of examples of semiconductor-based materials hosting helical channels, we expect the helical liquid to exist in a broader variety of materials.

	\begin{figure}
	\includegraphics[width=0.48\linewidth]{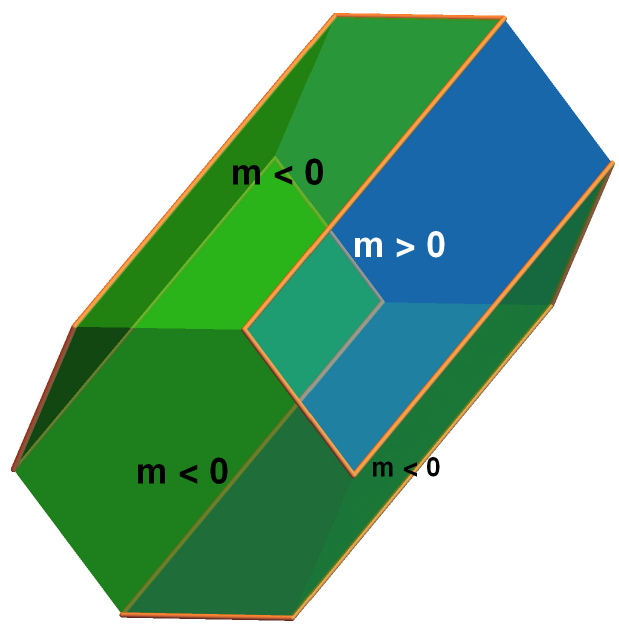} 
	\includegraphics[width=0.48\linewidth]{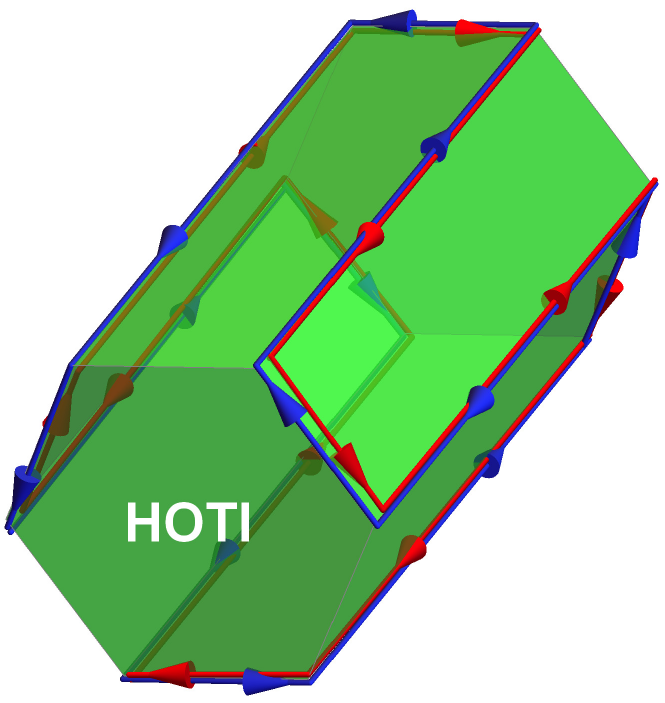} 
	\caption{Illustration of a three-dimensional second-order topological insulator, where the bulk and surface are gapped. (Left) The sign of the gap (Dirac mass) on a given surface, which depends on the surface orientation, is indicated by color. (Right) At the hinges between two surfaces with opposite signs, there are gapless helical states. The spin-up and -down hinge states are separated for clarity. 
	Here we illustrate a nanowire with a hexagonal cross section (realized in, for instance, Bi);
	 for a square cross section, see~\Fref{Fig:HOTI2}. 
	} 
 \label{Fig:HOTI}
	\end{figure}

Going beyond the two-dimensional nanostructures, platforms hosting helical channels include higher-order topological insulators (HOTI)~\cite{Benalcazar:2014,Slager:2015,Benalcazar:2017,Benalcazar:2017b,Ezawa:2018,Okugawa:2019}. 
Relevant to this review are three-dimensional helical second-order topological insulators preserving time-reversal symmetry~\cite{Langbehn:2017,Song:2017,Schindler:2018sa,Khalaf:2018,Geier:2018,Ezawa:2019,Calugaru:2019,Slager:2019,Plekhanov:2020,Fang:2020,Tanaka:2020}, 
where both the three-dimensional bulk and the two-dimensional surfaces are gapped.
One can generalize the above notion of Dirac mass to this three-dimensional system~\cite{Schindler:2020}, 
in which the low-energy theory can still be captured by the Dirac equation.
Each of the surfaces is described by the Dirac equation with a finite Dirac mass.
Distinct from a trivial insulator or a first-order topological insulator, the Dirac mass here depends on the surface orientation, as illustrated in~\Fref{Fig:HOTI}.
In this figure, the colors of the surfaces are assigned according to the sign of the Dirac mass. 
The mass has to change its sign, and therefore go through zero, at a hinge separating two neighboring surfaces of opposite mass.
In analogy to 2DTI, the sign change signifies closing the energy gap and appearance of gapless helical channels. 
In consequence, three-dimensional helical second-order topological insulators are characterized by one-dimensional gapless helical hinge states with opposite spin states propagating in opposite directions, similar to spin-momentum locked edge channels in quantum spin Hall insulators.
	
Experimental indications for HOTI materials have been reported in bismuth (Bi) nanodevices~\cite{Schindler:2018,Murani:2019}, van der Waals stacking of bismuth-halide (Bi$_4$Br$_4$) chain~\cite{Noguchi:2021} and multilayer WTe$_{2}$ in T$_d$ structure~\cite{Choi:2020,Wang:2021}.
The theory so far has not come to a consensus on the identity of the bulk topology in Bi, claiming 2DTI~\cite{Murakami:2006,Wada:2011}, HOTI~\cite{Schindler:2018}, topological crystalline insulator with multiple nontrivial topological invariants~\cite{Hsu:2019b}, or a system at the border between higher-order and first-order (strong) topological insulating phases in a combined theoretical and experimental study of~\cite{Nayak:2019}.
In contrast to the diverse theoretical results, experimental studies are more consistent, showing evidence in favor of edge or hinge channels:
An earlier STM study on locally exfoliated Bi(111) bilayer showed topologically protected transport over edges with length up to hundreds of nanometers~\cite{Sabater:2013}. Additional support on the existence of gapless hinge channels was seen in spectroscopic~\cite{Drozdov:2014,Takayama:2015} and transport~\cite{Murani:2017} experiments.

\subsection{Other variations}
Alternatively to these bulk topological materials, one can produce a spin-selective gap in a (quasi-)one-dimensional spin-degenerate semiconducting nanowire
combining Rashba spin-orbit interactions and magnetic field~\cite{Streda:2003,Pershin:2004,Devillard:2005,Zhang:2006,Sanchez:2008,Birkholz:2009,Rainis:2014,Cayao:2015}. The remaining gapless sector is then formed by a pair of pseudo-helical states.\footnote{In \cite{Braunecker:2012}, these pseudo-helical states are termed ``spiral'' states as opposed to ``helical'' states in a 2DTI.}
The difference between the pseudo-helical and helical states in their spectroscopy was pointed out by \cite{Braunecker:2012}. 
Later, similar approach was adopted to carbon nanotubes, graphene nanoribbons, or 2DTI constrictions~\cite{Klinovaja:2011,Klinovaja:2013x,Klinovaja:2015}.
We, however, do not cover these pseudo-helical states in systems where the time-reversal symmetry is explicitly broken by the magnetic field. We refer the interested reader to recent reviews on this topic~\cite{Prada:2020,Frolov:2020}.

Finally, it was proposed that hTLL can arise in a cylindrical nanowire made of a (strong) three-dimensional topological insulator threaded by a magnetic flux of a half-integer quantum~\cite{Egger:2010}. Alternatively, hTLL were proposed to occur due a nonuniform chemical potential induced by gating across the cross-section of nanowire in the presence of the Zeeman field~\cite{Legg:2021}. However, since we are interested in helical states arising without external magnetic fields, we do not cover this type of setups either. Besides, since we focus on solid-state systems, we do not cover realizations in other types of systems, such as photonic systems~\cite{Ozawa:2019}, nonequilibirum/Floquet systems~\cite{Harper:2020,Rudner:2020}, or magnonic systems~\cite{Nakata:2017}. 
Although we will not cover non-Hermitian systems, which provide an interesting new direction but are beyond the scope of this article, we point out recent reviews~\cite{MartinezAlvarez:2018,Bergholtz:2021} on the role of topology in non-Hermitian systems, including dissipative cold-atom systems, optical setups with gain and loss, and topological circuits.

\begin{figure}[t]
\centering
 \includegraphics[width=\linewidth]{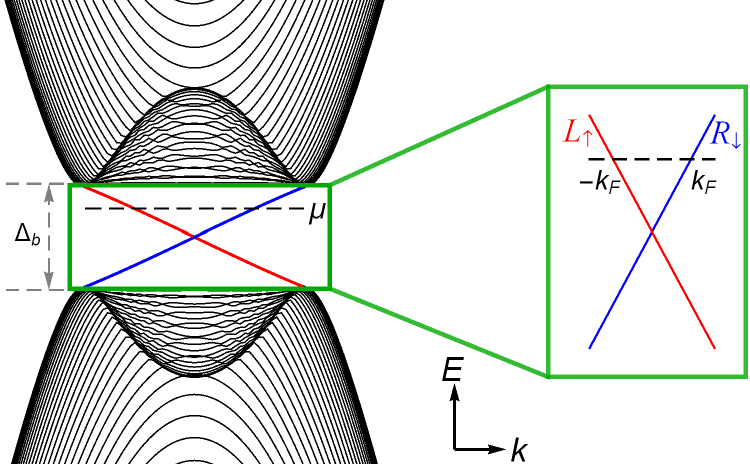}  
\caption{Energy spectrum of a 2DTI, where $k$ is the momentum along a given edge and $\mu$ is chemical potential.
In the energy window within the bulk gap $\Delta_{\rm b}$ near the Dirac point, one consider only the edge states and ignore the bulk states.
For the illustrated spectrum, we have helical edge states composed of right-moving spin-down and left-moving spin-up states.
 }
 \label{Fig:hTLL}
\end{figure}

\section{Interacting helical channels: helical Tomonaga-Luttinger liquid \label{Sec:hTLL}}
After discussing how helical states arise at the edges or the hinges of a topologically nontrivial system, we now turn to their own properties and how they can be characterized. 
Since the helical states in either 2DTI edges or HOTI hinges are spatially confined in one-dimensional channels, one expects strong interaction and correlation effects.
For the usual, nonhelical case, it is well known that elementary excitations in interacting one-dimensional systems are of bosonic nature. Such a system is known as Tomonaga-Luttinger liquid (TLL), which differs strongly from a Fermi liquid describing interacting fermions in higher dimensions~\cite{Haldane:1981,Giamarchi:2003}. 
Adding their helical nature, the edge or hinge states realize a special form of matter, which is named {\it helical Tomonaga-Luttinger liquid} (hTLL) in this review. 
Kane and Mele's proposal on the quantum spin Hall effect motivated investigations on such a helical liquid formed along the edge of the system. It was shown that it embodies a novel class of matter~\cite{Wu:2006,Xu:2006,Dolcetto:2016}, which is distinct from the spinful TLL (formed in spin-degenerate systems such as semiconductor quantum wires) or the chiral TLL (formed in the edge of a fractional quantum Hall system). 
To discuss the properties of the hTLL, we next introduce a description based on the bosonization formalism. 

Let us consider the hTLL located at an edge (a hinge) of a 2DTI (HOTI). At energy scales above the bulk gap, the hTLL description becomes invalid, as demonstrated in quantum Monte Carlo simulations on the Kane-Mele model with Hubbard interaction~\cite{Hohenadler:2012e,Hohenadler:2012}.
We therefore consider the illustrated spectrum in \Fref{Fig:hTLL} and restrict our discussions to energy scales within the bulk gap, where the bulk states are absent.
We note, however, the possibility to have coexisting gapless bulk and edge states in two-dimensional topological systems~\cite{Baum:2015}. 

With the above assumptions, we look at a single helical channel, which can be described as 
\begin{equation}
H_{\rm hel} = H_{\rm kin} + H_{\rm ee},
\label{Eq:H_hel}
\end{equation}
where $H_{\rm kin}$ and $H_{\rm ee}$ are the kinetic energy and electron-electron interactions, respectively.
In this section we assume that the electron spin along the $z$ axis $S^z$ is a good quantum number; we will discuss the generalization to a {\it generic helical liquid}, where $S^z$ is not conserved, in section~\ref{Sec:Scattering_TRS}. 
Here, we further assume that the helical edge or hinge states are formed by right-moving spin-down and left-moving spin-up electrons as in \Fref{Fig:hTLL}. We can thus write 
\begin{equation}
\psi_{\uparrow} (r) = e^{-i k_{F}r} L_{\uparrow} (r), \;\;
\psi_{\downarrow} (r) = e^{i k_{F}r} R_{\downarrow} (r),
\label{Eq:SlowlyVarying}
\end{equation}
with  the slowly varying right(left)-moving fermion field $R_{\downarrow}$ ($L_{\uparrow}$), the coordinate $r$ along the channel, and the Fermi wave vector $k_{F}$ of the helical states (measured from the Dirac point). From now on we will suppress the coordinate and the spin index unless it may cause confusion.
The kinetic-energy term reads
\begin{eqnarray}
H_{\rm kin} &=& -i \hbar v_{F} \int dr \, \left(R^{\dagger} \partial_{r} R - L^{\dagger} \partial_{r} L \right),
\label{Eq:H_kin}
\end{eqnarray}
with the Fermi velocity $v_{F}$. The electron-electron interaction term is given by 
\begin{eqnarray}
H_{\rm ee} &=& g_{2} \int dr \, R^{\dagger} R L^{\dagger} L \nonumber \\
&& + \frac{g_{4}}{2} \int dr \, \left[ \left(R^{\dagger}  R \right)^2 + \left( L^{\dagger}  L \right)^2 \right],  
\label{Eq:H_ee}
\end{eqnarray}
where $g_{2}$ and $g_{4}$ are the interaction strength describing the forward scattering processes. 
The interaction term \eref{Eq:H_ee} describes short-range interactions and is valid for systems in which Coulomb interactions between particles are screened by, for instance, electrons in a nearby metal gate. 

The fermion operators can be expressed in terms of the bosonic fields $(\theta,\phi)$,
\begin{eqnarray}
\eqalign{
R (r) &= \frac{U_{R}}{\sqrt{2\pi a}} e^{i[-\phi(r) + \theta(r)]}, \\
L (r) &= \frac{U_{L}}{\sqrt{2\pi a}} e^{i[\phi(r) + \theta(r)]},}
\label{Eq:bosonization}
\end{eqnarray}
where $U_{R/L}$ is the Klein factor and  $a = \hbar v_{F} / \Delta_{b}$ is the short-distance cutoff, which is associated with the high-energy cutoff set by the  bulk gap $\Delta_{b}$.
The bosonic fields satisfy
\begin{eqnarray}
\left[ \phi(r), \theta(r')\right] &=& i \frac{\pi}{2} \textrm{sign}(r'-r), 
\label{Eq:Commutation}
\end{eqnarray}
indicating that the field $\partial_r \theta / \pi$ is canonically conjugate to $\phi$. 
With \eref{Eq:bosonization}, the helical channel Hamiltonian can be bosonized as
\begin{eqnarray}
H_{\rm hel} &=& \frac{\hbar u}{2\pi} \int dr \, \left[  \frac{1}{K} \left( \partial_r \phi \right)^2 + K \left( \partial_r \theta \right)^2 \right],
\label{Eq:hTLL}
\end{eqnarray}
where the velocity $u$ and the interaction parameter $K$ are given by 
\numparts
\label{Eq:u&K}
\begin{eqnarray}
u &\equiv& \left[ \left(  v_{F} + \frac{g_{4}}{2\pi \hbar} \right)^2 - \left( \frac{g_{2}}{2\pi \hbar} \right)^2 \right]^{ \frac{1}{2}}, \\
K &\equiv& \left( \frac{2\pi \hbar v_{F} + g_{4} - g_{2}}{2\pi \hbar v_{F} + g_{4} + g_{2}} \right)^{ \frac{1}{2}}.
\end{eqnarray}
\endnumparts
For repulsive electron-electron interactions ($g_{2},~g_{4}>0$), we have $K<1$. For existing materials, the interaction parameter was estimated in theory: $K \approx 0.2$ for edge states in InAs/GaSb heterostructures~\cite{Maciejko:2009}, $K \approx 0.53$--$0.9$ for HgTe quantum wells~\cite{Hou:2009,Strom:2009,Teo:2009}, 
$K \approx 0.43$--$0.5$ for strained InAs/(Ga,In)Sb devices~\cite{Li:2017}, $K \approx 0.4$--$0.6$ for hinge channels in a bismuth HOTI~\cite{Hsu:2018} and $K \approx 0.4$--$0.6$ for bismuthene on SiC~\cite{Stuhler:2019}.

As we see in \eref{Eq:hTLL}, the bosonized Hamiltonian is quadratic in the bosonic fields and can, therefore, be exactly diagonalized. In the bosonic language, one can thus perform calculations that are nonperturbative in the electron-electron interaction strength, including the renormalization-group (RG) analysis.
Since the quantity $K$ parametrizes the strength of Coulomb interactions between electrons in the helical channel, it serves as a crucial parameter for the RG relevance of various interaction-induced and renormalized scattering processes, as well as for the interaction-stabilized topological bound states, which we will discuss in the following sections. It is, therefore, important to experimentally quantify this parameter in realistic settings. 
However, deducing the interaction parameter is tricky, especially from the---most common---dc transport measurements.
First, for a clean hTLL that is free from backscattering and adiabatically connected to Fermi-liquid leads, it was found that the dc conductance of the helical channel does not depend on $K$ \cite{Hsu:2018b}, resembling the ballistic conductance in a nonhelical channel~\cite{Maslov:1995,Ponomarenko:1995,Safi:1995}. 
Second, while in the presence of backscattering sources the conductance through a helical channel in general depends on the interaction strength, knowledge of the backscattering mechanism and resistance sources is required to extract $K$. This complication makes the extraction of the experimental $K$ value highly nontrivial, as pointed out by~\cite{Vayrynen:2016,Hsu:2017,Hsu:2019}.

One may, therefore, consider alternative probes. For instance, the ac conductivity $\sigma_{\rm ac} (\omega)$ of a hTLL can be measured optically without the influence of the leads. 
The real part of $\sigma_{\rm ac}$ shows a zero-frequency Drude peak with the weight depending on the interaction strength~\cite{Hsu:2018b,Meng:2014b}.
Alternatively, one can search for spectroscopic signatures by probing the local density of states, which exhibits a scaling behavior as a function of energy $E$ and temperature $T$~\cite{Stuhler:2019}, 
\begin{eqnarray}
\frac{
\rho_{\rm dos}  } {T^{\alpha_{\rm dos}} }
 &\propto & \cosh\left( \frac{ E }{2k_{\rm B} T} \right)  
 \left| \Gamma \left( \frac{ 1 + \alpha_{\rm dos}}{2} + i \frac{ E }{2 \pi k_{\rm B} T} \right)\right|^2, \nonumber \\ 
 \label{Eq:DOS} 
\end{eqnarray}
with the Boltzmann constant $k_{\rm B}$ and the interaction-dependent parameter $\alpha_{\rm dos} = (K + 1/K)/2 - 1$.
Remarkably, this formula not only provides spectroscopic signature for a hTLL, but also allows for the extraction of the interaction parameter $K$.
This universal scaling behavior was indeed observed on the edge of bismuthene on SiC through STS measurements~\cite{Stuhler:2019}, and the deduced value of $K$ was in good agreement with a theoretical estimation.
At $T=0$, the expression reduces to a power-law density of states depending on energy $\rho_{\rm dos} \propto |E|^{\alpha_{\rm dos}}$ as found earlier in the zero-temperature calculation~\cite{Braunecker:2012}. 

As an alternative, \cite{Ilan:2012} proposed a setup to extract $K$ by measuring the edge current through an artificial quantum impurity, which is realized by combining a local gate and an external magnetic field. Since the artificial impurity acts as a backscattering center with experimentally controllable strength, one can determine $K$ by fitting the edge current to the analytical expression.
Finally, \cite{Muller:2017} proposed a dynamical approach to determine the interaction parameter of the 2DTI edge states, from either time-resolved transport measurements with sub-nanosecond  resolution or the frequency dependence of the ac conductance.

While in principle the above proposals allow one to extract the $K$ value, using conventional---and thus well established---experimental probes seems more practical. 
To this end, \cite{Braunecker:2018} proposed the double-edge momentum conserving tunneling spectroscopy. It utilizes a setting analogous to the double-wire tunneling spectroscopy based on cleaved edge overgrowth GaAs quantum wires~\cite{Auslaender:2002,Tserkovnyak:2002,Tserkovnyak:2003,Patlatiuk:2018,Patlatiuk:2020}, which has been used to detect the spinful TLL in semiconductor quantum wires. Adopting it for helical channels, it requires a setup using a pair of controlled parameters: an applied bias voltage between edges of two 2DTIs and a flux induced by an external magnetic field penetrating in-between the edges. 
The bias voltage shifts the spectra of the two edges in energy, while the magnetic flux shifts them in momentum. A tunneling current then flows between the edges whenever the flux and bias meet the conditions for the energy and momentum conservation, leading to an oscillating tunneling conductance as a function of the magnetic field and the bias voltage.
The oscillation period allows one to deduce the parameter $K$ of a double-edge system with similar interaction strength.
\cite{Hsieh:2020} extended the calculation of \cite{Braunecker:2018} to finite temperature and the presence of disorder, providing a systematic analysis on the low-energy spectral function and the tunneling current. 

While the hTLL description \eref{Eq:hTLL} allows for investigations on strongly interacting systems and is adopted in many works, there exist studies that do not rely on the hTLL picture and bosonization. They focus on weakly interacting systems adequately described by the fermionic language, including generic helical liquids, which we will discuss in section~\ref{SubSec:ghl}.

\section{Charge transport of a helical channel \label{Sec:Transport}}
The presence of one-dimensional helical channels can be experimentally examined through their charge transport: A quantized conductance $G_0 = e^2 /h$ per channel is expected when the chemical potential lies within the bulk gap. 
In addition, the charge transport has direct implications for applications in electronics and spintronics. 
Below we first review the experimental progress, before discussing the theoretical results on charge transport. 

\subsection{Experiments on edge transport \label{Sec:Exp}}
There are a number of transport measurements on 2DTI edge channels.
 For sufficiently short channels, the expected ballistic value was observed in earlier studies on HgTe~\cite{Konig:2007}, InAs/GaSb~\cite{Knez:2011} and 1T'-WTe$_2$ monolayers~\cite{Wu:2018}, along with observations consistent with nonlocal edge transport in these materials~\cite{Roth:2009,Suzuki:2013,Fei:2017}.
Additional experimental features for helical channels include spin polarization of the edge states~\cite{Brune:2012} and real-space imaging of edge current in HgTe~\cite{Nowack:2013} and InAs/GaSb~\cite{Spanton:2014} based on microscopic superconducting quantum interference device (SQUID). 
However, in contrast to the well quantized conductance of chiral edge channels in quantum Hall states~\cite{Klitzing:1980,Klitzing:2017},  
 imperfect quantization of the edge conductance was seen in longer samples.
Moreover, the scanning gate microscopy identified individual scattering centers~\cite{Konig:2013}, which may originate from metallic puddles formed in inhomogeneous potential landscape.
These observations triggered further studies on charge transport of the potential 2DTI materials; such experiments were reviewed in~\cite{Gusev:2019,Culcer:2020}. 

In most settings, the low-temperature conductance or resistance was weakly temperature dependent,
 for both HgTe~\cite{Konig:2007,Gusev:2011,Grabecki:2013,Gusev:2014,Olshanetsky:2015,Bendias:2018} and InAs/GaSb~\cite{Suzuki:2013,Knez:2014,Suzuki:2015,Du:2015}. 
In addition, a peculiar fractional power-law conductance was observed in InAs/GaSb, which was attributed to hTLL signatures~\cite{Li:2015}. 
Furthermore, reproducible quasiperiodic fluctuations of both local and nonlocal resistance as functions of gate voltage observed in HgTe~\cite{Grabecki:2013} became less pronounced upon increasing the temperature, consistent with the expectations from charge puddles present in narrow-gap semiconductors with inhomogeneous energy landscape. 
A more recent study on HgTe demonstrated temperature-induced phase transition between the 2DTI and trivial insulating phases~\cite{Kadykov:2018}.
For a 100-nm channel of WTe$_2$ monolayer, the range for the temperature-insensitive conductance persists even up to 100~K~\cite{Wu:2018}, which is much higher than in the semiconductor heterostructures and is consistent with the theoretically predicted large topological gap.

Concerning nonlocal transport, measurements on HgTe have been performed for samples of different sizes.
On the one hand, the edge conductance is well described by the Landauer-B{\"u}ttiker formula for submicron-size samples in the ballistic regime~\cite{Roth:2009}.
On the other hand, nonquantized resistance was observed for larger samples in the diffusive regime~\cite{Grabecki:2013} and for even larger samples with channel lengths (perimeters of the samples) in the order of millimeter~\cite{Gusev:2011,Gusev:2013,Olshanetsky:2015}. 
By fabricating lateral $p$-$n$ junctions in wide HgTe quantum wells with thickness of 14~nm, \cite{Piatrusha:2017} observed highly linear current-voltage characteristics, indicating transport via ballistic edge states. 
For InAs/GaSb, \cite{Suzuki:2013} was able to observe dominant nonlocal edge transport in a micrometer-long device by optimizing the InAs layer thickness, along with reproducible resistance fluctuations in the gate voltage dependence, indicating multiple scatterers along the channel.  
Utilizing a dual-gate device, \cite{Suzuki:2015} monitored the transition from the semimetallic to the 2DTI phase through the nonlocal resistance measurements.  
\cite{Mueller:2015} observed nonlocal edge transport with the resistance systematically below the expected quantized values, indicating a residual bulk conduction.
For WTe$_2$, \cite{Fei:2017} demonstrated edge conductance in monolayer devices present over a wide range of gate voltage and temperature. On the other hand, the bilayer devices showed insulating behavior without a sign of edge conduction. 

As mentioned above, the helical states arise from a time-reversal-invariant system. 
To examine how they respond to broken time-reversal symmetry, one applies the external magnetic field. Measurements showed anisotropy with respect to the field orientation. 
For the ballistic regime of HgTe, \cite{Konig:2007} observed a sharp cusp-like conductance peak centered at zero for an out-of-plane magnetic field $B_{\perp}$ (perpendicular to the two-dimensional heterostructure) and a much weaker field dependence for an in-plane magnetic field $B_{||}$.
The observation was attributed to an anisotropic Zeeman gap opening in the edge spectrum.
\cite{Gusev:2011} examined a HgTe sample in the diffusive regime under both $B_{\perp}$ and $B_{||}$.  
They observed an increasing resistance with a small $B_{\perp}$, which developed a peak and eventually was suppressed by a strong field, probably due to a transition to a quantum Hall state. 
On the other hand, the same reference reports a large positive magnetoresistance for $|B_{||}| < 2~$T, in contrast to the ballistic samples.  
For $|B_{||}| > 6 ~$T, both local and nonlocal resistances quickly dropped with $|B_{||}| $. For  $|B_{||}| >10~$T, local resistance saturated while the nonlocal one vanished, suggesting the emergence of conductive bulk states and thus a field-induced transition to a conventional bulk metal state.
Similar results followed in their subsequent work~\cite{Gusev:2013}, which observed a decrease of the local resistance by $B_{||}$, accompanied by the complete suppression of nonlocal resistance by $B_{||} \approx 10~$T, consistent with a transition from 2DTI to a metallic bulk. 
A systematic investigation of the field dependence of the edge transport was carried out by~\cite{Ma:2015}, in combination with scanning microwave impedance probe allowing for detection of the local electromagnetic response. Unexpectedly, whether $B_{\perp}$ suppresses the edge conductance depends on the position of the chemical potential with respect to the charge neutrality point.   
On the $p$-doped side, the edge conduction is gradually suppressed by applying $B_{\perp}$, whereas the edge conduction on the $n$-doped side shows little changes up to 9~T, contradicting the theoretical expectation. 
A recent study showed an almost perfectly quantized edge conductance in a 6~$\mu$m-long edge at the zero magnetic field, which was strongly suppressed by a small field of $B_{\perp}=50~$mT~\cite{Piatrusha:2019}. 
With the field-induced broken time-reversal symmetry, they observed the exponential temperature dependence of conductance indicating Anderson localization for the edge states, along with reproducible mesoscopic resistance fluctuations as a function of gate voltage and the gap opening in the current-bias characteristics. 

Concerning different materials, surprisingly robust edge transport against magnetic fields was reported for InAs/GaSb.
A weak $B_{||}$-dependence of edge conductance was observed~\cite{Du:2015}, which remained approximately quantized up to 12~T.
Nonetheless, magnetic fields can still modify the edge transport upon combining them with electric fields.  
\cite{Qu:2015} demonstrated the tunability of the InAs/GaSb system through a dual-gate setting and magnetic fields. 
Utilizing the top and back voltage gates, which control the Fermi level and the relative alignment between the electron and hole bands, together with $B_{||}$, which shifts the two bands in momentum, they established the phase diagram by measuring the local resistance.  
In contrast to the semiconductor heterostructures, the edge conductance in WTe$_2$ can be suppressed exponentially by either $B_{||}$~\cite{Fei:2017} or $B_{\perp}$~\cite{Wu:2018}, demonstrating the Zeeman gap opening in the edge spectrum. 

The peculiar features in HgTe and InAs/GaSb under magnetic fields reported by \cite{Ma:2015,Du:2015} motivated theoretical works on ``Dirac point burial'' or ``hidden Dirac point''~\cite{Skolasinski:2018,Li:2018}.
Namely, detailed band structure calculations in these theoretical works revealed a Dirac point buried within the bulk valence or conduction bands.
It results in a hidden Zeeman gap responsible for the robust edge-state transport against magnetic fields.

In addition to the unexpected temperature and field dependencies, there exist other puzzles in experiments, including a more recent observation of the localization of HgTe edge channels with length of $O(1~\mu$m)--$O(10~\mu$m) in the absence of magnetic fields~\cite{Bubis:2021}.
Particularly for InAs/GaSb, the residual bulk conduction in parallel to the edge transport was observed since its first 2DTI demonstration~\cite{Knez:2011} and later on confirmed by scanning SQUID microscopy~\cite{Nichele:2016}.
It has motivated subsequent works on intentional impurity doping, either through Ga source materials with different impurity concentrations~\cite{Charpentier:2013} or Si doping~\cite{Knez:2014,Du:2015}.
By suppressing the bulk conduction with disorder, quantized edge conductance with a deviation of about 1~$\%$ was observed in a wide temperature range~\cite{Du:2015}. 
Alternatively, \cite{Couedo:2016} employed specific sample design, using a large-size device with asymmetric current path lengths, to electrically isolate a single edge channel and observed a conductance plateau close to the quantized value.  

There are also puzzles in the trivial regime of InAs/GaSb, where the energy bands are not inverted. In contrast to the theoretical expectation, \cite{Nichele:2016} observed edge channels even in the trivial regime in the scanning SQUID microscopy. 
This observation was further confirmed in a device with the Corbino geometry, in which the conduction through the bulk and edge states were decoupled~\cite{Nguyen:2016}.
Together with the resistive transport (the resistance growing with the edge length) reported in \cite{Nichele:2016}, the edge states are likely trivial and thus nonhelical;  
they might share the same origin as the counterpropagating edge transport observed in the quantum Hall regime of the InAs quantum well, as a result of the Fermi-level pinning and carrier accumulation at the surface~\cite{Akiho:2019}. 
The observation of \cite{Nichele:2016} motivated a systematic investigation~\cite{Mueller:2017} on length dependence of the edge resistances in the nominally topological InAs/GaSb and trivial InAs materials. 
Importantly, the two systems showed similar resistances with linear dependence on edge length, clearly indicating the presence of multiple resistance sources.
More recently, \cite{Shojaei:2018} assessed the disorder effects due to charge impurities and interface roughness of a dual-gate device by measuring the temperature and magnetic field dependence of resistance. 
Upon tuning the system from the band-inverted to the band-normal regime, the conduction resembles the behavior of a disordered two-dimensional system.
Therefore, even with the most advanced fabrication, potential fluctuations are sufficiently strong to destroy the topological state  as a result of the small hybridization gap. 

Motivated by a better understanding on the topological nature of the InAs/GaSb heterostructure, \cite{Akiho:2016,Du:2017} investigated strained InAs/(Ga,In)Sb composite quantum wells, aiming at engineering the band structure.
The lattice mismatch between the (Ga,In)Sb and AlSb layers in the heterostructure induces a strain in the former layer, which enhances the hybridization between the electron band of InAs and the hole band of (Ga,In)Sb.   
As a result, the bulk gap was increased by about two orders in comparison with the gap of InAs/GaSb.
The edge transport of the strained material was subsequently confirmed by~\cite{Du:2017}, revealing weak temperature dependence which can be fitted in a logarithmic form~\cite{Li:2017}.
Using the strained material, \cite{Irie:2020} demonstrated the gate-controlled topological phase transition. They also observed a tunable bulk energy gap, the magnitude of which can reach above the room temperature in highly strained quantum wells.
\cite{Irie:2020b} investigated the effects of epitaxial strain on the topological phase transition of InAs/GaSb by combining different substrates and buffer layers.
Upon inducing biaxial tensile strain in the GaSb layer, they observed that the strain can close the indirect gap of the bulk spectrum, resulting a semimetallic behavior even in the band-inverted regime. 

While these great experimental efforts finally led to the observation of well quantized edge conductance through gate training in macroscopic HgTe samples with edge length in the order of $100~\mu$m~\cite{Lunczer:2019}, there are apparent discrepancies among the above experimental results, as well as between the observations and theoretical predictions.
Motivated by these observations, numerous backscattering mechanisms were proposed.   
In addition to being resistance sources, since backscattering mechanisms might potentially localize the edge states and open a gap in the edge or hinge energy spectrum, it is also important to examine the stability of the helical states against them.  
In the following subsections, we group the mechanisms considered in the literature 
according to the time-reversal symmetry.
We discuss mechanisms that break the time-reversal symmetry in section~\ref{Sec:Scattering_noTRS}, and those that preserve it in section~\ref{Sec:Scattering_TRS}.

In theoretical works, the charge transport is discussed in terms of various physical quantities, including ac/dc conductivity, resistance, conductance correction, or backscattering current.
In order to make a systematic discussion, we focus on the dc limit and convert all the expressions in the following way.
When a given mechanism involves multiple scatterers with the scatterer number increasing with the channel length, we summarize the results in terms of the resistance $R$, including the linear-response resistance in the low-bias regime and differential resistance $dV/dI$ in the high-bias regime (we will use the same notation for simplicity). 
For mechanisms arising from a single backscattering source, the resistance does not scale with the edge length, which prompts us to express the results in terms of the (differential) conductance correction $\delta G<0$.\footnote{In an edge of intermediate length, the channel resistance $R$ might be comparable to the contact resistance $1/G_{0} = h/e^2$, with the total series resistance $h/e^2 + R$. 
The two quantities ($R$ and $\delta G$) are related through  $\delta G \propto - R G_0^2$.} 

Below we assume that the bulk samples are sufficiently large so that parallel edges or hinges of the system are well separated, allowing us to focus on a single channel. 
After discussing various backscattering mechanisms and the resulting formulas for $R$ or $\delta G$ as functions of the temperature $T$, bias $V$, and channel length $L$, we will summarize their temperature dependence in \Tref{Tab:mechanism} and \Tref{Tab:mechanism2} for time-reversal symmetry breaking processes and time-reversal-invariant processes, respectively.

\subsection{Mechanisms with broken time-reversal symmetry \label{Sec:Scattering_noTRS}} 
Since the nontrivial topology of 2DTI and helical HOTI relies on the time-reversal symmetry, it is natural to ask how the helical channels are affected by perturbations that explicitly break the time-reversal symmetry.  

\subsubsection{Single magnetic impurity \label{SubSec:Single} } 
~\\ 
A simple way to break the time-reversal symmetry is to introduce a single magnetic impurity in the helical channel.
\cite{Wu:2006,Maciejko:2009} considered a localized spin impurity (called Kondo impurity) with spin $I=1/2$ and an exchange coupling to the helical edge states in the following form,
\begin{eqnarray}
H_{\rm K}  &=& \sum_{\mu} \int dr \; J_{\mu} S^{\mu} (r) I^{\mu} \delta (r),  
\end{eqnarray} 
with the $\mu$ component of the electron spin ${\bf S}$ and the localized impurity spin ${\bf I}$ located at the origin. 
In terms of the right-and left-moving fields [see \eref{Eq:SlowlyVarying}], the electron spin operators read
\numparts
\begin{eqnarray}
 S^{x} = \frac{1}{2} \big( R^{\dagger} L + L^{\dagger}  R  \big), \;
 S^{y} = \frac{i}{2} \big( R^{\dagger} L - L^{\dagger}  R  \big), \label{Eq:Spin}  \\
 S^{z} = \frac{1}{2} \big( L^\dagger L - R^\dagger R \big),
\end{eqnarray} 
\endnumparts
where we omit the spatial coordinate.  
For isotropic transverse coupling $J_x=J_y$, the Kondo coupling can be expressed as
\begin{eqnarray}
H_{\rm K}  &=& \frac{1}{2} \int dr \; \delta (r) \Big[  J_{\perp } \big( I^{+} R^\dagger L +  I^{-} L^\dagger R \big) \nonumber \\
&& \hspace{52pt}  + J_z I^z \big( L^\dagger L - R^\dagger R \big) \Big],
\label{Eq:H_K}
\end{eqnarray}
with $I^{\pm} = I^x \pm  i I^y $. 
The transverse coupling $J_{\perp} = (J_x + J_y)/2$ leads to a spin-flip backscattering between the right- and left-moving states, and the $z$ component describes the forward scattering, which modifies the effective strength of the electron-electron interactions but does not affect the charge transport directly. 
The interaction $H_{\rm K}$, derived for $J_x = J_y$, conserves the $z$ component of the total spin of the electron and impurity. For anisotropic transverse coupling $J_x \neq J_y$, there can be additional terms breaking this conservation. 

In the strong-coupling regime, the formation of the Kondo singlet at the impurity site screens the spin and effectively removes the impurity from the helical channel at low temperature.
In consequence, the helical channel follows the deformed boundary of the underlying lattice due to its topological nature. One thus expects no conductance correction in this regime~\cite{Maciejko:2009}. 
In the weak-coupling regime where $J_{\perp}$ and $J_{z}$ are small parameters, \cite{Maciejko:2009} employed the Kubo formula and found that the Kondo scattering induces a power-law correction to the edge conductance at low temperature,
\begin{eqnarray}
\delta G_{\rm K}  & \propto & - \frac{e^2}{h} (\nu_0 J_{\perp})^2 \left( \frac{k_{\rm B} T}{\Delta_{\rm b}} \right) ^{2 \tilde{K} -2},
\label{Eq:G_Maciejko1}
\end{eqnarray}
where the modification in $K$ due to the forward-scattering term is given by 
\begin{eqnarray}
\tilde{K} & \propto & K \left(1 - \frac{\nu_0 J_z}{2K} \right)^2 , 
\end{eqnarray}
with $\nu_0 = a / (\pi \hbar u)$ being the density of states of the hTLL. 
When the temperature is above the Kondo temperature $T_{\rm K}$, one has, instead, a logarithmic correction
\begin{eqnarray}
\delta G_{\rm K}  & \propto & - \frac{e^2}{h} \left[ \eta_{\rm K}  + \gamma_{\rm K}  \ln \left( \frac{\Delta_{\rm b}}{ k_{\rm B} T} \right) \right],
\label{Eq:G_Maciejko2}
\end{eqnarray}
with $\eta_{\rm K}$ and $ \gamma_{\rm K}$ depending on the interaction parameter $K$ and the bare couplings $J_{\perp}$ and $J_{z}$.

In contrast, \cite{Tanaka:2011} pointed out that, for the isotropic transverse coupling \eref{Eq:H_K}, the result \eref{Eq:G_Maciejko1} does not hold in the dc limit $(\omega \to 0)$.
It is because the Kubo formula implemented in the calculation relies on a perturbation expansion in the Kondo couplings, which is valid for sufficiently high frequency where $J_{\perp,z} \ll \hbar\omega$ but not at zero frequency.  
Instead, \cite{Tanaka:2011} concluded that, for a single magnetic impurity which has an isotropic transverse coupling to the helical states, the dc conductance correction vanishes regardless of the formation of the Kondo singlet and the conductance quantization is preserved. 
 
The physical reason behind this property can be understood from the fact that in the isotropic limit the electron-impurity-spin coupling conserves the sum $S^z+I^z$ of the electron and the magnetic impurity spin components. An electron spin flip necessary for backscatterings must be accompanied by the flop of the magnetic impurity.   
Since a magnetic impurity with the spin $I$ can be flopped only up to $2I$ times, in a transport measurement with the electric current flowing through a helical channel, the magnetic impurity eventually becomes completely polarized and cannot allow for subsequent backscatterings without a depolarization mechanism. 
Therefore, a steady-state backscattering current cannot be sustained by a single magnetic impurity and thus the dc conductance remains quantized. 
To assess the influence of the single magnetic impurity on the helical states, one has to probe the system in the finite-frequency regime. Alternatively, one needs either a depolarization mechanism or a non-spin-conserving interaction (arising from, for instance, anisotropic transverse coupling $J_x \neq J_y$) in order to have observable effects in dc transport.

The situation changes when one includes the spin-orbit interaction (SOI). It can be induced, for instance, by an electric field of a gate on top of the heterostructure. 
\cite{Eriksson:2012} studied the effects of spatially uniform Rashba SOI on the helical states coupling to a Kondo impurity; 
a subsequent work considering additionally the Dresselhaus term reached the same conclusion~\cite{Eriksson:2013}.
Upon a unitary transformation that removes the SOI terms, the spin-spin interaction becomes non-spin-conserving (via an anisotropic coupling)
and generates additional noncollinear terms in the form of 
\begin{eqnarray}
S^{y'}I^{z'},\; S^{z'}I^{y'}, \; \cdots
\label{Eq:Eriksson}
\end{eqnarray}
 with $S^{y',z'}$ being the $y',z'$ component of the electron spin ${\bf S}$ in the transformed frame  
 at the impurity site.
As these Rashba-induced terms disfavor the formation of a Kondo singlet, the Kondo screening is suppressed.
As a consequence of the SOI, the Kondo impurity leads to a correction similar to \eref{Eq:G_Maciejko1}, but with the prefactor and the parameter $\tilde{K}$ modified by the SOI, along with subleading terms of higher order in $T$.
Interestingly, in contrast to the general expectation that the SOI does not affect the Kondo effect, here the Kondo screening gets obstructed by the Rashba-induced noncollinear coupling in certain parameter regimes.   
In consequence, in this setting the Rashba effect can be exploited to externally control the Kondo screening and $T_{\rm K}$.   

\cite{Zheng:2018} studied a single magnetic impurity with $I=1/2$ strongly coupled to the electrons above $T_{\rm K}$. 
Starting from the Kane-Mele model in two dimensions, they mapped the single magnetic impurity problem to a generalized Fano model, which describes an interacting one-dimensional channel coupled to resonant levels.
 As a result, the edge state electrons can tunnel into the resonant levels at the impurity site,
 thus causing a suppression of the transmission through the edge channel at low temperature.
Backscattering in this mechanism is allowed even without electron-electron interactions, in contrast to charge puddles, which we discuss in section~\ref{Sec:Scattering_TRS}.

More recently, \cite{Vinkler-Aviv:2020} made a detailed numerical analysis on the single magnetic impurity problem. 
Specifically, they employed time-dependent numerical RG method for a local spin $I=1/2$ coupling to helical states with a general exchange coupling tensor to compute the low-temperature conductance over a wide range of bias voltage and coupling strength.  
In the low-energy limit, the RG analysis shows that the exchange couplings tend to flow to the isotropic fixed point, restoring the $U(1)$ symmetry associated with the $S^z$ conservation. In this limit, the formation of the Kondo singlet protects the conductance quantization, consistent with earlier findings in the Kondo regime~\cite{Maciejko:2009} or the isotropic limit~\cite{Tanaka:2011}.  
In typical measurements, however, finite temperature or bias voltage can stop the RG flow earlier, with the characteristic energy scale set by $T_{\rm K}$, thus destroying the conductance quantization. 

In addition to these works on impurity $I=1/2$, there are theoretical works on a single magnetic impurity with large spin, particularly relevant for HgTe doped with Mn$^{2+}$ ions, which has a spin of $I = 5/2$~\cite{Furdyna:1988,Novik:2005}.
By extending the analysis of \cite{Eriksson:2012} to spatially non-uniform Rashba SOI,  
\cite{Kimme:2016} demonstrated that a single Kondo impurity with an anisotropic coupling that breaks the $S^z$ symmetry can result in backscattering. 
In particular, in the nonlinear regime where the bias voltage is larger than the thermal energy ($eV>k_{\rm B} T$), such a scattering causes a resistance increasing with $eV/ k_{\rm B} T$ for $I > 1/2$.
Furthermore, \cite{Kurilovich:2019b} considered a noninteracting channel coupled to an impurity with $I>1/2$, with a general form of the anisotropic exchange coupling.  In particular, they incorporated the electron-mediated indirect exchange interaction of the magnetic impurity with itself,  
\begin{eqnarray}
 D_x (I^x)^2 + D_z (I^z)^2 ,  
\label{Eq:Kurilovich}
\end{eqnarray}
with couplings $|D_z| > |D_x|$.
Such a local anisotropy term breaks the $I^z$ conservation.   
It can significantly affect the conductance for integer $I$ and easy-plane anisotropy, leading to a $T^4$ power-law correction.
In other cases, the correction has weak temperature dependence at low temperatures.

In summary, a single magnetic impurity can cause conductance correction when the Hamiltonian breaks the $U(1)$ axial rotational symmetry associated with the conservation of the spin component $S^z$ of the electrons and $I^z$ of the impurity. As this symmetry can be naturally broken in typical heterostructures, in general one expects nonvanishing effects on the charge transport.

\subsubsection{Ensemble of magnetic impurities \label{SubSec:Ensemble}}
~\\ 
Having discussed the effects of a single magnetic impurity, it is natural to ask how an ensemble of magnetic impurities influences the edge transport.
To this end, \cite{Altshuler:2013,Yevtushenko:2015} considered a one-dimensional array of spin-$1/2$ Kondo impurities interacting with the helical states with random anisotropic couplings. The total spin conservation along $z$ is violated by the random anisotropies in the coupling constants. Focusing on noninteracting electrons and at $T=0$, \cite{Altshuler:2013} concluded that in general the edge states can be localized and estimated the localization length as a function of the Kondo coupling $J_{\rm ka}$ and the strength of the anisotropy disorder $D_{\rm ka}$. 
As a side remark, \cite{Maciejko:2012} investigated also a Kondo lattice on the edge of a 2DTI. Instead of the transport, that work focused on the stability of the helical states themselves. It found that new phases emerge in the regime $\nu_0 J_z \ge 2K$, which is not accessible perturbatively in $\nu_0 J_z$, resulting in a rich zero-temperature phase diagram. 

\cite{Yevtushenko:2015} generalized the calculation of \cite{Altshuler:2013} to interacting electrons at finite temperature and showed that the conclusion holds also for interacting systems, with the renormalized gap $ \Delta_{\rm ka} \propto \Delta_{\rm b} ( J_{\rm ka} / \Delta_{\rm b} )^{1/(2-K)}$.  
In addition, their findings can be summarized in terms of the resistance in different temperature regimes.
Below the localization temperature, the channel is insulating, where the resistance $ R_{\rm ka}$ grows to infinity for long samples.
When the temperature is increased but still below the classical depinning energy 
\begin{eqnarray}
 E_{\rm pin} = \left[ \frac{ \sqrt{D_{\rm ka}} \Delta_{\rm ka}^2 }{ 4\pi K \Delta_{\rm b}^2 } \ln \left( \frac{\Delta_{\rm b}}{\Delta_{\rm ka } }  \right)  \right]^{2/3},
 \label{Eq:E_pin}
\end{eqnarray}
the channel exhibits transport features similar to glassy systems. For a transient system where the sample size is close to the localization length, 
the resistance is almost independent of the temperature, while the theory for long and short systems is still lacking. 
Further increasing the temperature gives
\begin{eqnarray}
R_{\rm ka}^{\rm low}  & \propto & \frac{h}{e^2}  \frac{L E_{\rm pin} }{\hbar v_F} \left( \frac{E_{\rm pin}}{k_{\rm B}T} \right)^{2} 
\label{Eq:R_Yevtushenko1}
\end{eqnarray}
for $ E_{\rm pin} < k_{\rm B} T \ll \Delta_{\rm ka}$, with the energy scale $E_{\rm pin}$ separating the quantum and semi-classical regimes.
Finally, for higher temperature $k_{\rm B} T > \Delta_{\rm ka}$, one has
\begin{eqnarray} 
R_{\rm ka}^{\rm hi}  & \propto & \frac{h}{e^2} \frac{L}{a}  \left( \frac{ \Delta_{\rm b} }{k_{\rm B}T} \right)^{2-2K} .
\label{Eq:R_Yevtushenko2}
\end{eqnarray}
The findings in \cite{Altshuler:2013,Yevtushenko:2015} motivated subsequent theoretical works on magnetic-impurity ensemble.

In addition to works on general magnetic impurities without specifying their physical origins~\cite{Jiang:2009,Hattori:2011,Vayrynen:2016}, \cite{Hsu:2017,Hsu:2018b} pointed out that nuclear spins embedded in the lattice forming 2DTI host materials is a natural source of magnetic impurities.  
Nuclear spins are typically present in semiconductor materials and can directly flip electron spins through the hyperfine interaction, leading to the main decoherence mechanism in spin qubits made of III-V semiconductor materials~\cite{Khaetskii:2002,Khaetskii:2003,Schliemann:2003,Coish:2004}. 

Concerning the helical channels of the topological materials, several remarks are in order. First, the dipolar interaction between nuclear spins enables the spin diffusion~\cite{Lowe:1967}  and thus provides the dissipation mechanism for the nuclear spin polarization, which would otherwise be accumulated during the backscattering process and eventually suppress the steady-state backscattering current~\cite{Lunde:2012,Kornich:2015}.
In the presence of the spin diffusion, the nuclear spins are able to (partially) retain their unpolarized states with random orientations, therefore allowing for subsequent backscattering events. 
In consequence, unlike isolated magnetic impurities, nuclear spins can lead to a substantial edge resistance in dc measurements.
Second, whereas for 2DTI edge states the hyperfine coupling is somewhat weaker, due to the mixed $s$- and $p$-wave orbitals of the electron and hole wave functions~\cite{Lunde:2013}, the quasi-one-dimensional nature of the edge channels drastically enhance the backscattering effects, making nuclear spins a potential resistance source. Third, the interplay between electron-electron interactions and the electron-nuclear-spin coupling further complicates the analysis. Namely, a spiral order of the nuclear spins can be stabilized at low temperature. In the spiral-order phase, magnons and a macroscopic magnetic field (that is, Overhauser field in the context of nuclear spins) lead to additional resistance sources.
In addition to nuclear spins, artificially doped magnetic moments might be a source of the localized spins. A relevant example is Mn-doped HgTe quantum well~\cite{Furdyna:1988,Novik:2005}, a platform of interest due to its potential in realizing the quantum anomalous Hall effect~\cite{Liu:2008b,Wang:2014}. 
In this case, the intentionally doped magnetic impurities interact with the helical states via the exchange coupling, which can be much stronger compared to the hyperfine coupling to nuclear spins.

	\begin{figure}
	\includegraphics[width=\linewidth]{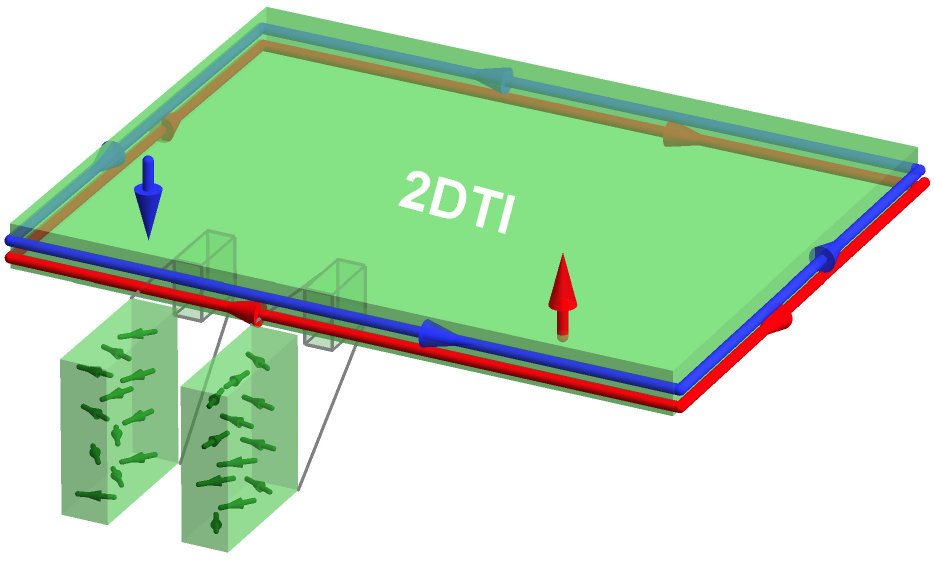}
		\caption{Edge states of a 2DTI and randomly oriented localized spins at the edge. The electron wave function of the edge states (red/blue for up/down spin) has a finite transverse extent, motivating us to define the cross section (gray) along the edge. In each of the cross sections (green blocks), the edge states interact with $N_{\perp}$ localized spins and can be spin-flopped by the latter.  
		}
 \label{Fig:2DTI_random}
	\end{figure}

Motivated by these observations, below we consider a helical channel coupled to a lattice of classical localized spins without specifying their origin, but having in mind that nuclear spins serve as a concrete example. While the gapless helical states move along one dimension only, the localized spins are distributed in all three dimensions and can interact with other localized spins in the bulk through, say, dipolar interaction.  
The coupling of the localized spins to the helical states can be written in an effective one-dimensional form,
\begin{equation}
\mathit{H}_{\rm rs} = \frac{1}{N_\perp} \sum_{\mu } \sum_{j} A_{\mu,j} S^{\mu}(r_{j}) \tilde{I}_{j}^{\mu},  
\label{Eq:H_rs_1d}
\end{equation}
with $\mu \in \{x,y,z\} $, the coupling constant $A_{\mu,j} $ (which is in general anisotropic), the number of localized spins per cross section in the transverse directions $N_{\perp}$, the electron spin ${\bf S}$, and the effective, classical spin $\tilde{{\bf I}}_{j}$ including all spins in a cross section labeled by $j$.
Since the $z$ component of $\tilde{{\bf I}}_{j}$ only leads to forward scattering terms and therefore does not affect the charge transport~\cite{Giamarchi:2003}, we focus on the transverse components of the spin operator.
In the continuum limit, they can be written in terms of the right- and left-moving fermion fields [see \eref{Eq:Spin}]. 
To proceed, we bosonize \eref{Eq:H_rs_1d} using \eref{Eq:bosonization},
\begin{eqnarray}
H_{\rm rs,b} &=& \int \frac{dr}{2\pi a} V_{\rm rs} (r) e^{2i \phi (r) } + \textrm{H.c.}
\label{Eq:H_rs_b}
\end{eqnarray}
The effective Hamiltonian now reads $ H_{\rm hel} + H_{\rm rs,b} $, with $H_{\rm hel} $ given in \eref{Eq:hTLL}.
In the above, electrons experience a random potential induced by the localized spins, which has a $2k_{F}$ component 
\begin{eqnarray}
V_{\rm rs} (r) &\equiv& \frac{1}{2 N_{\perp}} \left[ A_{x}(r) \tilde{I}^{x} (r) + i A_{y}(r)\tilde{I}^{y} (r) \right] e^{-2i k_{F} r}.
\end{eqnarray}
Considering independent and unpolarized spins (see \Fref{Fig:2DTI_random}), we assume that the disorder average satisfies 
\begin{eqnarray}
\left\langle V_{ \rm rs}^{\dagger} (r) V_{\rm rs} (r') \right\rangle_{\rm rs} &=& M_{\rm rs} \delta(r-r'), 
\label{Eq:random}
\end{eqnarray}
with  $\left\langle \cdots \right\rangle_{\rm rs}$ denoting the average over the random spin configurations and the strength $M_{\rm rs} \equiv a A_{0}^2 I(I+1)/(6 N_{\perp})$ depending on the typical coupling $A_0$. 
The effective action can be obtained using the replica method~\cite{Giamarchi:2003}, which leads to
\begin{eqnarray}
\frac{S_{\rm rs}}{\hbar} 
&=& - \frac{ \tilde{D}_{\rm rs} u^2}{8\pi a^3 }  \int  
dr d\tau d\tau' \, 
\cos \left[  2 \phi (r,\tau)- 2\phi (r,\tau') \right], \nonumber \\
\label{Eq:S_rs}
\end{eqnarray}
with  $\tilde{D}_{\rm rs} \equiv 2a M_{\rm rs} /(\pi \hbar^{2} u^{2})$. 

Using \eref{Eq:hTLL} and \eref{Eq:S_rs}, one can derive the RG flow equations upon changing the cutoff $a(l)=a(0) e^{l}$ with the dimensionless scale $l$ (see~\ref{App:RG}) and get
\numparts
\begin{eqnarray}
\frac{d \tilde{D}_{\rm rs}  }{d l } &=& (3-2K) \tilde{D}_{\rm rs} , \label{Eq:RG_rs1} \\
\frac{d K }{d l } &=& - \frac{K^2  }{2} \tilde{D}_{\rm rs}, \\
\frac{d u  }{d l } &=& -\frac{uK }{2} \tilde{D}_{\rm rs}. \label{Eq:RG_rs2}
\end{eqnarray}
\endnumparts 
The cosine term in \eref{Eq:S_rs} is RG relevant for $K< 3/2$, and the random-spin-induced backscattering is enhanced by the repulsive (or even weakly attractive) interactions.
Importantly, a gap is opened when the effective coupling $\tilde{D}_{\rm rs}$ flows to the strong-coupling limit, and thus localizes the helical states. 
The corresponding localization length and localization temperature are given by 
\begin{equation}
\xi_{\rm rs} = a \tilde{D}_{\rm rs} ^{-1/(3-2K)}, \; T_{\rm rs} = \frac{\hbar u} {k_{\rm B} \xi_{\rm rs}}.
\end{equation} 
For a channel longer than $\xi_{\rm rs}$, the helical states get localized below $T_{\rm rs}$, suppressing the conductance. The localization length and temperature strongly depend on the interaction parameter $K$, which enters the exponents of the above formulas. Considering nuclear spins in InAs/GaSb, the estimation on $\xi_{\rm rs}$ and $T_{\rm rs}$ suggest that the nuclear-spin-induced localization transition is experimentally accessible~\cite{Hsu:2017,Hsu:2018b}.

For a channel length $L$ comparable to $\xi_{\rm rs} $ at temperature $T$ comparable to $T_{\rm rs}$, the localized spins can in general lead to a substantial resistance. 
Before the channel gets localized, 
the resistance exhibits a power law, serving as a transport signature of this resistance mechanism. The functional form of the (differential) resistance depends on experimental conditions, which can be summarized as
\begin{eqnarray}
R_{\rm rs} &\propto& \frac{h }{e^2 } \frac{L}{a}  \tilde{D}_{\rm rs}  \left[ {\rm Max} \Big( \frac{k_{\rm B} T}{\Delta_{\rm b}}, \frac{eV}{\Delta_{\rm b}}, \frac{a}{  L} \Big) \right]^{2K-2},
\label{Eq:R_rs}
\end{eqnarray}
with the bias voltage $V$ across the channel. 
In the high-temperature regime, the temperature power law is the same as \eref{Eq:R_Yevtushenko2} from the Kondo impurity array in a purely one-dimensional model. 
On the other hand, when the localization length $\xi_{\rm rs}$ is shorter than the other length scales, $L$, $\hbar v_F/ (k_{\rm B} T)$, and $\hbar v_F/ (e V)$, we have a localized channel with an exponential resistance
\numparts
\begin{eqnarray}
R_{\rm rs} &\propto&  \frac{h }{e^2 } \frac{L}{a} \tilde{D}_{\rm rs}  e^{\Delta_{\rm rs} /(k_{\rm B}T)}, \\ 
 \Delta_{\rm rs} &=& \Delta_{\rm b} \big( 2K \tilde{D}_{\rm rs} \big)^{1/(3-2K)},
\label{Eq:R_rs_gapped}
\end{eqnarray}
\endnumparts
with the thermal activation gap $\Delta_{\rm rs}$. 

Before proceeding, we give several remarks on the above formulas.
First, for an ensemble of magnetic impurities along the channel, the resistance grows with the length $L$, in contrast to the single-impurity case where the conductance correction does not scale with $L$.  
Second, instead of assuming a random potential [see \eref{Eq:H_rs_b} and \eref{Eq:random}], one might start with a single weak impurity and then sum up the contributions from many alike impurities by assuming that they are independent~\cite{Vayrynen:2016}. 
However, the second approach does not capture the localization feature and leads to different RG flow equations, thus missing the renormalization of bulk quantities $u$ and $K$ as in nonhelical systems~\cite{Giamarchi:2003,Hsu:2019}.
Nonetheless, the approach employed by \cite{Vayrynen:2016} allows one to obtain the refined form of the resistance near the crossover $eV \approx k_{\rm B} T$, which was not captured by \eref{Eq:R_rs} due to the limitation of the RG approach. 
Interestingly,  
there appears an additional crossover $\frac{eV}{k_{\rm B} T} \approx \nu_0 A_{\rm eff}$ in the current-bias curve with the effective coupling $A_{\rm eff} \equiv (A_x^2+A_y^2+A_z^2 )/A_z$.
The second crossover occurs when one takes into account the dynamics of the local magnetic moment in the presence of the effective field $eV \nu_0 A_{\rm eff}$ generated by the electron spin imbalance due to a finite bias $V$. When $eV/k_{\rm B} T$ drops below $\nu_0 A_{\rm eff}$, the conductance is partially restored because the effective field partially polarizes the spin and thus reduces its ability to backscatter.
As a result, for $\frac{eV}{k_{\rm B} T} \le \nu_0 A_{\rm eff}$, the resistance is partially suppressed.

\begin{figure}
	\includegraphics[width=\linewidth]{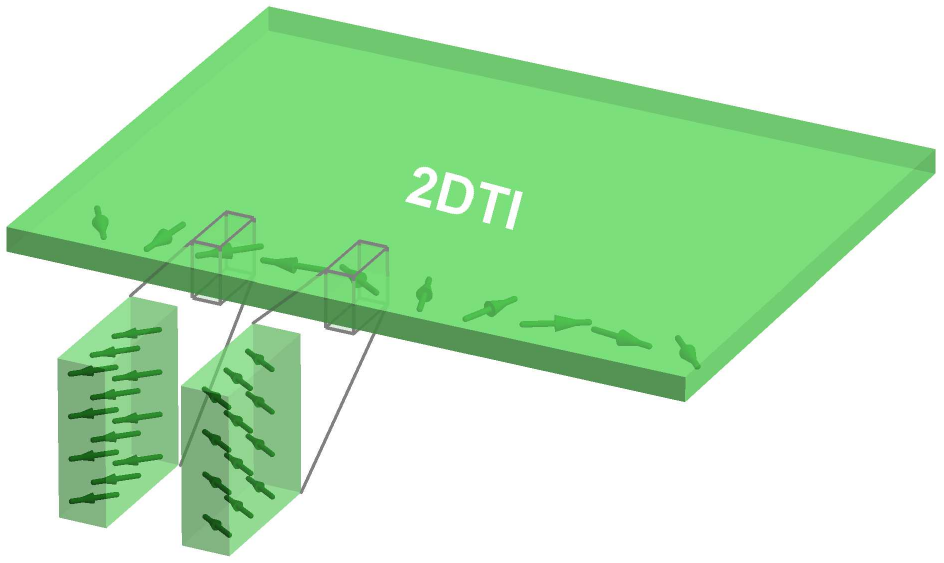}
		\caption{Illustration for the RKKY-induced spiral order of the localized spins along a 2DTI edge. The spins within a cross section have ferromagnetic alignment, while along the edge they rotate with the wavelength depending on $k_F$.
		For clarity, only spins along one edge are plotted.   }
 \label{Fig:2DTI_spiral}
	\end{figure}

While the backscattering on thermally randomized spins is rather straightforward, the situation is complicated by the possible ordering of these spins. 
It was shown that at low temperatures a spiral spin order can be stabilized by the Ruderman-Kittel-Kasuya-Yosida (RKKY) interaction, which is mediated by the edge states~\cite{Hsu:2017,Hsu:2018b}; we refer the interested readers to \cite{Maciejko:2012,Yevtushenko:2018} for discussion beyond the RKKY regime.
Since the RKKY interaction is mediated mainly by electrons at $\pm k_F$, in momentum space it develops a dip at $2k_F$ with the absolute magnitude
\numparts
\begin{eqnarray}
 J_{\rm RKKY} &\approx&  \frac{K A_{0}^2}{ \Delta_{ b} }
\left(\frac{\Delta_{b}}{2\pi K k_{\rm B} T} \right)^{2-2K} f_K, 
\label{Eq:JRKKY} \\ 
f_K&\equiv& \frac{  \sin (\pi K) }{8\pi^2} \left| \frac{ \Gamma\left( 1-K\right) \Gamma\left( \frac{K}{2} \right)} {\Gamma\left(1- \frac{K}{2} \right)} \right|^{2},
\end{eqnarray}
\endnumparts
with the Gamma function $\Gamma(x)$. 
As is clearly seen in \eref{Eq:JRKKY}, the magnitude of $J_{\rm RKKY} $ depends strongly on $K$ and is enhanced by the electron-electron interactions, in a similar fashion as in nonhelical channels~\cite{Braunecker:2009a,Braunecker:2009b,Klinovaja:2013a,Meng:2013,Meng:2014a,Stano:2014,Hsu:2015,Hsu:2020}. 
The localized spins minimize their RKKY energy if aligned along a direction which rotates as one moves along the edge. At low temperature, the spins are thus ordered into a spiral pattern as illustrated in \Fref{Fig:2DTI_spiral}, described by 
\begin{equation}
\langle {\bf \tilde{I}}(r) \rangle_{\rm s} = m_{\rm s} N_{\perp} I   \, \left[ \cos (2k_{F} r) {\bf e}_x - \sin (2k_{F}r) {\bf e}_y \right],
\label{Eq:I_order}
\end{equation}
where $\langle \cdots \rangle_{\rm s}$ indicates the expectation value of the nuclear spin state in the spiral phase and $m_{\rm s}$ is the temperature-dependent order parameter, which is normalized to unity at zero temperature. 
In the above, the rotation direction of the spiral spin order is fixed by the helicity of the electron edge state.
The order parameter and the ordering temperature $T_{\rm s}$, defined through $m_{\rm s}(T_{\rm s})=1/2$, are given by (valid for $T$ not much larger than $T_{\rm s}$)
\numparts
\begin{eqnarray}
m_{\rm s} (T)  &=& 1 - \frac{1}{2} \left( \frac{T}{T_{\rm s}} \right)^{3-2K} ,  \\
k_{\rm B}T_{\rm s} &=&  \left[ \frac{A_{0}^2 I^2  }{6\pi N_{\perp}}  
\left( \frac{\Delta_{\rm b}}{2\pi K} \right)^{1-2K} f_K \right]^{1/(3-2K)},
\label{Eq:Ts}  
\end{eqnarray} 
\endnumparts  
which strongly depend on the interaction strength parametrized by $K$.
Interestingly, the tendency to RKKY-induced spin order is stronger in a helical channel than in a nonhelical one~\cite{Braunecker:2009a,Braunecker:2009b}, as a result of the quench of the spin degree of freedom~\cite{Hsu:2018b}.

Concerning the edge transport, the presence of the spiral order can lead to two effects. 
First, the ordered spins are now polarized and generate a macroscopic magnetic field acting back on the electrons, which can break the time-reversal symmetry and mix the opposite spin states. As a result, the helical states become susceptible to charge impurities, which couple to the helical states through
\begin{eqnarray}
 H_{\rm imp}  &=& \int dr \, V_{\rm imp}(r) \rho (r)  ,
 \label{Eq:H_imp} 
\end{eqnarray}
where $\rho $ is the charge density and $V_{\rm imp}$ is the random potential induced by charge impurities, satisfying 
\begin{equation}
\eqalign{
\langle V_{\rm imp} (r) \rangle_{\rm imp} &= 0, \\ 
\langle V_{\rm imp} (r) V_{\rm imp} (r') \rangle_{\rm imp} &=  M_{\rm imp} (r-r'),   }
\label{Eq:Vimp_avg}
\end{equation}
with the disorder average $\langle \cdots \rangle_{\rm imp}$, the strength $M_{\rm imp} = n_{\rm imp} V_{\rm imp}^2$, the impurity density $n_{\rm imp}$ (the impurity number per length), and the typical magnitude $ V_{\rm imp}$ of the disorder potential. 
Second, at low but finite temperature, where the localized spins are not completely ordered, thermally excited magnons are present. They can interact with the electrons and induce additional backscattering processes. 

These additional backscattering mechanisms can be captured by the following contributions to the effective action, with the spiral-order-assisted backscattering on charge impurities, 
\begin{eqnarray}
\frac{S_{\rm sa} }{\hbar} 
&=& - M_{\rm sa} \int  
\frac{dr d\tau d\tau' }{(2\pi \hbar a)^2} \,  
\cos \left[ 2 \phi (r,\tau)- 2\phi (r,\tau') \right],  
\label{Eq:S_sa} 
\end{eqnarray}
and the magnon-induced backscattering,
\begin{eqnarray}
\frac{ S_{\rm m}}{\hbar} &=& - M_{\rm m} \int \frac{dr d\tau d\tau' }{(2\pi \hbar a)^2} \,  \cos \left[  2 \phi (r,\tau)- 2\phi (r,\tau') \right]
 \nonumber \\
&&  \times \Big[ e^{-\omega_{\rm m} |\tau-\tau'| } + 2 f_{B}(\hbar \omega_{\rm m}) \cosh (\omega_{\rm m} |\tau-\tau'| )\Big] . \nonumber \\
\label{Eq:S_mag}
\end{eqnarray}
In the above, $ M_{\rm sa} \equiv M_{\rm imp} B_{\rm s}^2/(8 \hbar v_{F} k_{F})^2$  
and $M_{\rm m} \equiv A_{0}^2 a I/(4N_{\perp})$ are the backscattering strengths, $\hbar \omega_{\rm m}$ is the magnon energy, $B_{\rm s} \equiv m_{\rm s} A_0 I $ is the spiral field, induced by the ordered spins, and  $f_{B}$ is the Bose-Einstein distribution.  
For a typical device with micrometer-long channels, the Mermin-Wigner theorem~\cite{Mermin:1966} and its extensions~\cite{Bruno:2001,Loss:2011} are not applicable. 
Instead of being Goldstone zero modes, the magnons have a finite energy, which can be approximated as
 \begin{eqnarray}
\hbar \omega_{ \rm m } (T)&=& \frac{2}{N_{\perp}} I J_{\rm RKKY}  (T) m_{\rm s} (T) .
\label{Eq:omega_mag_T}
\end{eqnarray}  
In the above, \eref{Eq:S_sa} is identical to \eref{Eq:S_rs} upon replacing the coupling $  \tilde{D}_{\rm rs} $ by $ \tilde{D}_{\rm sa} \equiv 2a M_{\rm sa} /(\pi \hbar^{2} u^{2})$. Therefore, one can derive the RG flow equations and show that \eref{Eq:S_sa} is also RG relevant for $K<3/2$ and can lead to localization of the helical channels for an edge longer than $ \xi_{\rm sa} = a \tilde{D}_{\rm sa}^{-1/(3-2K)}$ at temperature lower than $T_{\rm sa} = \hbar u / (k_{\rm B}  \xi_{\rm sa})$. 
Before entering the localized regime, the spiral-order-assisted backscattering induces a resistance in the form of \eref{Eq:R_rs} with a different prefactor $ \tilde{D}_{\rm sa} \propto B_{s}^2$, which depends on the temperature through the spiral field $B_{s}$. 
In the localized regime, on the other hand, the resistance becomes
$R_{\rm sa} (T) \propto  \tilde{D}_{\rm sa} e^{\Delta_{\rm sa} /(k_{\rm B}T)}$, 
with a gap $\Delta_{\rm sa} = \Delta_{\rm b} \big( 2 K \tilde{D}_{\rm sa} \big)^{1/(3-2K)}$ depending on the temperature.

The RG flow equations for the magnon-induced backscattering, derived from \eref{Eq:S_mag}, are given by 
\numparts
\label{Eq:RG_mag-bs}
\begin{eqnarray}
\frac{d \tilde{Y}_{\rm m} }{d l } &=& \left( 3-2K \right) \tilde{Y}_{\rm m} , \\
\frac{d K }{d l } &=& - \frac{K^2 \omega_{\rm m} a }{2u }  e^{- \omega_{\rm m} a/u} \tilde{Y}_{\rm m}, \\
\frac{d u}{dl} &=& -\frac{ K \omega_{\rm m} a }{2} e^{-\omega_{\rm m} a/u} \tilde{Y}_{\rm m}, 
\end{eqnarray} 
\endnumparts
with $\tilde{Y}_{\rm m} \equiv 2 M_{\rm m} / (\pi \hbar^2 u \omega_{\rm m})$.
In contrast to the spiral-order-assisted backscattering, here the coupling strength vanishes at zero temperature because it would cost large energy to excite the magnons [see \eref{Eq:omega_mag_T}]. The magnons therefore do not lead to localization.  
At low temperature $T<T_{\rm m}$, defined by $\omega_{\rm m} (T_{\rm m}) = \tilde{D}_{\rm m}^{1/(4-2K)} u/a$, the RG flow reaches the strong-coupling regime and the magnon-induced resistance is dominated by the magnon emission,
\begin{eqnarray}
R_{\rm m}^{\rm em} (T) &\propto&  \frac{h}{ e^2}  \frac{L}{ a} \tilde{D}_{\rm m} \left[ \frac{ \hbar \omega_{\rm m}(T) }{\Delta_{\rm b}} \right]^{2K-3},
\label{Eq:R_mag-em}
\end{eqnarray} 
with $\tilde{D}_{\rm m} \equiv 2a M_{\rm m} /(\pi \hbar^{2} u^{2})$. 
In addition, in the range $T_{\rm m} \le T \le T_{\rm s}$ the magnon absorption leads to a subdominant term, 
\begin{eqnarray}
R_{\rm m}^{\rm ab} (T) &\propto&  \frac{h}{ e^2}  \frac{L}{ a}  \tilde{D}_{\rm rs} \left( \frac{T}{T_{\rm s}} \right)^{3-2K}. 
\label{Eq:R_mag-ab}
\end{eqnarray} 
That the prefactor is set by the scale $\tilde{D}_{\rm rs}$ follows because the magnon absorption can be viewed as the resistance induced by the localized spins that remain disordered. 
Overall, we find that in the spiral-order phase the spiral-order-assisted backscattering on charge disorder dominates over the magnon-induced backscattering. 

As a summary for the discussion on localized spins embedded in the lattice, regardless of whether they are ordered by the RKKY interaction or not, they can localize helical states in a sufficiently long channel at a low temperature.  
Since nuclear spins are in general present in typical semiconductor-based 2DTI materials, they serve as the local magnetic moments discussed here, what places a fundamental limitation on exploiting helical channels in scalable quantum devices. 

In addition to the above mechanisms, new types of processes arise when taking into account the nonequilibrium dynamics of the localized (nuclear) spins. Namely, in a typical experiment, the applied current through the edge can induce the dynamic nuclear spin polarization (DNP)~\cite{Lunde:2013,Kornich:2015,Russo:2018}, which induces an Overhauser field and can influence the charge transport.
It was found that the DNP alone is not sufficient to maintain a steady-state backscattering current and to alter the transport~\cite{Lunde:2012}.
To cause a finite resistance, it requires the presence of a spin-flip mechanism that relaxes the polarization of the localized spins, in accordance with our discussion in section~\ref{SubSec:Single}. 
By introducing a phenomenological parameter  $\Gamma_{\rm sf}$ for the spin-flip rate per localized spin, \cite{Lunde:2012} computed 
the magnitude of the DNP as a function of the bias voltage and temperature in a noninteracting channel. 
It leads to a reduction of the current,
\begin{eqnarray}
\delta I_{\rm dnp} &\propto& - \frac{e^2}{h} \frac{\pi \hbar \Gamma_{\rm sf} }{eV \coth \big( \frac{eV}{2k_{\rm B} T} \big) + \frac{ 2\hbar }{\pi \eta_{\rm dnp} } \Gamma_{\rm sf} }, 
\end{eqnarray} 
where $\eta_{\rm dnp} =  (\nu_{0} A_{0} )^2 N_{\rm s} /N_{\rm tot}^2$ is a dimensionless constant with $N_{\rm s}$ being the total number of localized spins coupling to the helical states along the edge and $N_{\rm tot}$ being the total number of the lattice sites.
In the linear-response regime, it leads to a resistance 
\begin{eqnarray}
R_{\rm dnp}   &\propto& \frac{h}{e^2} \frac{\pi \hbar \Gamma_{\rm sf} }{ 2k_{\rm B} T  + \frac{ 2\hbar}{\pi \eta_{\rm dnp} } \Gamma_{\rm sf} }.
\label{Eq:R_temp1}
\end{eqnarray} 
Clearly, $R_{\rm dnp} $ reduces to zero when there is no spin flip to dissipate the current-induced polarization. 
\cite{DelMaestro:2013} investigated how the DNP can induce resistance in the presence of a random, spatially nonuniform Rashba spin-orbit coupling.
Since the spin-orbit coupling breaks the spin conservation, it can depolarize the DNP. 
For a short channel, the induced resistance is
\begin{eqnarray}
R_{\rm dnp+rso} (T) &\propto&  \frac{h}{ e^2}  \frac{L}{ a}   \frac{A_0^2 \langle I \rangle_{\rm dnp}^2 M_{\rm rso}}{\hbar^4 v_F^4}, 
\label{Eq:R_temp2}
\end{eqnarray} 
where $M_{\rm rso}$ denotes the disorder strength of the random spin-orbit coupling and the temperature-dependent spin polarization $\langle I \rangle_{\rm dnp}$ can be determined numerically. 
For a long channel, they obtained a nonlinear current voltage relation $R_{\rm dnp+rso} \propto (eV/k_{\rm B} T)^{2/3}$.

Similar to studies on a single magnetic impurity, there were also investigations on an ensemble of magnetic impurities with large spin $I$. For instance, with $I>1$ and uniaxial anisotropy, mesoscopic conductance fluctuations arise from the quantum interference of electron scattering amplitudes through multiple scatterers~\cite{Cheianov:2013}.
It leads to conductance fluctuations as a function of the Fermi energy, which are different to those of nonhelical channels. 
\cite{Wozny:2018} investigated the effects of magnetic adatoms (for example, Mn) on helical edge states, taking into account not only their local magnetic moments but also the random electrostatic potential that they induce. Whereas the magnetic moments can induce elastic backscatterings and open a gap for the helical channels, the accompanying electrostatic potential can have the opposite effect on the charge transport, by reducing the gap opened by the magnetic moments. Under certain experimental conditions it can  even close the gap, restoring the helical transport.
Finally, we note that a full first-principle study of edge transport was carried out~\cite{Vannucci:2020}, particularly focusing on several recently proposed quantum spin Hall materials with large bulk gaps (bismuth and antimony halides, binary compounds BiX and SbX with X $\in \{$F, Cl, Br, I$\}$).
As expected, magnetic impurities trapped at the vacancy defects were identified as crucial backscattering sources which break the time-reversal symmetry and, in general, lead to the conductance quantization breakdown.

\subsubsection{Other mechanisms with broken time-reversal symmetry}
~\\
Other than localized spins, one might consider single-particle backscattering off a charge impurity in the presence of an external magnetic field $B$~\cite{Lezmy:2012}. The high-temperature and high-bias limit of the resulting (differential) conductance correction can be summarized as 
\begin{eqnarray}
\delta G_{\rm B} \propto - \frac{e^2}{h} \left(\frac{B}{B_0}\right)^2  \left[ {\rm Max} \Big( \frac{eV}{\Delta_{\rm b}}, \frac{k_{\rm B} T}{\Delta_{\rm b}} \Big) \right]^{2K-2},
\end{eqnarray}
with the normalization parameter $B_0$ with the magnetic-field units. 
The correction has the same scaling as backscattering off localized spins \eref{Eq:G_Maciejko1}.
The zero-temperature edge conductance of a helical channel in the presence of an external magnetic field and quenched charge disorder was studied numerically by \cite{Maciejko:2010}. As expected, the conductance deviates from its quantized value when the disorder strength and the Zeeman energy are comparable to the bulk gap.

\cite{Tkachov:2010} studied a ballistic edge subject to an out-of-plane magnetic field at zero temperature, which can drive the system through a transition from quantum spin Hall to quantum Hall phases. Near the transition, the backscattering is enhanced, with a power-law suppression of the longitudinal conductance.
In addition, \cite{Delplace:2012} investigated effects of random magnetic field flux in a noninteracting helical channel at zero temperature. Based on the scattering theory, they concluded that the random magnetic field flux can localize the edge states, with a localization length which is inversely proportional to the square of the magnetic field for small fields and saturates for a sufficiently strong magnetic field. 
\cite{Vezvaee:2018} considered a helical edge channel formed at the outer boundary of a 2DTI ring threaded by a magnetic flux and coupled to magnetic impurities with arbitrary spin. 
They found a universal flux dependence of the energy spectrum independent of the details of the electron-impurity interaction and concluded that the magnetic impurities can lead to sizable energy gaps in the spectrum, thus affecting the edge transport.
Furthermore, \cite{Wang:2021b} investigated the effects of an out-of-plane magnetic field on a diffusive Na$_3$Bi sample, in which charge puddles with odd occupation numbers were modeled as magnetic impurities. Since the field can polarize the magnetic impurities, it can suppress the backscattering and thus enhance the mean free path. A sufficiently strong field can even drive the system into the ballistic regime. 

In addition to the RKKY-induced spiral spin order discussed above, there exist theoretical findings on other unconventional orderings that can be formed in the edge channels and influence the transport. 
For instance, due to the smooth (nonabrupt) edge confinement potential in a realistic setup, the system undergoes edge reconstruction and additional edge states emerge~\cite{Wang:2017}.
Due to a large exchange energy, the additional edge states are spin-polarized and thus break the time-reversal symmetry, inducing elastic single-particle backscattering and spontaneous breakdown of the topological protection. 
We note, however, that the opposite conclusion was drawn in \cite{John:2021}, which demonstrated the robust edge transport even in the presence of the edge reconstruction. 
As another example, \cite{Novelli:2019} studied nonmagnetic impurities in a combination with on-site electron-electron interactions in a 2DTI.
By incorporating the onsite interaction in the Kane-Mele model and employing the Hartree-Fock approximation, they demonstrated formation of noncollinear magnetic scatterers which break the time-reversal symmetry and thereby cause backscattering of edge electrons. 
The same conclusion holds as well in the presence of Rashba spin-orbit coupling, allowing them to conclude that, in general, the conductance quantization is not protected. 
This mechanism might be more relevant for monolayer systems such as WTe$_2$, where vacancy defects are naturally present.

\cite{Balram:2019} pointed out that in typical experiments on charge transport, the current-carrying state itself breaks the time-reversal symmetry, thus lifting the topological protection of the helical channels. Specifically, the applied current leads to an imbalance between the left and right-moving states, resulting in a momentum-dependent dynamic spin polarization in the helical channel. Through the electron-electron interactions, the spin polarization generates an internal magnetic field acting back on the electrons themselves. Similar to the Overhauser field from either the RKKY-induced spiral spin order or the DNP, this internal magnetic field allows for elastic backscattering and therefore generates a finite resistance.
They found a nonlinear contribution to current-bias curves, which remains finite even at  zero temperature and concluded that the internal-field-assisted backscattering can eventually open a gap in the edge-state spectrum. 

Instead of the static local magnetic moments discussed in the previous subsection,
\cite{Bagrov:2019} considered the coupling of the helical states to an external bath of itinerant spins. The interplay between the Coulomb and spin-spin interactions leads to a renormalization of the backscattering amplitude, and can even suppress backscattering processes in certain regimes. 
Finally, \cite{Yevtushenko:2019} studied a lattice of Kondo impurities. 
Similar to the conclusion drawn from studies on a single magnetic impurity in, for example, \cite{Tanaka:2011,Vinkler-Aviv:2020}, they emphasized the importance of the non-spin-conserving interaction. Namely, whereas the time-reversal symmetry is responsible for the formation of helical states, for a mechanism to have influence on the dc transport, one still needs to break the spin conservation. 

In addition to the above, there are also mechanisms that do not break the time-reversal symmetry explicitly, which we discuss below.

\subsection{Time-reversal-invariant mechanisms \label{Sec:Scattering_TRS}}
Whereas the time-reversal symmetry precludes the overlap of wave functions of counterpropagating degenerate time-reversal states, there is no such restriction for the counterpropagating states at {\it different energies}. 
Therefore, time-reversal-invariant perturbations can still lead to backscattering of the electrons in a helical channel through inelastic processes.  
Since the $U(1)$ symmetry associated with the electron spin plays an important role for inelastic scattering, we first introduce the generic helical liquid where this symmetry might be broken.
 
Breaking the $U(1)$ symmetry is generally expected in the presence of an out-of-plane electric field. 
In a heterostructure hosting two-dimensional electron gas,  the electric field generates the Rashba SOI term,
\begin{eqnarray}
H_{\rm R}^{\rm 2d} = \alpha_{\rm R} \big( \sigma^x k_y - \sigma^y k_x \big),
\label{Eq:H_R}
\end{eqnarray}
with the Rashba coupling strength $\alpha_{\rm R} $ depending on the out-of-plane electric field, the Pauli matrix $\sigma^{x,y}$ and the momentum $\hbar k_{x,y}$.
With \eref{Eq:H_R}, the electron spin along $z$ is no longer a good quantum number.
Concerning the edge states, their spin polarization direction is not necessarily fixed to any global axis and can change with the state energy. The only remaining symmetry is the time-reversal symmetry, which is not broken by the SOI. It requires that the left- and right-moving states at a given energy still form a Kramers pair. In the literature, such a system has been termed a {\it generic helical liquid}~\cite{Schmidt:2012}.
Below we first discuss backscattering mechanisms which rely on the $S^z$ symmetry being broken and then turn to those that do not require the broken $S^z$ symmetry.

To investigate the generic helical liquid, one can start from the Kramers doublet $\psi_{\pm,q}$ defined in momentum space, where the $\pm$ sign labels the moving direction. These helical eigenstates are related to the spin states $\psi_{\sigma,q}$ through
\begin{eqnarray}
\left( \begin{array}{c} \psi_{\downarrow,q} \\  \psi_{\uparrow,q} \end{array} \right) &=& U_{\rm so} (q) 
\left( \begin{array}{c} \psi_{+,q} \\  \psi_{-,q} \end{array} \right).
\label{Eq:ghl_transform}
\end{eqnarray}
Here, the SU(2) matrix $U_{\rm so}$ satisfies $ U_{\rm so}^{\dagger} (q) U_{\rm so} (q) =$ diag$(1,1)$ stemming in the orthonormality of the eigenstates and $  U_{\rm so} (q) = U_{\rm so} (-q)$ due to the time-reversal symmetry. 
The presence of the Rashba SOI leads to a momentum dependence of the $U_{\rm so}$ matrix, reflecting a momentum-dependent rotation of the spin-quantization axis~\cite{Rod:2015,Ortiz:2016}. 
In consequence, the right- and left-moving helical states are now a mixture of the up- and down-spin states. In momentum space, the density operator can be expressed as
\begin{eqnarray}
\rho_{q}  &=& \sum_{\sigma} \sum_{k} \psi_{\sigma,k}^{\dagger}  \psi_{\sigma,k+q} \nonumber \\
&=& \sum_{\alpha \beta }  [U_{\rm so}^{\dagger}(k) U_{\rm so}(k+q)]_{\alpha\beta} \psi_{\alpha, k}^\dagger \psi_{\beta, k+q},
\end{eqnarray} 
with the indices $\alpha,\beta \in \{ +, -\}$ labeling the direction of motion.
There appear backscattering terms ($\alpha \neq \beta$) in the density operator, which enter the Hamiltonian either through the electron-electron interactions or by coupling to the disorder potential induced by charge impurities.

In terms of the helical fermion fields $\psi_{\pm,q}$, the electron-electron interactions can be written as
\begin{eqnarray}
 H_{\rm ee}  &=& \int dr dr' \, U_{\rm ee}(r-r') \rho (r) \rho (r') \nonumber \\
&=&
\frac{1}{L} \sum_{qkp}  
\sum_{\alpha\alpha' \beta \beta'} U_{{\rm ee}, q} \psi_{\alpha, k}^\dagger \psi_{\beta, k+q} \psi_{\alpha', p}^\dagger \psi_{\beta', p-q}  
\label{Eq:H_ee_ghl} \\
&& \times [U_{\rm so}^{\dagger}(k) U_{\rm so}(k+q)]_{\alpha\beta} [U_{\rm so}^{\dagger}(p) U_{\rm so}(p-q)]_{\alpha'\beta'}, \nonumber
\end{eqnarray}
with the Fourier transform of the electron-electron interaction $U_{{\rm ee}, q}$.
The coupling to the disorder \eref{Eq:H_imp} now becomes
\begin{eqnarray}
 H_{\rm imp}  &=& \frac{1}{L} \sum_{q k}  V_{{\rm imp}, q - k}
\sum_{\alpha \beta }  [U_{\rm so}^{\dagger}(q) U_{\rm so}(k)]_{\alpha\beta} \psi_{\alpha, q}^\dagger \psi_{\beta, k} .  \nonumber \\
\label{Eq:H_imp_ghl} 
\end{eqnarray}
In the above, we model disorder by a random potential $V_{\rm imp}$ induced by charge impurities distributed along the entire channel.
As an alternative to such ``quenched disorder''~\cite{Wu:2006}, one might consider an isolated impurity localized at a single spatial point~\cite{Kane:1992}. While a single impurity might lead to negligible effects in weakly interacting systems, electron-electron interactions can enhance its backscattering effects. 
Therefore, in section~\ref{Sec:generic_hTLL} where we go beyond the weak-interaction regime, we will explicitly consider the conductance correction due to a single impurity. 
In any case, we assume that the electron-electron interactions in the helical channels are short-range due to the screening effect from a metallic gate in proximity to the helical channel, which allows us to approximate $U_{{\rm ee}, q} \approx U_{\rm ee}$. 
Finally, we also assume that the random potential due to impurities has roughly the same weight for all the Fourier components and hence $V_{{\rm imp}, q } \approx V_{\rm imp}$.

Clearly, there are several terms in \eref{Eq:H_ee_ghl} and \eref{Eq:H_imp_ghl}, which would otherwise be absent at zero SOI [that is, when $U_{\rm so} \to {\rm diag} (1,1)$]. 
This in turn leads to inelastic backscatterings without breaking the time-reversal symmetry. \cite{Kainaris:2014} made a systematic investigation on the backscatterings in both the weakly interacting regime (with the fermionic description) and beyond it (with the bosonization formalism).
In the following subsections, we discuss the two regimes separately. 

\subsubsection{Weak interactions: generic helical liquid \label{SubSec:ghl} }
~\\
When the electrons in a helical channel are weakly interacting, the perturbation terms can be described in the fermion picture~\cite{Schmidt:2012,Kainaris:2014}, which is valid for $K$ close to unity. 
To proceed, we specify the form of the matrix $U_{\rm so}$ by symmetry arguments and expanding around $q \approx 0$,
\begin{eqnarray}
 U_{\rm so} (q)  &=& 
\left( \begin{array}{cc} 
1  & - q^2/k_{\rm so}^2\\  
 q^2/k_{\rm so}^2 & 1 
\end{array} \right) + O(q^4),
\label{Eq:U_so}  
\end{eqnarray}
where the zero-SOI limit corresponds to $k_{\rm so} \to \infty$. 
It is simple to check that \eref{Eq:U_so} satisfies the requirements for $ U_{\rm so} $ described below~\eref{Eq:ghl_transform}. 
Physically, $\hbar k_{\rm so}$ can be understood as the momentum scale at which the quantization axis of the electron spin changes appreciably. 
The SOI-induced spin rotation enters \eref{Eq:H_ee_ghl} and \eref{Eq:H_imp_ghl} through the product of $U_{\rm so}$ at different momenta.
To the leading order, the product reads
\begin{eqnarray}
 U_{\rm so}^\dagger (q)  U_{\rm so} (k)  &=& 
\left( \begin{array}{cc} 
1  &  \frac{q^2 - k^2}{k_{\rm so}^2}\\  
\frac{k^2 - q^2}{k_{\rm so}^2} & 1 
\end{array} \right),
\end{eqnarray}
which can be plugged into \eref{Eq:H_ee_ghl}, leading to the following scattering terms~\cite{Kainaris:2014},
\numparts
\begin{eqnarray}
 H_{\rm ee,1} &=&
\frac{U_{\rm ee}}{k_{\rm so}^4 L} \sum_{qkp} \sum_{\alpha}  (q^2 +2qk) (q^2 - 2qp) \nonumber\\
&& 
 \times \psi_{\alpha, k}^\dagger \psi_{-\alpha, p}^\dagger \psi_{-\alpha, k+q} \psi_{\alpha, p-q} , \label{Eq:H_ee_1} \\
H_{\rm ee,2} &=&
\frac{U_{\rm ee}}{L} \sum_{qkp}  
\sum_{\alpha} \psi_{-\alpha, p}^\dagger \psi_{\alpha, k}^\dagger \psi_{\alpha, k+q} \psi_{-\alpha, p-q} , \label{Eq:H_ee_2} \\
H_{\rm ee,3} &=&
\frac{U_{\rm ee}}{k_{\rm so}^4 L} \sum_{qkp} \sum_{\alpha} (q^2 +2qk) (q^2 - 2qp) \nonumber\\
&& 
 \times \psi_{\alpha, p}^\dagger \psi_{\alpha, k}^\dagger \psi_{-\alpha, k+q} \psi_{-\alpha, p-q} , \label{Eq:H_ee_3} \\
 H_{\rm ee,4} &=&
\frac{U_{\rm ee}}{L} \sum_{qkp}  
\sum_{\alpha} \psi_{\alpha, p}^\dagger \psi_{\alpha, k}^\dagger \psi_{\alpha, k+q} \psi_{\alpha, p-q},  \label{Eq:H_ee_4} \\
 H_{\rm ee,5} &=&  -\frac{2U_{\rm ee}}{k_{\rm so}^2 L} \sum_{qkp} \sum_{\alpha} \alpha (k^2 - p^2) \nonumber\\
&&  \times \psi_{\alpha, k+q}^\dagger \psi_{-\alpha, p-q}^\dagger \psi_{\alpha, p} \psi_{\alpha, k}  + {\rm H.c.} .
  \label{Eq:H_ee_5} 
\end{eqnarray}
\endnumparts
In the above, the first four terms are in analogous to the electron-electron interactions in a nonhelical TLL in the literature~\cite{Giamarchi:2003}. The forward scattering terms $H_{\rm ee,2}$ and $H_{\rm ee,4} $ are present also at zero spin-orbit coupling [which gives \eref{Eq:H_ee} upon linearization], whereas the other terms are induced by the broken $S^z$ symmetry. 
Among these SOI-induced terms, $H_{\rm ee,1}$ describes backscattering of one left- and one right-moving particle, whereas $H_{\rm ee,3}$ describes umklapp backscattering of two right movers or two left movers. 
In comparison with a nonhelical channel, both $H_{\rm ee,1}$ and $H_{\rm ee,3}$ acquire momentum-dependent amplitude and are of fourth order in $1/k_{\rm so}$. 
What is unique to the generic helical liquid is the emergence of the $H_{\rm ee,5}$ term, which describes a single-particle backscattering. Since it is of lower order in  $1/k_{\rm so}$, it dominates over $H_{\rm ee,1}$ and $H_{\rm ee,3}$ for typical parameters. 

Concerning the impurity-induced scattering~\eref{Eq:H_imp_ghl}, it can be separated into forward ($H_{\rm imp,f}$) and backward ($H_{\rm imp,b}$) scattering terms,
\numparts
\begin{eqnarray}
 H_{\rm imp,f} &=& \frac{V_{\rm imp}}{L} \sum_{q k}   \sum_{\alpha }  \psi_{\alpha, q}^\dagger \psi_{\alpha, k} , 
 \label{Eq:H_imp_f}  \\
 H_{\rm imp,b} &=& \frac{V_{\rm imp}}{L} \sum_{q k}   \sum_{\alpha }  \alpha \frac{q^2 - k^2}{ k_{\rm so}^2} \psi_{\alpha, q}^\dagger \psi_{-\alpha, k} . 
\label{Eq:H_imp_b}
\end{eqnarray}
\endnumparts
Clearly, the backscattering term \eref{Eq:H_imp_b} is absent without the momentum-dependent spin rotation.  

Since the scatterings in $H_{\rm ee,1}$, $H_{\rm ee,2} $, $H_{\rm ee,4} $ and $ H_{\rm imp,f} $ do not change the number of the left and right movers, they alone do not cause resistance. 
Due to the restrictions from the momentum and energy conservation, the scattering processes described by the remaining electron-electron interaction terms, $H_{\rm ee,3}$ and $H_{\rm ee,5} $, are not allowed at zero temperature for finite $k_F$ in a clean system. Nonetheless, at finite temperatures the restrictions can be relaxed. Therefore, the scattering processes in general take place when the Fermi level is close to the Dirac point $k_F \approx 0$ or in the presence of disorder/lattice potential.\footnote{Although the backscattering due to $H_{\rm ee,3}$ could be allowed at commensurate fillings~\cite{Wu:2006}, for typical helical states the Fermi wavelength is much longer than the lattice period and achieving commensurate filling is unrealistic. We therefore do not consider the lattice potential here.}

	\begin{figure}
	\includegraphics[width=0.31\linewidth]{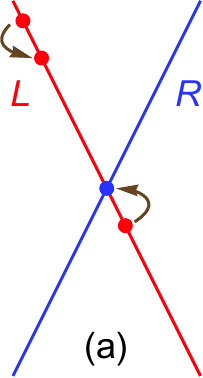}
	\hspace{0.01\linewidth}
	\includegraphics[width=0.32\linewidth]{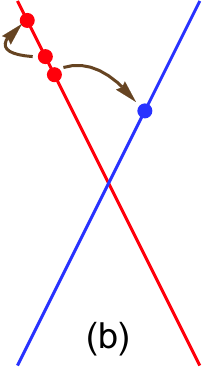}
	\hspace{0.01\linewidth}
	\includegraphics[width=0.29\linewidth]{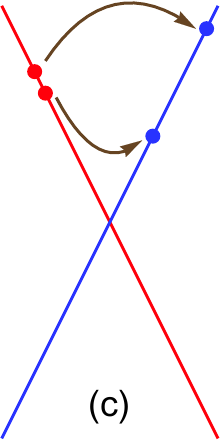}
		\caption{In a generic helical liquid, the interaction terms $H_{\rm ee,3}$  and $H_{\rm ee,5}$ lead to time-reversal-invariant processes which change the numbers of right- and left-moving particles.
		The term $H_{\rm ee,5}$ in \eref{Eq:H_ee_5} can back scatter one particle, with the energy difference compensated by the creation of a particle-hole pair. The term $H_{\rm ee,3}$ in \eref{Eq:H_ee_3} back scatters two particles. 
		When the Fermi level is sufficiently close to the Dirac point ($k_F \approx 0$), $H_{\rm ee,5}$ is allowed in a clean system, as illustrated in (a).
		For a general $k_F$, the presence of disorder allows for a one-particle backscattering (1PB) arising from $H_{\rm ee,5}$ (b) and a two-particle backscattering (2PB) induced by $H_{\rm ee,3}$ (c), with the momentum differences compensated by the disorder potential.
		 }
 \label{Fig:Scattering_ghl}
	\end{figure}

In a clean system, the scattering in $H_{\rm ee,3} $ does not conserve momentum, and only $H_{\rm ee,5} $ can directly lead to a finite edge resistance. 
In contrast, both $H_{\rm ee,3} $ and $H_{\rm ee,5} $ allow for backscatterings in a system with disorder, 
where the momentum difference between the initial and final states can be compensated by disorder either in the form of random potential in \eref{Eq:H_imp_f} and \eref{Eq:H_imp_b}  or a single impurity localized at one spatial point. 
In consequence, up to second order in $H_{\rm ee}$ and $H_{\rm imp}$, there can be various scattering processes, which back scatter one or two particles. We therefore call them {\it one-particle backscattering} (1PB) and {\it two-particle backscattering} (2PB), respectively. 
We illustrate the three main processes in \Fref{Fig:Scattering_ghl}.
The nomenclature for the SOI-induced mechanisms varies in the literature; for convenience we list them in \Tref{Tab:nomenclature}.

{\it 1PB process.}
Here we discuss first a clean weakly-interacting helical channel, where the resistance is exclusively due to $H_{\rm ee,5} $.
Since the momentum and energy conservation limit the process to only states near the Dirac point [see \Fref{Fig:Scattering_ghl}(a)], the resistance strongly depends on the position of the Fermi level~\cite{Schmidt:2012,Kainaris:2014}.
Omitting the numerical prefactors, for $\hbar v_F k_F \ll k_{\rm B} T$ it is given by 
\begin{eqnarray}
 R_{5}^{\rm hi} &\propto& \frac{h}{e^2} Lk_{\rm so} \left( \frac{U_{\rm ee}}{\hbar v_F} \right)^2  \left( \frac{ k_{\rm B} T}{\hbar v_F k_{\rm so}} \right)^5,
\label{Eq:R_5_highT}
\end{eqnarray}
whereas for $\hbar v_F k_F \gg k_{\rm B} T$ by
\begin{eqnarray}
 R_{5}^{\rm low} &\propto& \frac{h}{e^2} Lk_{\rm so} \left( \frac{U_{\rm ee}}{\hbar v_F} \right)^2  \left( \frac{ k_F }{ k_{\rm so}} \right)^5 e^{ \frac{-\hbar v_F k_F}{ k_{\rm B} T} }.
\label{Eq:R_5_lowT}
\end{eqnarray}
The result in \eref{Eq:R_5_lowT} describes thermal-activation behavior, due to the fact that the process is allowed only for states near the Dirac point which are deep below the Fermi level. 

In a system with disorder, $H_{\rm ee}$ and $H_{\rm imp}$ lead to various 1PB or 2PB processes~\cite{Kainaris:2014}. 
Among the 1PB processes, the combination of $H_{\rm ee,5}$ and $H_{\rm imp,f}$ described in \Fref{Fig:Scattering_ghl}(b) gives a dominating contribution to the resistance,
\begin{eqnarray}
 R_{\rm 5\times f} &\propto& \frac{h}{e^2} L n_{\rm imp} \left( \frac{U_{\rm ee} V_{\rm imp}}{\hbar^2 v_F^2} \right)^2  \left( \frac{ k_{\rm B} T}{\hbar v_F k_{\rm so}} \right)^4.
\label{Eq:R_5_imp}
\end{eqnarray}
There are also contributions from combining the disorder potential and $H_{\rm ee,1}$ or $H_{\rm ee,3}$, leading to different power laws in temperature or the $k_F / k_{\rm so}$ ratio.
Specifically, the combination of $H_{\rm ee,1}$ and $H_{\rm imp,b}$ gives 
\begin{eqnarray}
 R_{\rm 1 \times b} &\propto& \frac{h}{e^2} L n_{\rm imp} \left( \frac{U_{\rm ee} V_{\rm imp}}{\hbar^2 v_F^2} \right)^2   \left( \frac{ k_F }{ k_{\rm so}} \right)^6 \left( \frac{ k_{\rm B} T}{\hbar v_F k_{\rm so}} \right)^6, \nonumber \\
\label{Eq:R_1_imp}
\end{eqnarray}
whereas the combination of $H_{\rm ee,3}$ and $H_{\rm imp,b}$ leads to
\begin{eqnarray}
 R_{\rm 3 \times b} &\propto& \frac{h}{e^2} L n_{\rm imp} \left( \frac{U_{\rm ee} V_{\rm imp}}{\hbar^2 v_F^2} \right)^2  \left( \frac{ k_F }{ k_{\rm so}} \right)^8 \left( \frac{ k_{\rm B} T}{\hbar v_F k_{\rm so}} \right)^4. \nonumber \\
\label{Eq:R_3_imp}
\end{eqnarray} Among the three 1PB processes in disordered systems, since typically $k_F /k_{\rm so}<1$ and we are interested in the low-temperature regime, $ R_{\rm 5\times f}$ dominates. Therefore, we have $ R_{\rm 1pb} \approx R_{\rm 5\times f} \propto T^4$.  

{\it 2PB process.}
The 2PB process is dominated by the combination of $H_{\rm ee,3}$ and  $H_{\rm imp, f}$, illustrated in \Fref{Fig:Scattering_ghl}(c).
We obtain $ R_{\rm 2pb} \approx  R_{\rm 3 \times f}$ and therefore 
\begin{eqnarray}
 R_{\rm 2pb} &\propto& \frac{h}{e^2} L n_{\rm imp} \left( \frac{U_{\rm ee} V_{\rm imp}}{\hbar^2 v_F^2} \right)^2  \left( \frac{ k_F }{ k_{\rm so}} \right)^2 \left( \frac{ k_{\rm B} T}{\hbar v_F k_{\rm so}} \right)^6. \nonumber \\
\label{Eq:R_2pb}
\end{eqnarray}
The $T^6$ power law of $R_{\rm 2pb}$ leads to a much smaller resistance than the one from 1PB at low temperature. 
Generally speaking, in comparison with clean systems, here the energy and momentum conservation conditions for backscatterings are relaxed by the random disorder potential, so the resistance is not very sensitive to the Fermi level. 

Overall, both the 1PB and 2PB processes lead to resistance with high powers of $T$ stemming from the phase space necessary for energy conservation. 
Nonetheless, the weak temperature dependence observed in experiments called for alternatives.
The literature followed two strategies (other than the elastic scattering from broken time-reversal symmetry). The first one is to include the TLL effect, which is known to render fractional power laws in observables; we will turn to this in section~\ref{Sec:generic_hTLL}. 
The second one is to relax the restriction from energy and momentum conservation further.
The second strategy can be achieved by considering charge puddles, which we discuss below.

{\it Charge puddles.} \cite{Vayrynen:2013,Vayrynen:2014} pointed out that charge puddles can be formed in the 2DTI heterostructure as a result of nonuniform potential landscape. These charge puddles can be viewed as accidentally formed quantum dots which host the spin-degenerate states.  
Some of these accidental dots will be close enough to the helical edge channel to be tunnel coupled to it.  

Concerning edge conductance, charge puddles allow for inelastic processes, where an edge electron tunnels into the dot and interacts with another electron during the dwelling time, and then tunnels back to the channel with reversed moving direction. 
In addition, the presence of an accidental dot introduces new characteristic energy scales smaller than the 2DTI bulk gap $\Delta_{\rm b}$. As a result, the low-temperature transport is governed by the dot energy level $\delta_{\rm d}$, the charging energy $E_{\rm ch}$ and the level width $\Gamma_{\rm t}$ set by the tunneling between the dot and the edge state. 
By considering a single puddle occupied by an even number of electrons and weak interactions, \cite{Vayrynen:2013} computed the conductance correction due to the inelastic 1PB and 2PB with the corresponding corrections $\delta G_{\rm cp1pb}  $ and $\delta G_{\rm cp2pb}  $, respectively. 
In the low-temperature limit, one finds $ \delta G_{\rm cp1pb}  \propto  -  T^{4} $ and $ \delta G_{\rm cp2pb} \propto  -  T^{6} $, respectively. Therefore, typically, $\delta G_{\rm cp1pb}$ is dominant, and one can restrict the analysis to the 1PB process. Correspondingly, we drop the part ``1pb" from the subscripts below.

We summarize first the conductance correction due to a single puddle, which is relevant for relatively short channels.  
The correction strongly depends on the dot parity state, which is determined by the  occupation number in the dot (being even or odd) and can be adjusted through the gate voltage. 
When the dot is occupied by an even number of electrons, the low-temperature correction is
\begin{eqnarray}
\delta G_{\rm cp, e}^{\rm low}  & \propto & -  \frac{e^2}{h} \frac{  \Gamma_{\rm t}^{4} }{ g^2 \delta_{\rm d}^4 } \left(  \frac{k_{\rm B} T }{ E_{\rm ch} } \right)^4,
\label{Eq:G_Vayrynen1} 
\end{eqnarray}
with $g=E_{\rm Th}/\delta_{\rm d}$ and the Thouless energy $E_{\rm Th}$ of the dot. 
The $T^4$ dependence holds at low temperatures, $k_{\rm B} T \ll \delta_{\rm d} $. For the temperature range $  \delta_{\rm d} \ll k_{\rm B} T  \ll  E_{\rm ch}$, the dependence changes to 
\begin{eqnarray}
\delta G_{\rm cp, e}^{\rm mid}  & \propto & -  \frac{e^2}{h}  \left(  \frac{k_{\rm B} T }{ \delta_{\rm d} } \right)^2.  
\label{Eq:G_Vayrynen2}
\end{eqnarray} 
The correction eventually saturates at
 \begin{eqnarray}
\delta G_{\rm cp, e}^{\rm hi}  & \propto & -  \frac{e^2}{h} \frac{ \Gamma_{\rm t}  }{ \delta_{\rm d} } ,
\label{Eq:G_Vayrynen3}
\end{eqnarray}  
for $k_{\rm B} T \gg E_{\rm ch}$.

Upon changing the gate voltage, the magnitude of the conductance correction goes through a local maximum whenever a dot energy level passes through the Fermi energy, due to the enhanced dot-edge tunneling. At such a transition between the even- and odd-occupation states, the deviation has distinct temperature power laws, again different for different ranges. For $k_{\rm B} T \ll \Gamma_{\rm t}$, one has
\begin{eqnarray}
\delta G_{\rm cp,t}^{\rm low}  & \propto & -  \frac{e^2}{h} \frac{(k_{\rm B} T)^{4} }{ g^2 \Gamma_{\rm t}^2 \delta_{\rm d}^2 }, 
\label{Eq:G_Vayrynen4} 
\end{eqnarray}
which is also $T^4$ as the even-parity dot state. 
At higher temperatures, $\Gamma_{\rm t} \ll k_{\rm B} T \ll \delta_{\rm d}$, the correction becomes temperature independent,  
\begin{eqnarray}
 \delta G_{\rm cp, t}^{\rm mid} &\propto& -  \frac{e^2}{h} \frac{ \Gamma_{\rm t}^2  }{ g^2 \delta_{\rm d}^2 }.
 \label{Eq:G_Vayrynen5} 
\end{eqnarray}
Further increasing the temperature to $ k_{\rm B} T \gg \delta_{\rm d}$ gives the same corrections as \eref{Eq:G_Vayrynen2} and \eref{Eq:G_Vayrynen3} for the corresponding temperature ranges.
 
Finally, when the dot is in the odd-parity state,  
one has a single unpaired spin in the dot.
 As discussed in section~\ref{Sec:Scattering_noTRS}, since the Kondo effect can influence the transport, there emerges an additional energy scale set by the Kondo temperature $T_{\rm K}$.    
For $T \ll T_{\rm K} $, one gets
\begin{eqnarray}
\delta G_{\rm cp, o}^{\rm low}  & \propto & -  \frac{e^2}{h} \frac{ \Gamma_{\rm t}^2 }{ g^2 (\delta_{\rm d} + E_+ )^2 }   \left( \frac{T}{T_{\rm K}} \right) ^4 , 
\label{Eq:G_Vayrynen6} 
\end{eqnarray}
where $E_+ = 2 E_{\rm ch} (N_e - C_{\rm g} V_{\rm g} /e - 1/2)$ is the energy cost of adding an electron into the dot with the electron number $N_e$, the gate-dot capacitance $C_{\rm g}$ and the gate voltage $V_{\rm g}$. 
For $ k_{\rm B} T_{\rm K} \ll k_{\rm B} T  \ll  \delta_{\rm d} $, one gets 
\begin{eqnarray}
\delta G_{\rm cp, o}^{\rm mid}  & \propto & -  \frac{e^2}{h} \frac{ \Gamma_{t}^2  }{ g^2 (\delta_{\rm d} + E_+ )^2 } \frac{ \ln^2 \left( \frac{T}{T_{\rm K}} \right) }{ \ln^2 \left( \frac{\delta_{\rm d} }{k_{\rm B} T_{\rm K}} \right) }.  
\label{Eq:G_Vayrynen7} 
\end{eqnarray}
For $k_{\rm B} T \gg \delta_{\rm d} $, the corrections are the same as \eref{Eq:G_Vayrynen2} and \eref{Eq:G_Vayrynen3} in the corresponding temperature ranges. 
Whereas the corrections in \eref{Eq:G_Vayrynen1}--\eref{Eq:G_Vayrynen7} have $T^4$ dependence in the low-$T$ limit, similar to the 1PB in a generic helical liquid [see \eref{Eq:R_5_imp}], in the intermediate temperature range the $T$ dependence becomes weaker.
In consequence, charge puddles can be the dominating resistance source in weakly interacting channels at this temperature range.

Remarkably, since the dot level spacing is determined by the puddle size, there can be puddles of certain sizes with one of the dot levels aligned to the Fermi energy of the helical channel. In this case, electrons can tunnel into the puddle easily and get scattered. Such a ``resonant condition'' indicates that mostly puddles with particular sizes participate in the puddle-induced scattering.
In addition, the dwelling of carriers in the puddles can enhance inelastic backscattering if the dwelling time is longer than the scattering lifetime.  
Taking this into account for a long helical channel coupled to multiple puddles and assuming incoherent scattering off the puddles, 
one can find the resistance upon averaging over the puddle configuration (weighted by puddles with certain sizes and tunneling rates). 

The resistance for a long channel depends on the relative magnitude of the charging energy and the dot level spacing.
For $E_{\rm ch} \ll \delta_{\rm d}$, the charging effect is negligible and the scattering is dominated mostly by dots with $\Gamma_{\rm t}$ comparable to $k_{\rm B} T$, which leads to (for $k_{\rm B} T  \ll \delta_{\rm d}$)
\begin{eqnarray}
R_{\rm cp, \ll }  & \propto & \frac{h}{e^2}  L a  n_{\rm cp} \frac{ k_{\rm B} T }{ g^2 \delta_{\rm d} }, 
\label{Eq:R_Vayrynen1} 
\end{eqnarray}
with the dot density $n_{\rm cp}$ (puddle number per area).
For $E_{\rm ch} \approx \delta_{\rm d}$, one gets  (for $k_{\rm B} T  \ll \delta_{\rm d}$)
\begin{eqnarray}
R_{\rm cp, \approx }  & \propto & \frac{h}{e^2}  L a  n_{\rm cp} \frac{1 }{ \ln^2 \left( \frac{\delta_{\rm d} } { k_{\rm B} T} \right) }.  
\label{Eq:R_Vayrynen2} 
\end{eqnarray}
For $E_{\rm ch} \gg \delta_{\rm d}$, the system is in the Coulomb-blockade regime. The resistance is dominated by puddles with an odd occupation number, which reads 
\begin{eqnarray}
R_{\rm cp, \gg }^{\rm low}  & \propto & \frac{h}{e^2}  L a  n_{\rm cp}  \frac{1 }{ g^2  \ln^2  \left( \frac{\delta_{\rm d} } { k_{\rm B} T} \right) } 
\label{Eq:R_Vayrynen3} 
\end{eqnarray}
for $k_{\rm B} T  \ll \delta_{\rm d} e^{- \pi E_{\rm ch} / (2 \delta_{\rm d})}$, and 
\begin{eqnarray}
R_{\rm cp, \gg }^{\rm hi}  & \propto & \frac{h}{e^2}  L a  n_{\rm cp} \frac{ \delta_{\rm d} }{ g^2 E_{\rm ch} } \frac{1 }{ \ln  \left( \frac{\delta_{\rm d} } { k_{\rm B} T} \right)}  
\label{Eq:R_Vayrynen4} 
\end{eqnarray}
for $\delta_{\rm d} e^{- \pi E_{\rm ch} / (2 \delta_{\rm d})} \ll k_{\rm B} T  \ll \delta_{\rm d}$.
As a consequence, the resistance in the Coulomb-blockade regime has a weak $T$ dependence for $k_{\rm B} T  \ll \delta_{\rm d}$. 

So far we have discussed various mechanisms arising from the SOI-induced terms in \eref{Eq:H_ee_ghl}. 
In contrast, whether the SOI-induced terms in \eref{Eq:H_imp_ghl} [or, more precisely, the backscattering terms in \eref{Eq:H_imp_b}] induce a finite resistance by themselves has led to a lively debate.  
Since their role in transport was discussed in the context of bosonized models, we will come back to this topic in section~\ref{Sec:generic_hTLL}. 
Before that, we discuss how a term similar to \eref{Eq:H_imp_ghl} can arise in a weakly interacting channel when we go beyond the static disorder potential. 

{\it Noise-induced backscattering.} 
Remarkably, apart from static disorder, the restrictions from the momentum and energy conservation can be lifted by charge noise which includes not only spatial but also temporal fluctuations.  
Specifically, instead of a static potential in \eref{Eq:H_imp_ghl}, one can consider the coupling to noise-induced potential~\cite{Vayrynen:2018},
\begin{eqnarray}
 H_{\rm ns}  (t) &=& \frac{w (t) }{L} \sum_{q k} V_{{\rm ns}, q - k} 
 \nonumber \\
&& \hspace{20pt} \times
 \sum_{\alpha \beta }  [U_{\rm so}^{\dagger}(q) U_{\rm so}(k)]_{\alpha\beta} \psi_{\alpha, q}^\dagger \psi_{\beta, k}.
\label{Eq:H_ns_ghl} 
\end{eqnarray}
Here, the noise is modeled by a product of functions of the time coordinate $w(t)$ and the spatial coordinate $V_{\rm ns}(r)$, which has Fourier components $V_{{\rm ns}, k}$.
The backscattering terms $\alpha \neq \beta$ in \eref{Eq:H_ns_ghl} lead to a conductance correction
\begin{eqnarray}
 \delta G_{\rm ns}  &=& - \frac{ e^2  }{ h } \int dq  \int dk \, \left\{ [U_{\rm so}^{\dagger}(q) U_{\rm so}(k)]_{12} \right\}^2
 \nonumber \\
&& \hspace{20pt} \times \frac{1}{ 2 \pi \hbar k_{\rm B} T} \big| V_{{\rm ns}, k+q + 2k_F} \big|^2 S_{w} (v_F  |q-k|)
 \nonumber \\
&& \hspace{20pt} \times  f_F (\hbar v_F k) [1-f_F (\hbar v_F k)].
\label{Eq:G_ns1}
\end{eqnarray}
Here, $f_F$ is the Fermi-Dirac distribution and the noise enters through its spectrum,
\begin{eqnarray}
 S_{w} (\omega) &=& \int_{-\infty}^{\infty}  d\tau \, e^{i \omega \tau} \langle w(t) w (t+\tau) \rangle_{\rm st},
 \end{eqnarray}
which is the Fourier transform of the noise time-correlator. The latter is defined as a statistical (long-time) average
\begin{eqnarray}
 \langle w(t) w (t+\tau) \rangle_{\rm st} &=& \lim_{T_0 \to \infty} \frac{1}{T_0} \int_0^{T_0} dt \, w(t) w (t+\tau).
 \end{eqnarray}
The general formula \eref{Eq:G_ns1} displays two important points. 
First, for a helical channel preserving the $S^z$ symmetry, the matrix $U_{\rm so}$ is diagonal and there is no conductance correction from any kind of charge noise.
Second, for static disorder, described by a constant $w(t)$, we have  $ S_{w} (\omega) \propto \delta (\omega)$, which imposes $q=k$ in the integrand and thus zero correction. Therefore, for the noise-induced term \eref{Eq:H_ns_ghl} to have nonvanishing effects on the conductance, one has to remove the $S^z$ conservation and to involve fluctuating disorder in time.\footnote{This result supports the conclusion of \cite{Kainaris:2014} that the static perturbation \eref{Eq:H_imp_b} alone does not influence the charge transport, which will be discussed in section~\ref{Sec:generic_hTLL}.} 

To this end, one incorporates the spin-orbit-induced rotation [see \eref{Eq:U_so}] and obtains~\cite{Vayrynen:2018}
\begin{eqnarray}
 \delta G_{\rm ns}  &\propto& - \frac{ e^2  }{ h }   \frac{ |V_{{\rm ns}, 2k_F}|^2 n_{{\rm so} \perp}^2 }{v_F^2 D_{\rm so}^2 } \int \frac{d \omega}{2\pi} \, \omega^2  S_{w} (\omega) .
 \label{Eq:G_ns2}
\end{eqnarray} 
Here $n_{{\rm so} \perp} \approx 1$ depends on the direction of spin rotation and $D_{\rm so}$ denotes the energy scale over which the spin rotates; using the notation in \eref{Eq:U_so}, we have $D_{\rm so} = \hbar v_F  k_{\rm so}$.   
Since \eref{Eq:G_ns2} holds for a general noise, one can consider, in particular, telegraph noise from fluctuating two-level charge states in puddles, which have Lorentzian noise spectra.
For a long channel with multiple charge puddles, averaging over the random parameters of the two-level fluctuators yields
\begin{eqnarray}
R_{\rm tls}  &\propto&  \frac{h}{ e^2  }  L n_{\rm imp}   \frac{ |V_{{\rm ns}, 2k_F}|^2 n_{{\rm so} \perp}^2 }{\hbar v_F^2  E_{\rm ch} \tau_{\rm tls} }  \frac{ (k_{\rm B} T)^2 } { D_{\rm so}^2 } \tanh \left( \frac{E_{\rm ch}} {2 k_{\rm B} T} \right) , \nonumber \\
 \label{Eq:G_tls_Vayrynen2}
\end{eqnarray}
where $\tau_{\rm tls} $ is the relaxation time of the excited state of the two-level fluctuator.  
Focusing on a regime where $\tau_{\rm tls}$ is independent of temperature, one can see that $R_{\rm tls} \propto T^2$ for $k_{\rm B} T \ll E_{\rm ch}$, whereas $R_{\rm tls} \propto T$ for $k_{\rm B} T \gg E_{\rm ch}$.

Finally, for $1/f$ noise, which is omnipresent in solid-state devices and can be viewed as an ensemble of two-level fluctuators,
one can average over the distribution of $\tau_{\rm tls}$ and get $R_{1/f}^{\rm low}  \propto T^2$ for $k_{\rm B} T \ll E_{\rm ch}$, and
\begin{eqnarray}
R_{1/f}^{\rm hi}  &\propto&  \frac{h}{ e^2  }  L n_{\rm imp}   \frac{ |V_{{\rm ns}, 2k_F}|^2 n_{{\rm so} \perp}^2 }{\hbar^2 v_F^2 }  \frac{\delta_{\rm d}  k_{\rm B} T }{ D_{\rm so}^2 }
 \label{Eq:G_inv-f_Vayrynen4}
\end{eqnarray}
 for $k_{\rm B} T \gg E_{\rm ch}$. 
In comparison with $T^4$ or $T^6$ power-law resistance of a generic helical liquid due to static disorder, the noise-induced resistance has a weaker $T$ dependence. 

Interestingly, applying an external ac gate voltage described by $V_{\rm ac}(r) \cos (\omega_{\rm ac} t)$ results in a deviation of the dc conductance for $\hbar \omega_{\rm ac}, ~k_{\rm B} T \ll D_{\rm so}, ~\hbar v_F k_F$,
\begin{eqnarray}
\delta G_{ac} &\propto&  \frac{ e^2 }{h}   \frac{ |V_{{\rm ac}, 2k_F}|^2 n_{{\rm so} \perp}^2 }{v_F^2 D_{\rm so}^2 }  \omega_{\rm ac}^2, 
 \label{Eq:G_ac_Vayrynen}
\end{eqnarray}
where $V_{{\rm ac}, 2k_F}$ is the spatial Fourier transform of the ac voltage.
In consequence, by verifying the quadratic dependence of $\delta G$ on the voltage amplitude $|V_{\rm ac}| $ and frequency $\omega_{\rm ac}$, one can demonstrate the existence of the momentum-dependent spin texture (that is, a nonzero $1/D_{\rm so} \propto 1/k_{\rm so}$), which is important for characterizing the generic helical liquid~\cite{Vayrynen:2018}.

Having discussed time-reversal-invariant perturbations in a weakly interacting channel, we now move on to a helical channel with an arbitrary interaction strength.
In particular, we aim to discuss how the electron-electron interactions, known for enhancing the backscatterings in one dimension, affect the transport in a helical channel.

\subsubsection{Generic helical Tomonaga-Luttinger liquid \label{Sec:generic_hTLL}}
~\\
As discussed in section~\ref{Sec:hTLL}, since electrons in a helical channel are spatially confined, one expects substantial effects from the electron-electron interactions. It is thus natural to go beyond the weakly interacting fermion picture described in the previous subsection. Here, we look into a different regime, where electron-electron interactions are no longer weak perturbations. Since we consider a generic helical liquid beyond the weak-interaction regime, we adopt an analogous name for it: {\it generic helical Tomonaga-Luttinger liquid} (generic hTLL). 

It is convenient to bosonize \eref{Eq:H_ee_ghl} and \eref{Eq:H_imp_ghl} and establish the bosonic form of the generic hTLL. It will allow us to investigate the resistance for arbitrary interaction strength. For convenience, we first linearize the model, which amounts to modifying the up- and down-spin fermion operators in \eref{Eq:SlowlyVarying} into
\numparts
\begin{eqnarray}
\psi_{\uparrow}(r) &=& e^{-i k_F r} L(r) - i \zeta_{\rm so}^{*} e^{i k_F r} \partial_r R(r),\\
\psi_{\downarrow}(r) &=& e^{i k_F r} R(r) - i\zeta_{\rm so} e^{-i k_F r} \partial_r L(r).
\end{eqnarray}
\endnumparts
In this form, the fermion operators mix the left- and right-moving helical states. Also, we parametrize the spin-orbit strength with a complex parameter $\zeta_{\rm so}$;
with the notation used in section~\ref{SubSec:ghl}, it reduces to $\zeta_{\rm so} = 2k_F / k_{\rm so}^2$.   
Neglecting terms of the second order in $|\zeta_{\rm so}|$, the density operator is
\begin{eqnarray}
\rho  & \approx&  R^{\dagger} R + L ^{\dagger} L \nonumber\\
&& + \Big\{ i \zeta_{\rm so} e^{-2 i k_F r}  \left[ \big( \partial_r R^{\dagger} \big) L  - R^{\dagger} \partial_r L \right] + {\rm H.c.} \Big\},
\end{eqnarray}
exhibiting in the second line the spin-orbit-induced correction proportional to $\zeta_{\rm so}$. 
These additional terms allow backscattering processes which would be forbidden in a helical channel conserving the spin component $S^z$.
Guided by the discussion on the weakly interacting fermion systems, we group the processes into three types below. 

The first two types, corresponding to  \eref{Eq:H_ee_3} and \eref{Eq:H_ee_5}, arise from the screened Coulomb interaction, $\int dr \, U_{\rm ee} (r) \rho(r)  \rho (r)$. The interaction includes the usual $g_2$ and $g_4$ forward-scattering processes described in \eref{Eq:H_ee}, as well as $\zeta_{\rm so}$-induced terms, 
\begin{eqnarray}
H_{\rm ee,3} &\approx&
\frac{8k_F^2 U_{\rm ee}}{k_{\rm so}^4 } \int dr \, e^{-4ik_F r} (\partial_r R^\dagger )R^\dagger (\partial_r L) L + {\rm H.c.}, \label{Eq:H_ee_3_linear} \nonumber \\
&& \\
H_{\rm ee,5} &\approx& \frac{4k_F U_{\rm ee}}{k_{\rm so}^2 } \int dr \, \Big[ i e^{-2ik_F r} L^\dagger  R^\dagger L  (\partial_r L)   \nonumber \\
&& \hspace{48pt} + i e^{2ik_F r} R^\dagger L^\dagger R (\partial_r R)  + {\rm H.c.} \Big].
  \label{Eq:H_ee_5_linear}  
\end{eqnarray}
These additional terms can also be derived by directly linearlizing \eref{Eq:H_ee_3} and \eref{Eq:H_ee_5}. 
We note that here we do not include $H_{\rm ee,1}$ as it does not change the number of left nor right movers and has higher-order derivatives, which makes it less relevant in the RG sense.   
As already mentioned, for in-gapped helical states the Fermi wavelength is much longer than the underlying lattice constant, so the condition for a commensurate lattice would be unrealistic.
Therefore, the oscillating factors in the above integrands indicate vanishing contributions for $k_F \neq 0$ due to momentum mismatch.
However, as in weakly interacting channels, each of \eref{Eq:H_ee_3_linear} and \eref{Eq:H_ee_5_linear} can lead to backscatterings when disorder is present.

The third type, corresponding to \eref{Eq:H_imp_b}, arises from the coupling of the disorder potential $V_{\rm imp}$ to the anomalous $\zeta_{\rm so}$ term in the density operator,
\begin{eqnarray}
H_{\rm rso} &=&  - \int dr \; V_{\rm imp}(r) \Big\{ i \zeta_{\rm so} e^{-2 i k_F r}   \nonumber \\
&& \hspace{42pt}  \times \left[ R^{\dagger} \partial_r L - \big( \partial_r R^{\dagger} \big) L  \right] + \textrm{H.c.} \Big\}, 
\label{Eq:H_rso}
\end{eqnarray} 
with $V_{\rm imp}$ characterized by the averages in \eref{Eq:Vimp_avg}. 
The expression \eref{Eq:H_rso} can be derived by directly linearlizing \eref{Eq:H_imp_b}.
Next, we discuss the resistance arising from $H_{\rm ee,3}$, $H_{\rm ee,5}$ and $H_{\rm rso}$.

{\it 1PB process in a generic hTLL.} 
In a clean channel, the backscattering by $H_{\rm ee,5}$ is allowed when the Fermi level is sufficiently close to the Dirac point~\cite{Kainaris:2014}.
This 1PB process, illustrated in \Fref{Fig:Scattering_ghl}(a), leads to the resistance for $\hbar v_F k_F \ll k_{\rm B} T$,
\begin{eqnarray}
R_{\rm 5}^{\rm hi} &\propto& \frac{h}{e^2} Lk_{\rm so} \left( \frac{U_{\rm ee}}{\hbar u} \right)^2  \left( \frac{ k_{\rm B} T}{\hbar u k_{\rm so}} \right)^{2K+3}, 
\label{Eq:R_5_hiT_hTLL}
\end{eqnarray}
which reverts to the $T^5$ power for $K=1$.   
For $\hbar v_F k_F \gg k_{\rm B} T$, we again get a thermal activation behavior, 
\begin{eqnarray}
 R_{5}^{\rm low} &\propto& \frac{h}{e^2} Lk_{\rm so} \left( \frac{U_{\rm ee}}{\hbar u} \right)^2  \frac{ (k_F a)^{2K+3} }{ (k_{\rm so} a)^5}  e^{ \frac{-\hbar u k_F}{ k_{\rm B} T} },
\label{Eq:R_5_lowT_hTLL}
\end{eqnarray}
due to the energy and momentum conservation. 

In a typical system, disorder is present and can relax the restriction from momentum conservation, by compensating the oscillating factors in  \eref{Eq:H_ee_3_linear} and \eref{Eq:H_ee_5_linear}. 
As a consequence, similar to the weak-interaction regime, the combination of $H_{\rm ee,5} $ and disorder leads to 1PB, whereas $H_{\rm ee,3} $ and disorder jointly induce 2PB,
with the momentum difference compensated by the corresponding Fourier components of the random disorder potential.   

We now consider a generic hTLL with disorder. The results obtained in the weakly-interacting-fermion picture showed that the 1PB processes involving $H_{\rm imp,b}$ are parametrically suppressed by higher powers of a small factor $k_F / k_{\rm so}$. Therefore, the 1PB process  is dominated by the combination of $H_{\rm ee,5}$ and the forward scattering term of the quenched disorder $H_{\rm imp,f}$, which we illustrate in \Fref{Fig:Scattering_ghl}(b).
In the bosonic form, it leads to a contribution to the action~\cite{Kainaris:2014}
\numparts
\begin{eqnarray}
\frac{S_{\rm 1pb} }{\hbar}  &=& - \tilde{D}_{\rm 1pb,1} u^2a \sum_{\eta \eta'} \int dr d\tau d\tau' \,
\partial_r^2 \theta_{\eta}  (r,\tau) \nonumber \\
&&\hspace{10pt} \times \partial_r^2 \theta_{\eta'}  (r,\tau') \cos \left[  2 \phi_{\eta} (r,\tau)- 2\phi_{\eta'} (r,\tau') \right] \nonumber \\
&& + \tilde{D}_{\rm 1pb,2} u^2 \sum_{\eta \eta'} \int dr d\tau d\tau' \,
\partial_r^2 \theta_{\eta}  (r,\tau) \nonumber \\
&&\hspace{10pt} \times \partial_r \theta_{\eta'}  (r,\tau') \sin \left[  2 \phi (r,\tau)- 2\phi (r,\tau') \right], 
\label{Eq:S_1pb}
\end{eqnarray}
with the replica indices $\eta$, $\eta'$ and the dimensionless backscattering strengths
\begin{eqnarray}
  \tilde{D}_{\rm 1pb,1} \equiv \frac{2 U_{\rm ee}^2 K^2 M_{\rm imp} }{\pi^6 a^3 k_{\rm so}^4 \hbar^4 u^4} ,\; 
	\tilde{D}_{\rm 1pb,2} \equiv \frac{8 k_F U_{\rm ee}^2 K^2 M_{\rm imp} }{\pi^6 a^2 k_{\rm so}^4 \hbar^4 u^4} . \nonumber \\
\end{eqnarray}
\endnumparts
Due to the higher-order derivative, \eref{Eq:S_1pb} is RG irrelevant for any interaction strength and cannot lead to localization. 
It induces a resistance,
\begin{eqnarray}
R_{\rm 1pb} &\propto& \frac{h}{e^2} L n_{\rm imp} \left( \frac{U_{\rm ee} V_{\rm imp}}{\hbar^2 u^2} \right)^2  \left( \frac{ 1 }{ k_{\rm so} a} \right)^4 \left( \frac{ k_{\rm B} T}{\Delta_{\rm b}} \right)^{2K+2}, \nonumber \\
\label{Eq:G_5f_hTLL}
\end{eqnarray}
which reduces to $\propto T^4$ in the noninteracting limit. 

Instead of the random potential  $V_{\rm imp}$ from quenched disorder, one might consider the combination of $H_{\rm ee,5}$ and a single scatterer due to a local impurity. The latter can be modeled by a Dirac delta function at the origin, $H_{\rm imp}^{\rm loc} = V_{\rm imp}^{\rm loc} \delta (r)$. 
Together with \eref{Eq:H_ee_5_linear}, one obtains
\begin{eqnarray}
H_{\rm 1pb}^{\rm loc}  
&=&  \tilde{g}_{1pb}  \hbar u a  \Big[  \partial_r^2 \theta  (0) e^{2 i \phi (0)} + {\rm H.c.} \Big], 
\end{eqnarray}
with the effective coupling $ \tilde{g}_{1pb}$ and the boson fields $\phi$ and $\theta$ at $r=0$.
The local perturbation $H_{\rm 1pb}^{\rm loc} $ was investigated by \cite{Lezmy:2012} who found the following correction to the (differential) conductance, 
\begin{eqnarray}
\delta G_{\rm 1pb} \propto - \frac{e^2}{h}  \tilde{g}_{1pb}^2  \left[ {\rm Max} \Big( \frac{eV}{\Delta_{\rm b}}, \frac{k_{\rm B} T}{\Delta_{\rm b}} \Big) \right]^{2K+2},
\label{Eq:G_5_loc}
\end{eqnarray}
in the high-temperature and high-bias limits.  
 
{\it 2PB process in a generic hTLL.} 
We now turn to the 2PB process due to $H_{\rm ee,3} $ and quenched disorder~\cite{Xu:2006,Wu:2006,Kainaris:2014}.
The scattering process is plotted in \Fref{Fig:Scattering_ghl}(c).
It contributes the following term in the action,
\numparts
\begin{eqnarray}
\frac{S_{\rm 2pb} }{\hbar}  &=& - \tilde{D}_{\rm 2pb} \frac{u^2}{a^3} \sum_{\eta \eta'} \int dr d\tau d\tau' \, \nonumber \\
&&\hspace{10pt} \times \cos \left[  4 \phi_{\eta} (r,\tau)- 4\phi_{\eta'} (r,\tau') \right], 
\label{Eq:S_2pb}
\end{eqnarray}
with 
\begin{eqnarray}
  \tilde{D}_{\rm 2pb} \equiv \frac{2 k_F^2 U_{\rm ee}^2 K^2 M_{\rm imp} }{\pi^6 a^5 k_{\rm so}^8 \hbar^4 u^4}.
\end{eqnarray}
\endnumparts
In contrast to the 1PB process, the 2PB process due to quenched disorder becomes relevant at $K<3/8$ and can localize the helical states at low temperature in a long channel.
Before entering the localization regime, the resistance induced by the 2PB process is given by
\begin{eqnarray}
R_{\rm 2pb} &\propto&  \frac{h}{e^2} L n_{\rm imp} \left( \frac{U_{\rm ee} V_{\rm imp}}{\hbar^2 u^2} \right)^2  \frac{ (k_F a)^2 }{ (k_{\rm so} a)^8} \left( \frac{ k_{\rm B} T}{\Delta_{\rm b}} \right)^{8K-2} , \nonumber \\
\label{Eq:G_3f_hTLL}
\end{eqnarray}
which recovers the $T^6$ dependence for $K \to 1$. 
In the localization regime, $R_{\rm 2pb}$ has an exponential form due to thermal activation. 
By analyzing \eref{Eq:S_2pb}, \cite{Chou:2018} pointed out that the localized phase for $K<3/8$ exhibits a glassy, insulating edge state, which breaks the time-reversal symmetry locally, but preserves the global time-reversal symmetry upon disorder average.  

In addition to quenched disorder, a single scatterer can also induce the 2PB~\cite{Wu:2006,Maciejko:2009,Lezmy:2012}.
Combining $H_{\rm imp}^{\rm loc}$ with \eref{Eq:H_ee_3_linear} leads to
\begin{eqnarray}
H_{\rm 2pb}^{\rm loc} & = &  \tilde{g}_{2pb} \frac{\hbar u}{a} \cos [4 \phi (0)], 
\end{eqnarray}
with the effective coupling $ \tilde{g}_{2pb}$.
It gives rise to a correction to the edge (differential) conductance,
\begin{eqnarray}
\delta G_{\rm 2pb}  & \propto & - \frac{e^2}{h}  \tilde{g}_{2pb}^2  \left[ {\rm Max} \Big( \frac{eV}{\Delta_{\rm b}}, \frac{k_{\rm B} T}{\Delta_{\rm b}} \Big) \right]^{8K -2},
\label{Eq:G_2pb_Maciejko3}
\end{eqnarray} 
assuming $K>1/4$. In the regime $K<1/4$, where the 2PB is relevant, $ \tilde{g}_{2pb}$ flows to the strong-coupling limit and blocks the current. 
In this case, the charge transport at low temperature proceeds by tunneling through the single impurity. The tunneling conductance, $G_{\rm tun}   \propto   T^{1/(2K) -2}$~\cite{Maciejko:2009}, is analogous to a nonhelical channel~\cite{Kane:1992}, though the power law is different.

We have two remarks regarding the 2PB process. 
First, a fractional power-law conductance was observed in InAs/GaSb edge channels by~\cite{Li:2015}. To extract the value of $K$, they fitted the measured curves to the above $T^{1/(2K) -2}$ power law.  
However, it remains to be clarified whether the conductance was indeed due to the tunneling through a single impurity or other resistance sources, which would invalidate the $K$ extraction since the power law might be different.   
Second, while the 2PB process from both a local impurity and quenched disorder result in similar power-law resistance, they differ in the crossover value of the interaction parameter below which the channel becomes insulating, namely $K<1/4$ and $K < 3/8$, respectively.
 This difference, arising from the different scaling dimensions of the corresponding operators, has been pointed out in the context of nonhelical channels~\cite{Giamarchi:2003,Hsu:2019}. 
Generally speaking, the additional time integral in the disorder-averaged effective action [see $\eref{Eq:S_2pb}$] makes the backscattering due to quenched disorder more RG relevant than the one due to a single impurity. 

 {\it Random SOI}.
We now revisit the spin-orbit-induced term given in \eref{Eq:H_rso}.
As explained there, it does not involve $H_{\rm ee}$ but arises from the combination of the anomalous $\zeta_{\rm so}$ term in the density operator and the quenched disorder $V_{\rm imp}$. 
In this review, we refer to \eref{Eq:H_rso} as {\it random spin-orbit interaction} (random SOI); see \Tref{Tab:nomenclature} for the nomenclature for the same term in the literature. 
 
By performing the disorder average, one obtains the following term in the action~\cite{Strom:2010},
\numparts
\begin{eqnarray}
\frac{ S_{\rm rso} }{\hbar}  &=& - \frac{\tilde{D}_{\rm rso} u^2}{a} \sum_{\eta \eta'} \int dr d\tau d\tau' \,
\partial_r \theta_{\eta}  (r,\tau) \partial_r \theta_{\eta'}  (r,\tau') \nonumber \\
&&\hspace{24pt} \times \cos \left[  2 \phi_{\eta} (r,\tau)- 2\phi_{\eta'} (r,\tau') \right],  
\label{Eq:S_rso}
\end{eqnarray}
with the dimensionless coupling
\begin{eqnarray}
  \tilde{D}_{\rm rso} \equiv \frac{4 k_F^2 M_{\rm imp} }{\pi^2 a k_{\rm so}^4 \hbar^2 u^2} + \frac{32 k_F^2 U_{\rm ee}^2 K^2 M_{\rm imp} }{\pi^6 a k_{\rm so}^4 \hbar^4 u^4}.
\end{eqnarray}
\endnumparts
The effects of this random SOI has been studied in \cite{Strom:2010,Geissler:2014,Kainaris:2014,Xie:2016,Kharitonov:2017} and, in a combination with dynamic nuclear spin polarization, in~\cite{DelMaestro:2013}. We now discuss these results.

By computing the scattering time due to \eref{Eq:S_rso}, \cite{Kainaris:2014} demonstrated that the random SOI does not lead to a finite scattering time to the first order in $ \tilde{D}_{\rm rso}$ for any interaction strength. They thus concluded that, to the lowest order, the ballistic transport of the edge states is protected against perturbations given by \eref{Eq:H_rso} alone, contradicting the conclusions of \cite{Strom:2010,Geissler:2014}. 
To resolve the paradox, \cite{Xie:2016} analyzed \eref{Eq:H_rso} by mapping the problem onto a free boson theory with inhomogeneous density-density interactions. 
Upon the mapping, the random SOI is translated into modifications of the velocity $u(r)$ and the interaction parameter $K(r)$ that are random in space.  
As a result, the conductance of a helical channel in contact with Fermi-liquid leads is quantized irrespective of the random SOI. This property is analogous to nonhelical channels~\cite{Maslov:1995,Ponomarenko:1995,Safi:1995}, where the ballistic conductance does not depend on the interaction strength and the velocity in the channel. 
This result should not be surprising, given that the linear-in-momentum SOI terms in strictly one-dimensional systems (without Zeeman terms) can be gauged away and does not affect the charge transport~\cite{Braunecker:2010,Meng:2014c,Kainaris:2015}.
A detailed analysis on the random SOI combined with backscattering current calculation at zero temperature~\cite{Kharitonov:2017} confirmed the conclusion from \cite{Kainaris:2014,Xie:2016}, together with the criterion for which a generalized backscattering beyond the simple form \eref{Eq:H_rso} is allowed.
Specifically, the generalized backscattering is not forbidden in a realistic system, since it requires only one of the following ingredients: nonlocal $U(1)$-symmetry breaking terms, nonlocal/SU(2)-symmetry-breaking electron-electron interactions, or the high energy cutoff in the edge-state spectrum. 
However, one expects that the resistance is parametrically suppressed.

{\it Higher-order random SOI}. 
\cite{Crepin:2012} investigated the higher-order contributions of \eref{Eq:H_rso} to the effective action. In combination with the forward scattering terms from $H_{\rm ee}$, it generates an effective 2PB process similar to \eref{Eq:S_2pb}.
To distinguish it from the 2PB from $H_{\rm ee,3}$, we refer to it as {\it higher-order random SOI} in this review. 

With a single impurity $H_{\rm imp}^{\rm loc}$, the higher-order random SOI leads to a conductance correction for $T<T^{*}_{\rm rso}$,  
\begin{eqnarray}
\delta G_{\rm horso}^{\rm low}  
& \propto & 
\left\{ \hspace{-5pt}
\begin{array}{l}
- \frac{e^2}{h}   \Big( \frac{k_{\rm B} T}{\Delta_{\rm b}} \Big)^{8K-2}  ~{\rm for }~ 1/4 < K < 1/2, \\
- \frac{e^2}{h}  
 \Big( \frac{k_{\rm B} T}{\Delta_{\rm b}} \Big)^{4K}  ~{\rm for }~ K > 1/2, 
 \end{array}
 \right. 
\label{Eq:G_ho-rso_Crepin-low}
\end{eqnarray}
multiplied by an overall prefactor $\tilde{g}_{\rm horso}^2 \propto \alpha_R^4/(\hbar u)^4 $. 
The crossover temperature is given by $T^{*}_{\rm rso} = \Delta_{\rm b} [2K/(4K-1)]^{1/(2K-1)}$.
For $T>T^{*}_{\rm rso}$ the correction has a logarithmic dependence for any $K$ value,
\begin{eqnarray}
\delta G_{\rm horso}^{\rm hi}  & \propto & 
- \frac{e^2}{h}  \tilde{g}_{\rm horso}^2   \Big( \frac{k_{\rm B} T}{\Delta_{\rm b}} \Big)^{4K}  \ln^2 \Big( \frac{k_{\rm B} T}{\Delta_{\rm b}} \Big).
 \label{Eq:G_ho-rso_Crepin-hi}
\end{eqnarray}
In consequence, whereas for sufficiently strong interactions the higher-order random SOI leads to the same power law as the 2PB \eref{Eq:G_2pb_Maciejko3}, for weaker interactions the two processes lead to distinct resistance. In particular, in the weak-interaction limit $K \approx1$, the above temperature dependence predicts a $T^4$ power law instead of $T^6$ from \eref{Eq:G_2pb_Maciejko3}. However, one should notice that the leading-order contribution here is of second order in $H_{\rm imp,b}$, so the  conductance correction is of fourth order in a typically small parameter $\alpha_{\rm R}/(\hbar u)$. 

Before proceeding, we briefly summarize the SOI-induced mechanisms discussed so far. To this end,  in \Tref{Tab:nomenclature} we list the backscattering processes and the nomenclature in the literature.

\begin{table*}[t]
\centering
\caption{
\label{Tab:nomenclature}
SOI-induced backscattering mechanisms discussed in section~\ref{Sec:Scattering_TRS}. The mechanisms involve the electron-electron interactions $H_{\rm ee}= \sum_{i=1}^5 H_{{\rm ee},i}$ [see \eref{Eq:H_ee_1}--\eref{Eq:H_ee_5}] and disorder. The latter is modeled either in the form of quenched disorder potential $H_{\rm imp} = H_{\rm imp,f} + H_{\rm imp,b}$ [see \eref{Eq:H_imp_f}--\eref{Eq:H_imp_b}] or a single local impurity $H_{\rm imp}^{\rm loc} = V_{\rm imp}^{\rm loc} \delta (r)$. 
For a given mechanism, we list the involved terms in $H_{\rm ee}$, $H_{\rm imp} $ and $H_{\rm imp}^{\rm loc}$, the equation number(s) for the induced resistance $R$ or conductance correction $\delta G$, the corresponding references to the original works, and their names or notations therein. 
}
\footnotesize\rm
\begin{indented}
\item[]\begin{tabular}[c]{  ll  l  l }
\br
Involved terms & Equation & Reference &  Notation or name in the original work    \\
\mr
 $H_{\rm ee,1} $ and $ H_{\rm imp,b}$   & \eref{Eq:R_1_imp} &   \cite{Kainaris:2014}  &   $g_{1} \times b $ process   \\
\mr
                                                      & ~ & \cite{Wu:2006} & $H_{\rm dis}$ or two-particle backscattering due to quenched disorder \\
 $H_{\rm ee,3} $ and $  H_{\rm imp,f}$     & \eref{Eq:R_2pb}, \eref{Eq:G_3f_hTLL}  &\cite{Xu:2006} & Scattering by spatially random quenched impurities \\
                                                         & ~ &\cite{Kainaris:2014} & $g_3 \times f$ process (in their class of two-particle processes) \\ 
\mr
$H_{\rm ee,3} $ and $   H_{\rm imp,b}$ &  \eref{Eq:R_3_imp} &  $ \hspace{-6pt}  \begin{tabular}{l}  \cite{Schmidt:2012}   \\ \cite{Kainaris:2014}  \end{tabular} $ &
$ \hspace{-6pt}  \begin{tabular}{l}    $H_{{V}, {\rm int}}^{\rm eff}$ \\ $g_3 \times b$ process (in their class of one-particle processes)   \end{tabular} $ \\
\mr
                       & ~ & \cite{Wu:2006} & $H_{\rm bs}^\prime$ or impurity-induced two-particle correlated backscattering \\
 $H_{\rm ee,3}  $ and $  H_{\rm imp}^{\rm loc}$   & \eref{Eq:G_2pb_Maciejko3} & \cite{Maciejko:2009} & $H_{2}$ or local impurity-induced two-particle backscattering  \\
                         & ~ & \cite{Lezmy:2012} & $g_{\rm 2p}$ process or two-particle scattering \\
\mr
$H_{\rm ee,5}$$^{\rm a}$  & 
$ \hspace{-6pt}  \begin{tabular}{l}   \eref{Eq:R_5_highT}--\eref{Eq:R_5_lowT},  \\ \eref{Eq:R_5_hiT_hTLL}--\eref{Eq:R_5_lowT_hTLL} \end{tabular} $ &
$ \hspace{-6pt}  \begin{tabular}{l}   \cite{Schmidt:2012} \vspace{10pt}  \\  \cite{Kainaris:2014} \\  \cite{Chou:2015}  \end{tabular} $ &
$ \hspace{-6pt}  \begin{tabular}{l}   $H_{\rm int}$ or inelastic backscattering of a single electron with \\ energy transfer to another particle-hole pair \\ 
$g_5$ process \\  $\hat{H}_{W}$ or one-particle spin-flip umklapp term  \end{tabular} $  
 \\
\mr
 $H_{\rm ee,5} $ and $   H_{\rm imp,f}$  & \eref{Eq:R_5_imp}, \eref{Eq:G_5f_hTLL}   & 
  $ \hspace{-6pt}  \begin{tabular}{l}   \cite{Kainaris:2014}  \\  \cite{Chou:2015}  \end{tabular} $ 
 &   
 $ \hspace{-6pt}  \begin{tabular}{l}    $g_{5} \times f $  process (in their class of one-particle processes)  \\ $\hat{H}_{W}$ (same notation for clean and disordered systems)  \end{tabular} $ \\
\mr
 $H_{\rm ee,5} $ and $  H_{\rm imp,b}$ & N/A & \cite{Kainaris:2014}  &  $g_{5} \times b $ (in their class of one-particle processes) \\
\mr 
$H_{\rm ee,5} $ and $   H_{\rm imp}^{\rm loc}$  & \eref{Eq:G_5_loc} & \cite{Lezmy:2012}&  $g_{ie}$ process or inelastic scattering \\ 
\mr
Random SOI$^{\rm b}$ & N/A & $ \hspace{-6pt}  \begin{tabular}{l}   \cite{Strom:2010} \\ \cite{Geissler:2014} \\ \cite{Kainaris:2014}  \\  \cite{Xie:2016} \end{tabular} $ & 
$ \hspace{-6pt}  \begin{tabular}{l}   $H_{R}$ or randomly fluctuating Rashba spin-orbit coupling \\ Random Rashba spin-orbit coupling \\ 
$g_{\rm imp,b}$ process   \\   Random Rashba backscattering \end{tabular} $ \\
\mr
Random SOI$^{\rm c}$ & N/A & \cite{Kharitonov:2017} & $\hat{H}_R$ or $U(1)$-asymmetric single-particle backscattering field \\
\mr
$ \hspace{-6pt}  \begin{tabular}{l}   $H_{\rm rso} $ and $  H_{\rm imp}^{\rm loc}$$^{\rm d}$   \\  \end{tabular} $ & \eref{Eq:G_ho-rso_Crepin-low}--\eref{Eq:G_ho-rso_Crepin-hi} &  \cite{Crepin:2012}  & Inelastic two-particle backscattering from a Rashba impurity \\ 
\br
\end{tabular}
\item[] $^{\rm a}$ For clean systems, $H_{\rm ee,5}$ can lead to a finite resistance by itself.
\item[] $^{\rm b}$ The random SOI $H_{\rm rso}$ arises from the backscattering terms $H_{\rm imp,b} $ due to the quenched disorder potential. 
\item[] $^{\rm c}$ \cite{Kharitonov:2017} considered a nonlocal generalization of $H_{\rm rso}$.
\item[] $^{\rm d}$ \cite{Crepin:2012} considered the second-order contribution from $H_{\rm rso}$ in the presence of a local impurity.
\end{indented}
\end{table*}
   

{\it Phonon-induced backscattering}. 
In addition to disorder, phonons can also compensate the momentum and energy difference for scatterings. 
\cite{Budich:2012} investigated electron-phonon coupling in a one-dimensional helical channel. They considered spatially-dependent linear and cubic Rashba terms,
\numparts
\begin{eqnarray}
H_{\rm R1}  & =  & \frac{1}{2}  \int dr \, \Psi^\dagger (r)\left\{ \alpha_{\rm R1} (r) , \frac{ 1}{i} \partial_r     \right\} \sigma^y \Psi (r), \\ 
H_{\rm R3}  & =  & \frac{1}{2}  \int dr \, \Psi^\dagger (r) \left\{ \alpha_{\rm R3} (r) , i  \partial_r^3  \right \} \sigma^y \Psi (r),
\end{eqnarray}
\endnumparts
with the linear  $\alpha_{\rm R1}$ and cubic $\alpha_{\rm R3}$ Rashba spin-orbit coupling strengths and the two-component vector $\Psi (r)\equiv [\psi_+ (r), \psi_- (r) ]^{\rm T}$. In the latter, ${\rm T}$ is the transpose operator and $\psi_{\pm}$ is the inverse Fourier transform of the fermion fields introduced in \eref{Eq:ghl_transform}.
Finally, the helical states are linearly coupled to longitudinal acoustic phonons~\cite{Martin:1995}, 
\begin{eqnarray}
H_{\rm e-ph}  & =  &\lambda_{\rm ep} \int dr \, \Psi^\dagger (r) \Psi (r) \partial_r  d_{\rm ph}, 
\label{Eq:H_e-ph}
\end{eqnarray}
where $\lambda_{\rm ep}$ is the electron-phonon coupling strength and $d_{\rm ph}$ is the displacement field of the phonon.

\cite{Budich:2012} found that the first- and second-order contributions from the linear Rashba term to the backscattering matrix element vanish for any interaction strength.
To obtain a nonvanishing matrix element, one needs to involve either the cubic Rashba term or the third-order contribution from the linear Rashba term.
For instance, in a noninteracting channel with a local cubic Rashba impurity they found a nonlinear conductance correction $\delta G_{\rm e-ph} \propto -V^6 $ at zero temperature.
However, due to the higher-order origin of the prefactor and high power of $V$, one expects a negligible resistance for typical conditions.


\begin{table*}[t]
\centering
\caption{
\label{Tab:mechanism}
Temperature ($T$) dependence of the resistance ($R$) or the conductance correction  $(\delta G) $ due to backscattering mechanisms that break the time-reversal symmetry discussed in section~\ref{Sec:Scattering_noTRS}. 
We give the expressions for general $K$, if present in the literature; the noninteracting expressions follow upon using $K = 1$. 
In realistic settings, multiple resistance sources are present; they should be added in series, including the contact resistance, $h/e^2$, per channel.  
Here and in \Tref{Tab:mechanism2}, the abbreviations 1D, 1PB, 2PB, DNP, SOI and TRS stand for one-dimensional, one-particle backscattering, two-particle backscattering, dynamic nuclear spin polarization, spin-orbit interaction, and time-reversal symmetry, respectively. 
}
\footnotesize\rm
\begin{indented}
\item[]\begin{tabular}[c]{  l  l  l   }
\br
 TRS breaking mechanism & $R$ or $ - \delta G$ & Remark \\
\mr
Single magnetic impurity   &   
$ \left\{ \hspace{-6pt} 
\begin{tabular}{ll}  $T^{2K-2}$ & for $T \ll T_{\rm K}$ \\ 
 const. $ + \ln \left( \frac{\Delta_{\rm b}}{ k_{\rm B} T} \right) $ &  for $T > T_{\rm K}$
 \end{tabular} \right. $ 
& See Footnote ${\rm a}$ for the regime of validity. \\
\mr
$ \hspace{-6pt}  \begin{tabular}{l} 
Single charge impurity    \\
(with a finite magnetic field)
\end{tabular} $
&
$ T^{2K-2}$ 
& See Footnote ${\rm b}$. \\
\mr
Kondo lattice (1D Kondo array)   &   
$ \left\{ \hspace{-6pt} \begin{tabular}{ll}  
 $  T^{-2}$ & for $E_{\rm pin} < k_{\rm B} T \ll \Delta_{\rm ka}$,  \\
 $T^{2K-2} $ & for $k_{\rm B} T > \Delta_{\rm ka}$ 
\end{tabular} \right. $ & Localization at low $T$; see Footnote ${\rm c}$.  \\
\mr
$ \hspace{-6pt}  \begin{tabular}{l} 
Magnetic-impurity ensemble  \\ 
(with spin diffusion into the bulk)
\end{tabular} $
&
$ \left\{ \hspace{-5pt} \begin{tabular}{ll}   $e^{\Delta_{\rm rs} / (k_{\rm B} T) }$ & for $T < T_{\rm rs}$  \\ 
$ T^{2K-2}$  & for $T > T_{\rm rs}$ \end{tabular} \right. $ 
 & Localization for $K<3/2$; see Footnote ${\rm d}$.    \\
$ \hspace{-6pt}  \begin{tabular}{l} 
Spiral-order-induced field \\
(below spiral ordering $T_{\rm s}$)
\end{tabular} $
&  
$ \left\{ \hspace{-5pt} \begin{tabular}{ll}   $m_{\rm s}^2  e^{\Delta_{\rm sa} / (k_{\rm B} T) }$ & for $T < T_{\rm sa}$  \\ 
 $ m_{\rm s}^2   T^{2K-2}$   & for $T > T_{\rm sa}$ \end{tabular} \right. $ 
& Localization for $K<3/2$;  see Footnote ${\rm e}$.  \\
$ \hspace{-6pt}  \begin{tabular}{l} 
Magnon \\
(below spiral ordering $T_{\rm s}$)
\end{tabular} $
 & 
$ \left\{ \hspace{-5pt} \begin{tabular}{ll}   $\omega_{\rm m}^{2K-3} $ & for magnon emission  \\ 
$   T^{3-2K} $ & for magnon absorption \end{tabular} \right. $ & 
See Footnote ${\rm e}$. \\
\mr
$ \hspace{-6pt}  \begin{tabular}{l} 
DNP$^{\rm f}$  \\
(for $K \approx 1$ and finite spin-flip rate)
\end{tabular} $
 & $ (T + {\rm const.} )^{-1} $  & \\ 
 $ \hspace{-6pt}  \begin{tabular}{l} 
DNP with random SOI$^{\rm g}$  \\
(for $K \approx 1$ and long channels)
\end{tabular} $
 & $ T^{-2/3}$  &  \\
\br
\end{tabular}
\item[] $^{\rm a}$ While the correction vanishes in the presence of Kondo screening~\cite{Maciejko:2009} or an isotropic coupling~\cite{Tanaka:2011}, in general it is nonzero in the presence of an anisotropic coupling~\cite{Tanaka:2011} or SOI-induced non-spin-conserving interaction~\cite{Eriksson:2012,Eriksson:2013}. In the latter case, the correction takes the form listed here, with  $K$ modified by SOI.
See also \cite{Kimme:2016,Zheng:2018,Kurilovich:2019,Vinkler-Aviv:2020} for more general cases including $I>1/2$ and nonlinear response, where the conductance is beyond a simple power law.   
\item[] $^{\rm b}$ See~\cite{Lezmy:2012} for a refined expression for general $V$ and $T$.
\item[] $^{\rm c}$ See~\cite{Altshuler:2013} for a discussion on the localization at $T \to 0$ in a noninteracting channel and \cite{Yevtushenko:2015} for a detailed discussion on transport at a finite temperature.
\item[] $^{\rm d}$  In addition to the listed power law obtained from the RG analysis~\cite{Hsu:2017,Hsu:2018b}, one can refer to \cite{Vayrynen:2016} for a refined expression, which includes an additional crossover below which the conductance is partially restored.
\item[] $^{\rm e}$ In the spiral-ordered phase, the expressions contain the magnon energy $\hbar \omega_{\rm m}$ and the spiral-order parameter $m_{\rm s}$ with their temperature dependence given by $\omega_{\rm m} \propto T^{2K-2} m_{\rm s}$ and $ (1- m_{\rm s}) \propto (T/T_{\rm s})^{3-2K}$~\cite{Hsu:2017,Hsu:2018b}.
\item[] $^{\rm f}$ From~\cite{Lunde:2012}. 
\item[] $^{\rm g}$ From~\cite{DelMaestro:2013}. 
\end{indented}
\end{table*}
   

\subsubsection{Other time-reversal-invariant perturbations}
~\\
While most of the time-reversal-invariant mechanisms stem from the broken $S^z$ symmetry, there are exceptions.  
For instance, \cite{Mani:2016,Mani:2019} investigated the effects of disordered contacts on the edge transport.  
Moreover, additional perturbations arise when considering transverse phonons instead of longitudinal ones~\eref{Eq:H_e-ph}.

{\it Backscattering induced by spin-phonon coupling}. 
In contrast to \cite{Budich:2012}, \cite{Groenendijk:2018} proposed that phonon-induced backscatterings can arise even in the absence of broken $S^z$ symmetry. In particular, they pointed out that whereas the helical channels have one-dimensional character, phonons are inherently three-dimensional. By modeling the lattice deformations induced by phonons, they demonstrated a direct coupling of transverse phonons to the electron spins in a helical channel without Rashba spin-orbit coupling. 
Such a direct spin-phonon coupling induces inelastic spin-flip scattering without breaking the time-reversal symmetry. 

For short edges, it leads to a conductance deviation $\delta G_{\rm s-ph} \propto - T^5$ for $k_F =0$.
For a finite but small $k_F$ where $\hbar v_F k_F/(k_{\rm B} T) \ll v_F / c_{{\rm ph}, t}$ with the transverse-phonon velocity $c_{{\rm ph}, t} $,
an additional energy window opens for backscattering, leading to an additional term $\propto - T^3$ in $\delta G_{\rm s-ph} $.
For $\hbar v_F k_F/(k_{\rm B} T) \gg v_F / c_{{\rm ph}, t} $, the conductance correction is suppressed exponentially. 
For long edges, the spin-phonon coupling leads to distinct power laws in different temperature ranges determined by the Debye temperature $\theta_{\rm D}$. Specifically, for $k_{\rm B} T \ll  \hbar v_F k_F,~  k_{\rm B} \theta_{\rm D}$,  
\begin{eqnarray}
R_{\rm s-ph}^{\rm low}  & \propto & \frac{h}{e^2} \frac{L \hbar c_{{\rm ph}, t} k_F^2 }{ v_F^2 \rho_m} \left( \frac{ k_{\rm B} T }{ \hbar c_{{\rm ph}, t} } \right)^3, 
\label{Eq:R_sp_Groenendijk1} 
\end{eqnarray}
where $\rho_m$ is the material density.
For  $ \hbar v_F k_F \ll k_{\rm B} T \ll k_{\rm B} \theta_{\rm D}$,
\begin{eqnarray}
R_{\rm s-ph}^{\rm mid}  & \propto & \frac{h}{e^2} \frac{L \hbar c_{{\rm ph}, t}  }{ v_F^2 \rho_m} \left( \frac{ k_{\rm B} T }{ \hbar c_{{\rm ph}, t} } \right)^5 ,
 \label{Eq:R_sp_Groenendijk2} 
 \end{eqnarray}
and for $k_{\rm B} T \gg  k_{\rm B} \theta_{\rm D} \gg \hbar v_F k_F$,
 \begin{eqnarray}
R_{\rm s-ph}^{\rm hi}  & \propto &  \frac{h}{e^2} \frac{L \hbar c_{{\rm ph}, t}  }{ v_F^2 \rho_m} \left( \frac{ k_{\rm B} \theta_{\rm D} }{ \hbar c_{{\rm ph}, t} } \right)^5 \frac{T}{\theta_{\rm D}} .
\label{Eq:R_sp_Groenendijk3} 
\end{eqnarray}
The above results are from a noninteracting model, and it might be interesting to explore the resistance due to the spin-phonon coupling in an interacting channel.

 
\begin{table*}[pt]
\centering
\caption{
\label{Tab:mechanism2}
Similar to \Tref{Tab:mechanism}, but for time-reversal-invariant backscattering mechanisms introduced in section~\ref{Sec:Scattering_TRS}. 
}
\footnotesize\rm
\begin{indented}
\item[]\begin{tabular}[c]{  l  l  l   }
\br
TRS preserving mechanism 
& $R$ or $ - \delta G$ &  Remark \\
\mr
$ \hspace{-5pt}  \begin{tabular}{l} 
1PB by $H_{\rm ee,5}$ \\ 
(for clean systems) 
\end{tabular} $
&  $ \left\{ \hspace{-5pt}
\begin{tabular}{ll} 
$ e^{-\hbar v_F k_F / (k_{\rm B}T)} $ & for $ k_{\rm B} T \ll \hbar v_F k_F$   \\
$T^{2K+3}$ & for $ k_{\rm B} T \gg \hbar v_F k_F$
\end{tabular}   \right. $
& See Footnote ${\rm a}$. \\
1PB by $H_{\rm ee,5} ~\&~ H_{\rm imp,f}$ $^{\rm b,c}$  &  $T^{2K+2}$  &  \\ 
1PB by $H_{\rm ee,5} ~\&~H_{\rm imp}^{\rm loc}$   
&  $T^{2K+2}$  & See Footnote ${\rm d}$.  \\ 
1PB by $H_{\rm ee,1} ~\&~ H_{\rm imp,b}$ $^{\rm c}$  &  $T^{6}$ for $K \approx 1$  &  \\ 
1PB by $H_{\rm ee,3} ~\&~ H_{\rm imp,b}$ $^{\rm b,c}$  &  $T^{4}$ for $K \approx 1$ & \\ 
2PB by $H_{\rm ee,3} ~\&~  H_{\rm imp,f}$  $^{\rm c, e}$  &$T^{8K-2}$  & Localization for $K<3/8$. \\  
2PB by $H_{\rm ee,3}  ~\&~H_{\rm imp}^{\rm loc}$   
  & $T^{8K-2}$  &  See Footnote ${\rm d}$ and Footnote ${\rm f}$.  \\ 
\mr
Random SOI & 0  & See Footnote ${\rm g}$.  \\
\mr
$ \hspace{-5pt}  \begin{tabular}{l} 
Higher-order random SOI$^{\rm h}$ \\
(single scatterer)
\end{tabular} $ 
 &
$ \hspace{-5pt}  \begin{tabular}{l} 
For $K>1/2$: \\
\hspace{5pt} $ \left\{ \hspace{-5pt}  \begin{tabular}{l} $T^{4K}$ for $ T < T^{*}_{\rm rso} $ \\
 $ T^{4K} \ln^2  (k_{\rm B}T/\Delta_{\rm b})$ for $ T > T^{*}_{\rm rso} $ \end{tabular}  \right. $\\
For $1/4<K<1/2$: \\
\hspace{5pt} $ \left\{ \hspace{-5pt}  \begin{tabular}{l}  $ T^{8K-2}$ for $ T < T^{*}_{\rm rso} $ \\ $ T^{4K} \ln^2 (k_{\rm B}T/\Delta_{\rm b})$ for $ T > T^{*}_{\rm rso} $ \end{tabular}  \right. $\\
\end{tabular} $ 
 &  
$ \hspace{-5pt}  \begin{tabular}{l} 
Crossover temperature: \\
$T^{*}_{\rm rso} = \Delta_{\rm b} \big(\frac{2K}{4K-1}\big)^{1/(2K-1)}$. 
~\\
~\\
\end{tabular} $
 \\
 \mr
$ \hspace{-5pt}  \begin{tabular}{l} 
1PB in charge puddles$^{\rm i}$ \\
(for $K \approx 1$)
\end{tabular} $
& 
$ \hspace{-5pt}  \begin{tabular}{l} 
Short channel (a single puddle): \\
~even valley: 
$ \hspace{0pt}  \left\{ \hspace{-5pt}  \begin{tabular}{ll} $T^{4}$ &  \hspace{-10pt}  for $k_{\rm B} T \ll \delta_{\rm d}  $  \\ 
$ T^{2}$ &  \hspace{-10pt} for $\delta_{\rm d} \ll k_{\rm B} T \ll E_{\rm ch}$  \\
const. &  \hspace{-10pt}  for $k_{\rm B} T \gg E_{\rm ch} $
\end{tabular}  \right. $ \\
$ \hspace{-5pt}  \begin{tabular}{l} 
~even-odd \\
~transition: 
\end{tabular} $
$ \hspace{-2pt}  \left\{ \hspace{-5pt}  \begin{tabular}{ll} $T^{4}$ &  \hspace{-10pt}  for $k_{\rm B} T \ll \Gamma_{\rm t}  $  \\ 
const. &  \hspace{-10pt} for $\Gamma_{\rm t} \ll k_{\rm B} T \ll \delta_{\rm d}  $ \\
$ T^{2}$ &  \hspace{-10pt}  for $\delta_{\rm d} \ll k_{\rm B} T \ll E_{\rm ch}$ \\
const. &  \hspace{-10pt}  for $k_{\rm B} T \gg E_{\rm ch} $
\end{tabular}  \right. $ \\ 
~odd valley: 
 $ \hspace{2pt}  \left\{ \hspace{-5pt}  \begin{tabular}{ll} $T^{4}$ &  \hspace{-10pt}  for $T \ll T_{\rm K} $ \\ 
$\ln^2 (T/T_{\rm K})$ & \hspace{-10pt}  for $k_{\rm B}  T_{\rm K} \ll k_{\rm B} T \ll \delta_{\rm d} $  \\
$ T^{2}$ &  \hspace{-10pt}  for $\delta_{\rm d} \ll k_{\rm B} T \ll E_{\rm ch}$  \\
const. &  \hspace{-10pt}  for $k_{\rm B} T \gg E_{\rm ch} $
\end{tabular}  \right. $ \\
Long channel (averaged over puddle configurations): \\ 
~ $E_{\rm ch} \ll \delta_{\rm d}$:  
$   \hspace{-5pt}  \begin{tabular}{ll} $T $ & for $k_{\rm B} T  \ll \delta_{\rm d}$
 \end{tabular}$ \\
~ $E_{\rm ch} \approx \delta_{\rm d}$:  
$  \hspace{-5pt}  \begin{tabular}{ll} $1/\ln^2 [ \delta_{\rm d} / (k_{\rm B} T ) ]  $ & for $k_{\rm B} T  \ll \delta_{\rm d}$
 \end{tabular} $ \\
~ $E_{\rm ch} \gg \delta_{\rm d}$:  
$    \left\{ \hspace{-5pt}  \begin{tabular}{ll} $1/ \ln^2 [ \delta_{\rm d} / (k_{\rm B} T ) ] $  &     \hspace{-10pt}  for  $ k_{\rm B} T  \ll \delta_{\rm d}'  $  \\
$1/ \ln [ \delta_{\rm d} / (k_{\rm B} T ) ]  $ &   \hspace{-10pt}  for $\delta_{\rm d}' \ll k_{\rm B} T  \ll \delta_{\rm d}$ \hspace{-64pt} 
 \end{tabular}  \right. $
\end{tabular} $
 ~
  &
\hspace{-5pt} \begin{tabular}{l} 
The 2PB in charge puddles \\
is dominated by 1PB and \\
thus neglected. \\
 ~ \\
For short channels, the \\
expressions vary with the dot \\
parity states (tuned by gate \\
voltage) and the temperature \\
ranges set by the dot level \\ 
spacing $\delta_{\rm d}$, the dot-edge \\
tunneling $\Gamma_{\rm t}$ and the charging \\
energy $E_{\rm ch}$. \\ 
 ~\\
For long channels, the  \\
resistance depends on the \\
additional energy scale \\
$\delta_{\rm d}' \equiv \delta_{\rm d} e^{- \pi E_{\rm ch} / (2 \delta_{\rm d})} $. 
\end{tabular} \\
\mr
$ \hspace{-5pt}  \begin{tabular}{l} 
Noise$^{\rm j}$  \\
(for $K \approx 1$, long channels)
\end{tabular} $
&  
$ \hspace{-5pt}  \begin{tabular}{ll} 
Telegraph noise:  &  
\hspace{-5pt} $T^2 \tanh \left( \frac{E_{\rm ch} }{ 2 k_{\rm B} T} \right) $ \\
$1/f$ noise: &
$ \hspace{-5pt}  \left\{ \hspace{-5pt}  \begin{tabular}{ll}  
$T^2$  & for $k_{\rm B} T \ll E_{\rm ch}$ \\
$T$ & for $k_{\rm B} T \gg E_{\rm ch}$
\end{tabular}  \right. $ \\
\end{tabular} $
&  \\
\mr
Acoustic longitudinal phonon  & 0 & See Footnote {\rm k}.    \\
\mr
$ \hspace{-5pt}  \begin{tabular}{l} 
Transverse phonon$^{\rm l}$ \\
(for $K \approx 1$)
\end{tabular} $
& 
$ \hspace{-5pt}  \begin{tabular}{l} 
Short channel: \\
$ \hspace{5pt}  \left\{ \hspace{-5pt}  \begin{tabular}{ll} 
$e^{-\hbar v_F k_F / (k_{\rm B} T)}$  & \hspace{-10pt}  for $ k_{\rm B} T \ll \hbar c_{{\rm ph},t} k_F  $ \\
const. $ T^{3} + $ const. $ T^5$  & \hspace{-10pt}  for $ k_{\rm B} T \gg  \hbar c_{{\rm ph},t} k_F  $ \\ 
\end{tabular}  \right. $ \\
Long channel: \\
 $ \hspace{5pt}  \left\{ \hspace{-5pt}  \begin{tabular}{ll} 
 $T^{3}$ & for  $k_{\rm B} T \ll  \hbar v_F k_F,~  k_{\rm B} \theta_{\rm D}$   \\ 
$T^5$ & for $ \hbar v_F k_F \ll k_{\rm B} T \ll k_{\rm B} \theta_{\rm D}$ \\
$ T $ & for $k_{\rm B} T \gg  k_{\rm B} \theta_{\rm D} \gg \hbar v_F k_F$
\end{tabular}  \right. $ 
\end{tabular} $ & 
  \hspace{-5pt} \begin{tabular}{l} 
The spin-phonon coupling \\
arises from deformation and \\
does not require Rashba SOI. \\
~\\
The expressions depend on \\
the transverse phonon velocity  \\ 
$c_{{\rm ph},t} $ and the Debye \\
temperature  $\theta_{\rm D}$. 
\end{tabular} \\
\br
\end{tabular}
\item[] $^{\rm a}$ There exist discrepancies in the literature. For low temperatures, \cite{Kainaris:2014} obtained the listed expression and explained its difference from~\cite{Schmidt:2012} as a result of Pauli blocking at the Dirac point.  
In the high-temperature regime, \cite{Chou:2015} found $T^{2K+1}$ but did not comment on the difference from the listed expression in \cite{Kainaris:2014}.
\item[] $^{\rm b}$ \cite{Schmidt:2012} investigated the $K \approx 1$ limit of the conductance correction. 
\item[] $^{\rm c}$ \cite{Kainaris:2014} obtained the conductivity for 1PB by combining $H_{\rm ee,5}$ and $H_{\rm imp,f}$, and for 2PB for a general $K$, as well as others for $K \approx 1$.
\item[] $^{\rm d}$ See~\cite{Lezmy:2012} for a refined expression for general $V$ and $T$.
\item[] $^{\rm e}$ \cite{Xu:2006,Wu:2006} investigated the localization due to the 2PB induced by quenched disorder.  
\item[] $^{\rm f}$ \cite{Wu:2006,Maciejko:2009,Lezmy:2012} investigated 2PB off a single scatterer. For $K<1/4$, the edge channel is cut by 2PB, leading to a power-law tunneling conductance with $T^{1/(2K)-2}$~\cite{Maciejko:2009}.
\item[] $^{\rm g}$ \cite{Kainaris:2014,Xie:2016} concluded that the random SOI has no influence on the transport, contradicting \cite{Strom:2010,Geissler:2014}; see \cite{Kharitonov:2017} for discussions on the generalized random SOI. 
\item[] $^{\rm h}$ From \cite{Crepin:2012}.
\item[] $^{\rm i}$ From~\cite{Vayrynen:2013,Vayrynen:2014}.  
\item[] $^{\rm j}$ From \cite{Vayrynen:2018}.
\item[] $^{\rm k}$ The linear Rashba term gives a vanishing electron-phonon backscattering matrix element in the first order~\cite{Budich:2012}.
 \item[] $^{\rm l}$ From~\cite{Groenendijk:2018}. 
 \end{indented}
\end{table*}
   

\subsection{Discussion on the charge transport}
Since the theoretically proposed backscattering mechanisms generally lead to different bias, length, and temperature dependences of the resistances, these predictions can be verified by experiments. For convenience, we summarize the temperature dependences, giving \Tref{Tab:mechanism} for time-reversal symmetry breaking mechanisms and \Tref{Tab:mechanism2} for time-reversal symmetry preserving ones. 
In these tables, we also summarize three sets of theoretical results discussed above and the corresponding references. 
For Kondo impurities, they are given in Footnote a of \Tref{Tab:mechanism}; 
for $ H_{\rm ee,5} $-induced 1PB process, the discrepancies are summarized in Footnote a of \Tref{Tab:mechanism2};
for random SOI, we point out the contradicting results in Footnote g of \Tref{Tab:mechanism2}. 

Comparing the two tables, the following major difference stands out. The mechanisms arising from time-reversal invariant inelastic processes lead to conductance corrections which vanish at zero temperature (due to the vanishing phase space), whereas those due to broken time-reversal symmetry might not (depending on the material parameters). In addition, mechanisms involving hTLL physics in general lead to fractional power-laws with exponents decreasing upon increasing the strength of repulsive interactions. Therefore, electron-electron interactions enhance the backscattering and thus the low-temperature resistance, a feature common with nonhelical channels~\cite{Giamarchi:2003}.

In addition to the dc resistance reviewed above, there are also studies on other transport-related quantities. For instance, finite-frequency conductivity, backscattering current noise, or shot noise in 2DTI edges were studied \cite{Lezmy:2012,DelMaestro:2013,Aseev:2016,Kurilovich:2019,Pashinsky:2020}.
The influence of disorder on the propagation of the edge-state wave function and the induced momentum broadening were investigated in~\cite{Gneiting:2017}. 

In this review, we focus on an isolated helical channel and its transport properties. For a nanoribbon or stripe-like geometry in which the interchannel coupling becomes nonnegligible, the stability of multiple parallel helical liquids has been investigated \cite{Xu:2006,Tanaka:2009b,Santos:2019}. In addition, there is a first-principle study on interedge scattering due to a single nonmagnetic bulk impurity, a relevant backscattering source for narrow ribbons~\cite{Vannucci:2020}, as well as works on Coulomb drag in two parallel interacting helical edge modes~\cite{Zyuzin:2010,Chou:2015,Kainaris:2017,Chou:2019,Du:2021}.
Furthermore, \cite{Hou:2009} proposed corner junctions as a probe of 2DTI helical edge states. 
\cite{Posske:2013,Posske:2014} studied two 2DTI helical edges coupled to a spin-$\frac{1}{2}$ magnetic impurity in a gate-defined quantum antidot. Employing a two-channel Kondo model, they investigated correlation functions between the impurity spin and the electron spin, which form a Kondo screening cloud. Utilizing the spin-momentum locking property of the helical edges, they proposed a setup to detect the Kondo cloud through the space- and time-resolved current cross correlation functions. 
There are also studies on quantum point contacts~\cite{Teo:2009,Strom:2009,Lee:2012,Dolcetto:2016}; we note that 2DTI quantum point contacts have been achieved experimentally using HgTe~\cite{Strunz:2020}. Surprisingly, in addition to the expected conductance plateaus at multiples of $2e^2/h$ through the point contact, they observed anomalous conductance plateau at the half-quantized value $e^2/h$. Attributed to the opening of a spin gap, it urges further theoretical and experimental efforts on reproducing the results and examining alternative interpretations.  

We have shown that helical channels bring about unusual phenomena of charge transport, distinct from nonhelical channels. We now discuss how even more exotic states of matter can arise in helical channels when superconductivity is added.

\section{Topological superconductivity in helical channels \label{Sec:TSU}}
Analogous to topological insulators, there are superconducting materials with nontrivial topology. 
They are topological superconductors~\cite{Hasan:2010,Qi:2011,Ando:2015,Sato:2017}. 
Theoretical proposals for their realizations include various materials or settings, ranging from unconventional superconducting phase in stoichiometric compounds such as chiral $p$-wave superconductors Sr$_{2}$RuO$_4$~\cite{Kallin:2016} to artificially engineered structures such as proximitized three-dimensional topological insulators~\cite{Fu:2008}, Cu-doped three-dimensional topological insulators~\cite{Hor:2010} or LaAlO$_3$/SrTiO$_3$ bilayer with interface superconductivity~\cite{Nakosai:2012}.

As their insulating counterparts, topological superconductors are characterized by their bulk topology and in-gap modes at their boundaries.
Similar to  the analogy between the low-energy edge theory of topological insulators and the Dirac Hamiltonian, here the in-gap modes find their analogues in Majorana's theory on real solutions of the Dirac equation~\cite{Majorana:1937}. These real solutions, representing exotic particles identical to their own antiparticles, are neither ordinary fermions nor bosons. While such exotic particles have so far not been found as elementary particles~\cite{Wilczek:2009}, they can appear as quasiparticles in certain superconducting systems. 

Mathematically, an ordinary fermion can be expressed as a combination of two Majorana modes~\cite{Kitaev:2001}. To realize a single Majorana mode in a realistic system, one has to separate the modes in space. This task is challenging in conventional solid-state systems hosting spin-1/2 electron or hole states, where the fermion modes have intrinsic spin degeneracy. In helical channels, the spin degeneracy is lifted, making them potentially suitable for Majorana modes, either in the form of zero-energy modes or dispersive (propagating) modes.
In this review, we focus on Majorana zero modes, which are bound states with exponentially localized wave functions in all three spatial dimensions.

In addition to an academic interest, zero-energy Majorana bound states (MBS) can have technological applications, providing building blocks for advanced quantum computation~\cite{Ivanov:2001,Kitaev:2003,Tanaka:2009,Sato:2009,Alicea:2011,Crepin:2014,Landau:2016,Hoffman:2016,Karzig:2017}. 
There are existing reviews and tutorials on this topic~\cite{Nayak:2008,Alicea:2012,Beenakker:2013,DasSarma:2015,Sato:2017,Beenakker:2020,Laubscher:2021}. Remarkably, while the initial work by \cite{Fu:2008} motivated subsequent works on MBS localized at vortices in topological superconductors in three-dimensional structures~\cite{Wang:2018c,Kong:2019,Machida:2019,Chiu:2020,Liu:2020,Zhu:2020},
one can realize MBS by assembling existing ingredients in quasi-one-dimensional nanoscale systems. Along this line, great efforts were made on semiconductor nanowires combining Rashba spin-orbit coupling, magnetic field, and proximity superconductivity~\cite{Lutchyn:2010,Oreg:2010,Alicea:2011,Klinovaja:2012a,Mourik:2012,Das:2012,Deng:2012,Rokhinson:2012,Finck:2013,Churchill:2013,Rainis:2013,Cayao:2015,Albrecht:2016,Gul:2018,Prada:2020,Frolov:2020}, though a smoking-gun evidence for the existence of MBS in this setup is still missing~\cite{Castelvecchi:2021,Frolov:2021}. In spite of being relatively accessible in experiments, the Rashba setup has disadvantages, including the detrimental effects of the external magnetic fields on the superconductivity and the sensitivity of topological bound states to the field orientation, which might affect their scalability for practical applications.  

The disadvantages motivated works on alternative platforms, including semiconductor-ferromagnet-superconductor hybrid devices~\cite{Sau:2010}, spontaneous helical spin textures in either magnetically doped nanowires or atomic chains~\cite{Klinovaja:2013a,Braunecker:2013,Vazifeh:2013,Nadj-Perge:2013,Nadj-Perge:2014,Pientka:2014,Hsu:2015,Ruby:2015,Pawlak:2016,Kim:2018,Pawlak:2019} and nanodevices with micromagnets~\cite{Klinovaja:2012b,Kjaergaard:2012,Klinovaja:2013x,Maurer:2018,Desjardins:2019}.  

There are also proposals on time-reversal-invariant settings~\cite{Haim:2019}, such as iron-based superconductors with $s_{\pm}$ pairing~\cite{Zhang:2013}, nonhelical channels in multisubband quantum wires~\cite{Haim:2014,Gaidamauskas:2014}, or multiple Rashba wires~\cite{Klinovaja:2014a,Klinovaja:2014b,Ebisu:2016,Schrade:2017,Thakurathi:2018}. Some of these quasi-one-dimensional systems involve interaction-induced MBS and might be able to realize parafermions~\cite{Alicea:2016}, which provide an even more advanced quantum computation scheme~\cite{Hutter:2016}. 

A common trait of the just mentioned alternative setups is that they do not require magnetic fields and that they exploit nonhelical channels. Interacting helical channels provide additional advantages. Namely,  repulsive interactions in helical channels can have stronger effects on suppressing the local pairings than the nonhelical channels~\cite{Hsu:2018}, thus favoring nonlocal pairing that is crucial for topological bound states. Below, we point out that MBS arise in helical channels, taking coupled pairs of helical channels in 2DTI and HOTI as examples.

	\begin{figure}[t]
	\includegraphics[width=0.48\linewidth]{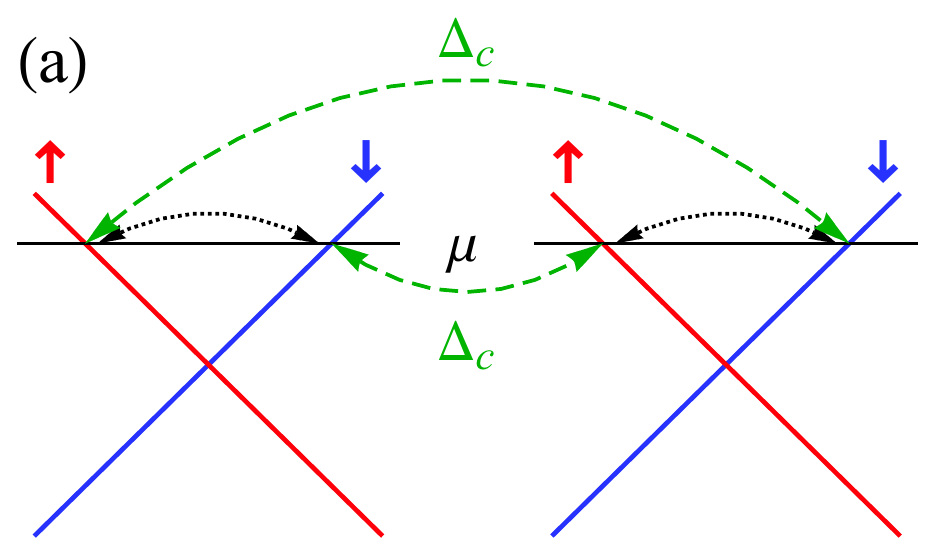}
	\includegraphics[width=0.48\linewidth]{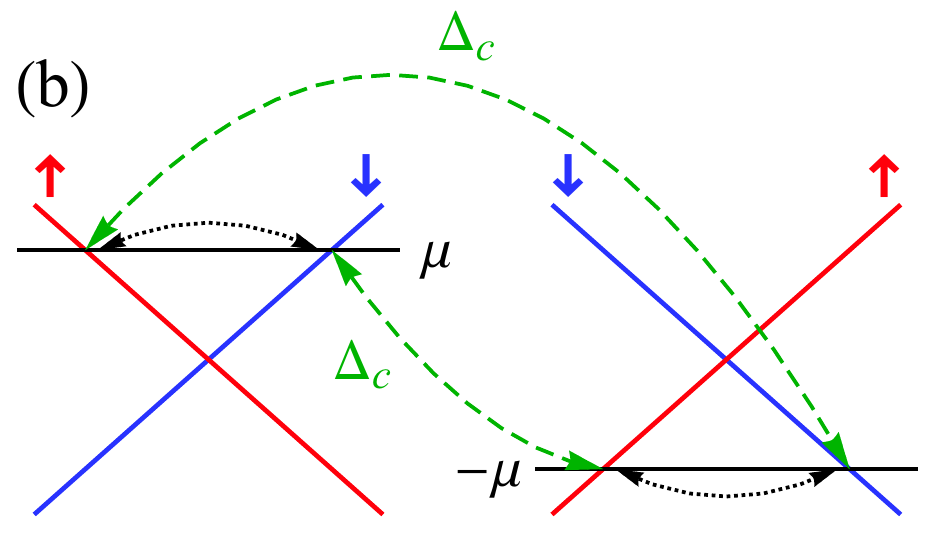}
		\caption{Proximity-induced $s$-wave local (black arrows) and nonlocal (green arrows) pairings in helical channels. For simplicity we illustrate a system where $S^z$ is a good quantum number. (a) For two channels with the same helicity (parahelical), the chemical potentials should be the same for the crossed Andreev pairing between the two channels. (b) For two channels with the opposite helicity (orthohelical), the chemical potentials should be opposite.
		 }
 \label{Fig:CAP_1}
	\end{figure}

\subsection{Setups using double helical channels \label{SubSec:TSU-db}}
Utilizing the proximity effect, one can induce pairings in the helical channels to realize topological superconductivity.  
Majorana or parafermion bound states can appear at domain walls between the band-inverted and non-inverted regimes.
Creating such a domain wall requires a second gap opening (that is, mass inducing) mechanism, in addition to the proximity-induced direct pairing. 
Some works considered the Zeeman gap induced by 
either external magnetic fields~\cite{Fu:2009,Mi:2013,Fleckenstein:2021} or ferromagnetism~\cite{Crepin:2014b,Keidel:2018,Jack:2019} serving this purpose. 
However, similar to the prototype setup with Rashba nanowires, the magnetism or magnetic fields are detrimental to the superconductivity. 
In this section, we discuss theories proposing an alternative, albeit less familiar, mechanism relying on a nonlocal superconducting pairing.

	\begin{figure}[t]
	\includegraphics[width=1\linewidth]{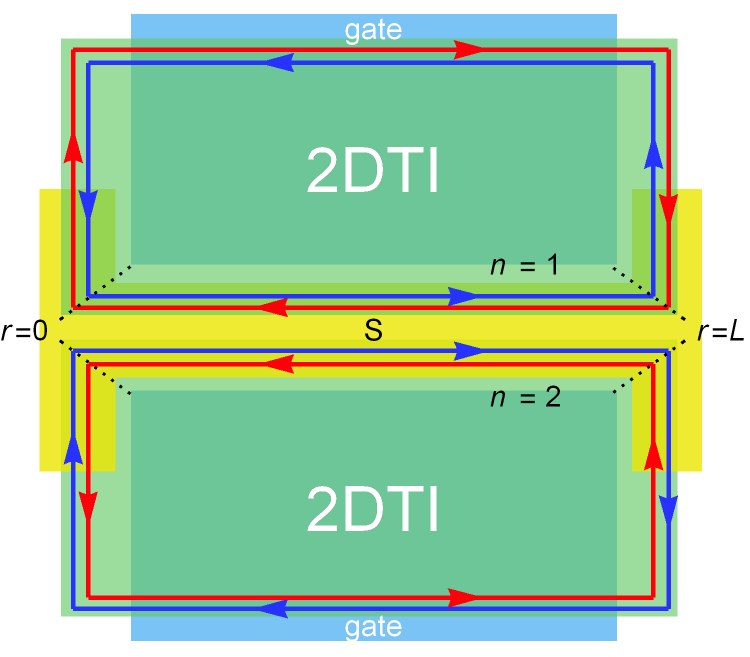}
	\caption{Parahelical setup for stabilizing MBS in proximitized edge channels of two 2DTI layers in a horizontal alignment. A superconductor (yellow) is deposited to proximitize the edge channels, while gates (blue) control their chemical potentials. With identical chemical potential, both local and nonlocal pairings take place in the segment $0<r<L$ with the edge coordinate $r$, whereas there is only local pairing in other segments $r<0$ and $r>L$.
		}
 \label{Fig:2DTI-setup_para1}
	\end{figure}

To dispose the magnetic field, one can use a pair of helical channels, which can be formed by either 2DTI edge channels or HOTI hinge channels. 
When bringing the two helical channels into proximity of an $s$-wave superconductor, two types of superconducting pairing arise. 
One is the local, intrachannel pairing, where the Cooper pairs in the parent superconductor tunnel into one of the two channels.
Since each helical channel consists of Kramers partners with the opposite spin and momentum, the local pairing is allowed for the $s$-wave superconductivity.
The other pairing type is the nonlocal, interchannel pairing [also known as crossed Andreev pairing~\cite{Klinovaja:2014b,Reeg:2017}], where the two partners of a Cooper pair tunnel into different channels. For the momentum- and spin-conserving tunneling process to take place, we have different conditions on the chemical potential, depending on the relative helicity between the two channels. For convenience, we adopt the terminology of \cite{Klinovaja:2014} and name the combination as {\it parahelical} ({\it orthohelical}) when the two channels have the same (opposite) helicity. In \Fref{Fig:CAP_1}, we illustrate the local and nonlocal pairings for different combinations of helical channels. For the parahelical setup shown in \Fref{Fig:CAP_1}(a), the nonlocal pairing is in effect when the chemical potentials in the two channels are the same. For orthohelical channels [see \Fref{Fig:CAP_1}(b)] the nonlocal pairing is effective when the chemical potentials (tuned by local gates) are opposite. 

\begin{figure}[t]
	\includegraphics[width=0.99\linewidth]{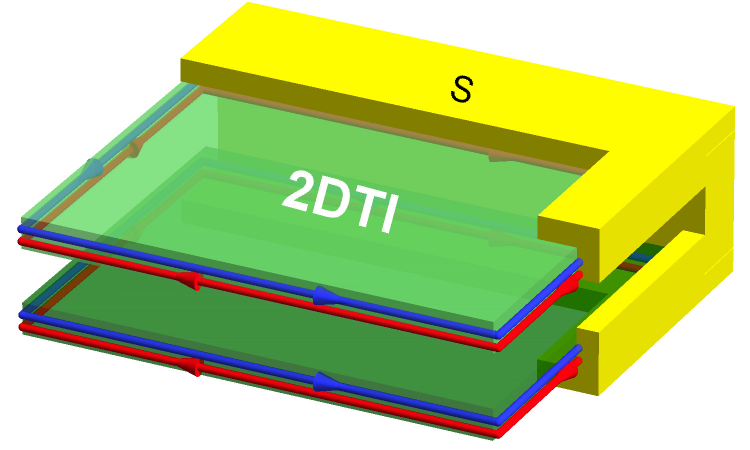}	
		\caption{Similar to \Fref{Fig:2DTI-setup_para1} but with a vertical alignment, which might be suitable for van der Waals materials. 
				}
 \label{Fig:2DTI-setup_para2}
	\end{figure}

After introducing the basic concept for the interchannel pairing, we now discuss how to achieve the crossed Andreev pairing using the helical channels of  topological materials. In \Fref{Fig:2DTI-setup_para1} and \Fref{Fig:2DTI-setup_para2}, we show two structures for the parahelical setup using two 2DTIs, one in a vertical and the other in a horizontal arrangement. In both structures, a superconductor is deposited in such a way that it induces a crossed Andreev pairing in two edge segments (for instance, in the region $0 <r < L$ in \Fref{Fig:2DTI-setup_para1}) and only local pairing in other segments. As a result, at a corner (either $r=0$ or $r=L$) where the edges change their direction, a domain wall separates two regions, one with a finite nonlocal pairing and the other without it. 

\begin{figure}[t]
	\includegraphics[width=0.45\linewidth]{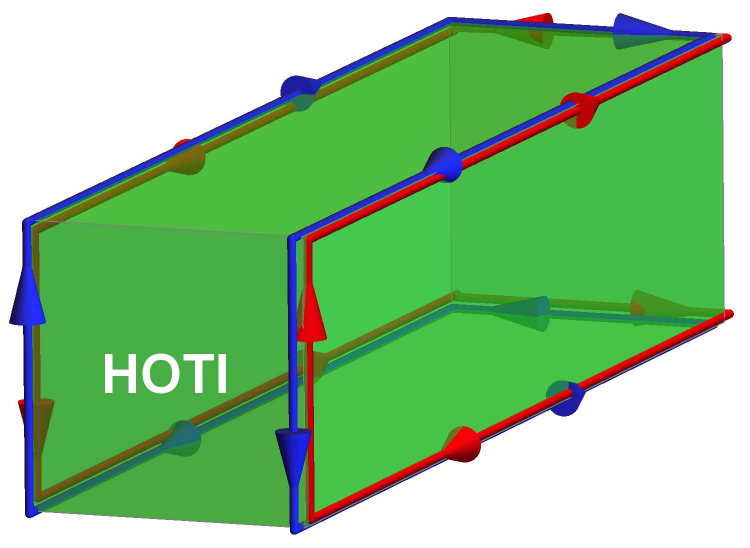} 
	\hspace{0.02\linewidth}
	\includegraphics[width=0.52\linewidth]{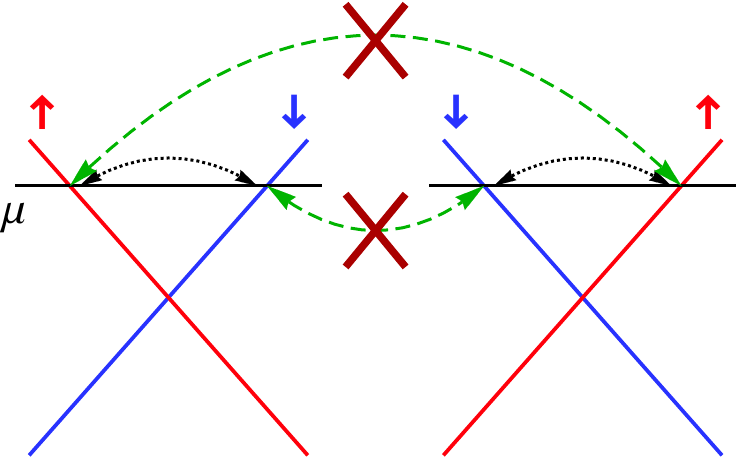}
	\caption{Illustration of a nanowire made of a three-dimensional second-order topological insulator with square cross section (left) and hinge-state spectrum of two parallel hinges sharing a side surface (right). 
	Since the two parallel hinges are orthohelical, the nonlocal pairing is forbidden.
	} 
 \label{Fig:HOTI2}
	\end{figure}

Alternatively, one can use a three-dimensional helical second-order topological insulator (see \Fref{Fig:HOTI} and \Fref{Fig:HOTI2} for HOTI nanowires with a hexagonal or a tetragonal cross section, respectively). In contrast to 2DTI, the hinge channels in a single HOTI bulk can have both parahelical and orthohelical arrangement. Since the two parallel hinges on a single side surface are orthohelical and the chemical potential is uniform along the hinges, the nonlocal pairing between the two hinges is forbidden (see the right panel of \Fref{Fig:HOTI2}). However, one can cover two side surfaces with a superconducting layer (see \Fref{Fig:HOTI_MKP}), where the nonlocal pairing is allowed in two hinges of the same helicity. Then, the crossed Andreev pairing takes place between a pair of parallel hinges of the same helicity, but vanishes at the short hinges perpendicular to them, thus creating a domain wall separating two regions, with and without the nonlocal pairing.

For both the 2DTI and HOTI setups, the competition between the gap opening mechanisms (local vs nonlocal pairing) leads to the band inversion and bound states at the domain walls which are located at corners, as we demonstrate below. The HOTI has an advantage that it does not require local gates to adjust the chemical potential. The latter is uniform since all the hinge channels are connected. 
	
The parahelical setting is described by the following single-particle Hamiltonian, $H_{\rm sp} = H_{\rm kin}^{\rm d} + H_{\rm loc} + H_{\rm cap}^{\rm }$. The first term is the kinetic energy,
\begin{eqnarray}
H_{\rm kin}^{\rm d} = -i \hbar v_F  \sum_{n} \int dr \; \big( R_{n}^\dagger \partial_r R_{n} - L^\dagger_{n} \partial_r L_{n}   \big),
\label{Eq:H_kin_d}
\end{eqnarray}
with the channel index $n \in \{ 1,2\}$ and $r$ is the spatial coordinate along the channel. 
The second term is the local pairing,
\begin{eqnarray}
H_{\rm loc} =  \sum_{n} \int dr \; \Big[ \frac{\Delta_{n}}{2} (R_{n}^\dagger L_{n}^\dagger - L^\dagger_{n} R_{n}^\dagger)  + {\rm H.c.} \Big],
\end{eqnarray}
with spatially uniform pairing amplitude $\Delta_{n}$.
The last term is the nonlocal pairing,
\begin{eqnarray}
H_{\rm cap} &=&  \frac{1}{2} \int dr \; \Delta_{\rm c}(r) \left[  (R_1^\dagger L^\dagger_{2} - L^\dagger_{2} R_{1}^\dagger) \right.  \nonumber \\
&& \hspace{52pt} \left. + (R_2^\dagger L^\dagger_{1} - L^\dagger_{1} R_{2}^\dagger) \right] + \textrm{H.c.} ,
\label{Eq:H_CAP}
\end{eqnarray}
with a spatially dependent pairing amplitude
\begin{eqnarray}
 \Delta_{\rm c}(r) &=&  \left\{
\begin{array}{ll}
\Delta_{\rm c} , & ~~{\rm for}~~ 0 < r < L. \\ 
0, &~~{\rm otherwise}
\end{array}
\right.  
\end{eqnarray}
In the above, we consider real pairing amplitudes $\Delta_{n}$ and $\Delta_{\rm c}$. 
Apart from the single-particle terms, there are forward scattering interaction terms corresponding to \eref{Eq:H_ee}, which will be included later when we discuss the interaction effects. 
The Hamiltonian \eref{Eq:H_kin_d}--\eref{Eq:H_CAP} describes noninteracting double channels in either 2DTI or HOTI depicted in \Fref{Fig:2DTI-setup_para1}, \Fref{Fig:2DTI-setup_para2} and \Fref{Fig:HOTI_MKP}.

\begin{figure}[t]
	\includegraphics[width=0.99\linewidth]{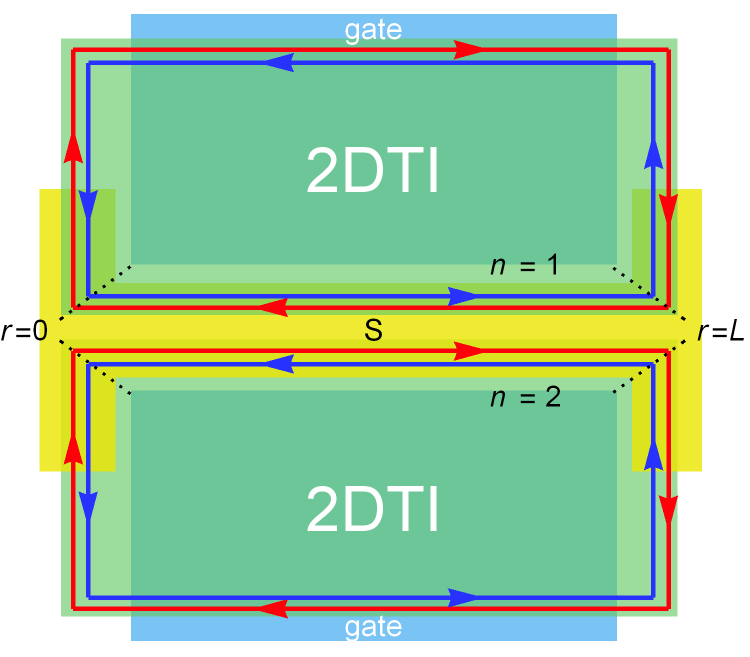}
	\hspace{0.0\linewidth}
		\caption{Orthohelical setup for creating MBS in proximitized edge channels of two 2DTI layers.
		Upon adjusting the chemical potentials of the two edges with local gates, one can realize orthohelical configuration with the energy spectrum shown in \Fref{Fig:CAP_1}(b).
		}
 \label{Fig:2DTI-setup_ortho}
	\end{figure}

On the other hand, the orthohelical arrangement can be achieved using two layers of 2DTI as displayed in~\Fref{Fig:2DTI-setup_ortho}, where the local gates adjust the chemical potentials in different edges. Again, at both $r=0$ and $r=L$, a domain wall separates two regions. While both parahelical and orthohelical setups can produce MBS in 2DTI~\cite{Klinovaja:2014}, only the parahelical setup can be exploited in a HOTI-based setup. Therefore, below we take the parahelical setup as an example to demonstrate how MBS can be stabilized in both systems and its topological criterion.

\subsection{Topological criterion for Majorana zero modes \label{SubSec:Criterion}}
We now examine the topological criterion considering the single-particle Hamiltonian \eref{Eq:H_kin_d}--\eref{Eq:H_CAP}. We rewrite it as
\numparts
\begin{eqnarray}
H_{\rm sp } &=& \frac{1}{2} \int dr \; \Psi^{\dagger} (r) \mathcal{H}_{\rm sp} (r) \Psi (r),
\label{Eq:H_sp}
\end{eqnarray}
using $\Psi = (R_1, L_1, R_2, L_2, R_1^\dagger, L_1^\dagger, R_2^\dagger, L_2^\dagger)^{\rm T}$ where  ${\rm T}$ is the transpose operator, and the Hamiltonian density
\begin{eqnarray}
\mathcal{H}_{\rm  sp}(r) &=& -i \hbar v_F \eta^0 \tau^0 \sigma^z \partial_r  -\Delta_{+} \eta^y \tau^0  \sigma^y \nonumber \\
 && -\Delta_{-}  \eta^y \tau^z \sigma^y  - \Delta_{\rm c}  (r) \eta^y \tau^x \sigma^y,
\label{Eq:H_para}
\end{eqnarray}
\endnumparts
with $\Delta_{\pm} = (\Delta_1 \pm \Delta_2)/2$ and the Pauli (identity) matrices for superscripts $\mu \in \{ x,y,z\}$ ($\mu =0$). In the above, $\eta^{\mu}$, $\tau^{\mu}$, and $\sigma^{\mu}$ act in the particle-hole, channel, and spin space, respectively.

		\begin{figure}[t]
	\includegraphics[width=0.48\linewidth]{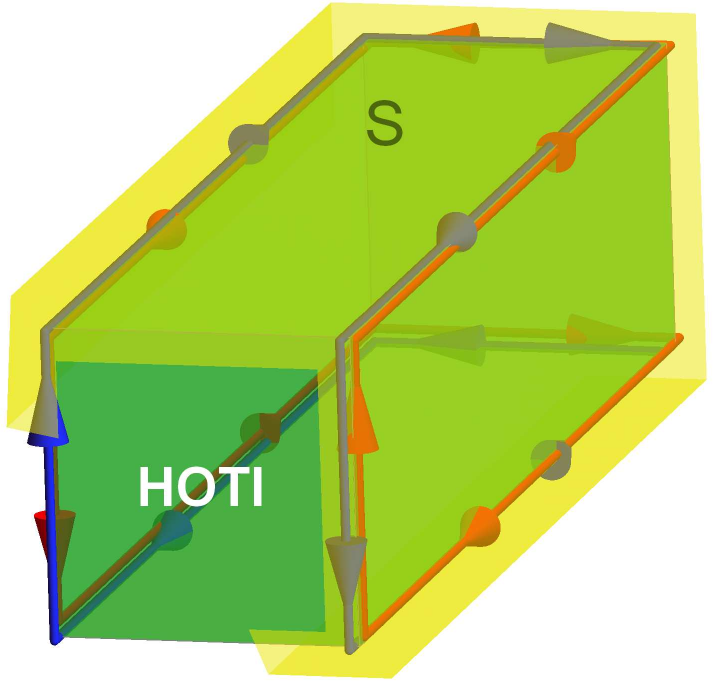}
	\includegraphics[width=0.48\linewidth]{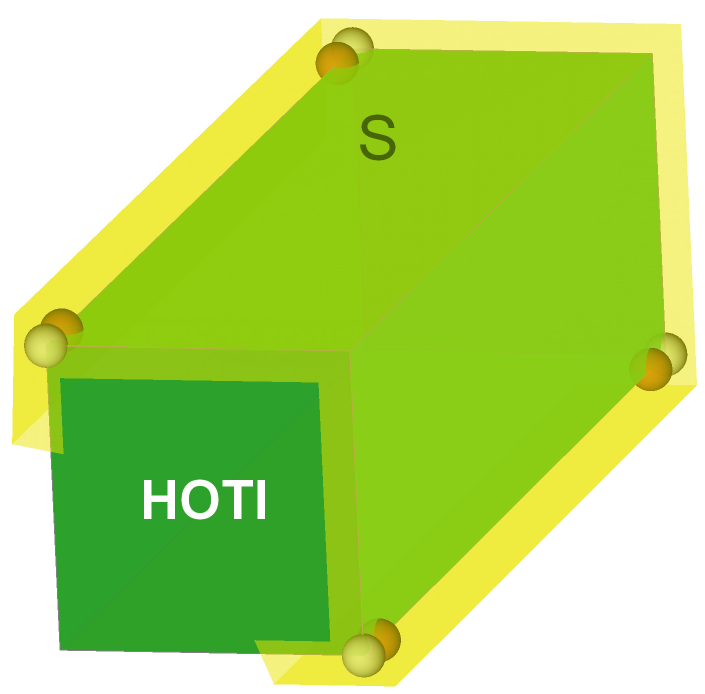}
		\caption{ (Left) MBS setup using a HOTI nanowire with two side surfaces covered by a superconductor.  
		 (Right) A pair of MBS (dots) emerges at each end of the nanowire. 
		 Each MBS wave function has a finite weight on {\it both channels} (a single solution is marked by the same color), showing its composite nature. We emphasize that, while we intentionally separate the two MBS at one end for clarity, they are in fact located at the same position.  
		}
 \label{Fig:HOTI_MKP}
	\end{figure}

Assuming a periodic boundary condition along the $r$ coordinate and performing Fourier transform on $H_{\rm sp }$, it can be shown that the system is time-reversal-invariant with the time-reversal operator $\mathcal{T}=i \sigma^y\mathcal{K}$ introduced in \eref{Eq:TRS} and the complex conjugate operator $\mathcal{K}$.
Since the operator $\mathcal{T}$ squares to -1, the Kramers degeneracy theorem applies, indicating a time-reversal pair of states at a given energy.
Indeed, the corresponding ``bulk'' spectrum deep inside the $0 < r < L$ region is doubly degenerate and given by 
\begin{eqnarray}
E_{\rm pbc}^{(\pm,\pm)} (k) =& \pm \sqrt { (\hbar v_F k)^2 + \Big( \Delta_{+} \pm \sqrt{\Delta_{-}^2 + \Delta_{\rm c}^2 } \Big)^2 }. \nonumber \\
\end{eqnarray}
It has a gap $\Delta_{\rm pbc}$ at $k=0$, \begin{eqnarray}
\Delta_{\rm pbc} &\equiv & E_{\rm pbc}^{(+,-)} (k=0) - E_{\rm pbc}^{(-,-)} (k=0) \nonumber \\
&=& 2 \Big( \Delta_{+} - \sqrt{\Delta_{-}^2 + \Delta_{\rm c}^2 } \Big).
\label{Eq:Delta_pbc}
\end{eqnarray}
Assuming $\Delta_{1}, ~\Delta_{2} >0$, the sign of $\Delta_{\rm pbc}$ becomes negative when 
\begin{eqnarray}
\Delta_{1} \Delta_{2} - \Delta_{\rm c}^2 < 0.
\label{Eq:criterion}
\end{eqnarray}
Therefore, a band inversion can be achieved by reversing the relative strength of the local and nonlocal pairings, which gives rise to domain walls at $r=0$ and $r=L$.
		
Before demonstrating that MBS can indeed appear at the domain walls, we remark that, for $\Delta_1$ and $\Delta_2$ with general signs, the band inversion can occur in the absence of the nonlocal pairing. Indeed, setting $\Delta_{\rm c} = 0$, $\Delta_{\rm pbc}$ is negative when the signs of $\Delta_1$ and $\Delta_2$ are opposite (dubbed a $\pi$-junction configuration).
In the above setup with a single superconducting layer, the phases of $\Delta_1$ and $\Delta_2$ are the same and do not form a $\pi$ junction.
Alternatively, one can achieve the $\pi$-junction setup by proximitizing the two helical channels with two separate superconducting layers and controlling the phase between them through a magnetic flux~\cite{Laubscher:2020}. 
Below we focus on the setting with a single superconducting layer and the case where $\Delta_1$ and $\Delta_2$ have the same sign.

To examine the presence of MBS, we solve the Bogoliubov-de~Gennes equation at zero energy near the origin $r=0$ (assuming a sufficiently long channel so that $r=L$ is far away).
We find that two Majorana solutions $\Phi_{\rm mbs, 1}$ and $\Phi_{\rm mbs, 2}$ localized at $r=0$ emerge for $\Delta_{\rm c}^2 > \Delta_{1} \Delta_{2}$. 
The wave function of the first solution is given by 
 $\Phi_{\rm mbs, 1}(r) = \Phi_{>}(r)\Theta(r) + \Phi_{<}(r)\Theta(-r)$ with the step function $\Theta(r)$ and 
\numparts
\begin{eqnarray}
	\Phi_{>}(r) = e^{-\kappa_{\rm c} r} \left( \begin{array}{c}
	i \eta_{\rm m} \\ - \eta_{\rm m} \\ -i \\ 1 \\ -i \eta_{\rm m} \\ - \eta_{\rm m} \\ i \\ 1
	\end{array}
	\right), \;
	\Phi_{<}(r) =
	 \left( \begin{array}{c}
	i \eta_{\rm m} e^{\kappa_1 r} \\ - \eta_{\rm m} e^{\kappa_1 r} \\ -i e^{\kappa_2 r} \\ e^{\kappa_2 r} \\ -i \eta_{\rm m} e^{\kappa_1 r} \\ - \eta_{\rm m} e^{\kappa_1 r}\\ i e^{\kappa_2 r}\\  e^{\kappa_2 r}
	\end{array}
	\right), \nonumber \\ 
\end{eqnarray}
where we omit the normalization constants and introduce the parameters,
\begin{eqnarray}
	\eta_{\rm m} &= & \frac{ \sqrt{\Delta_{-}^2 + \Delta_{\rm c}^2} - \Delta_{-} }{\Delta_{\rm c}},	\\
	\kappa_{\rm c} &=& \frac{ \sqrt{\Delta_{-}^2 + \Delta_{\rm c}^2} - \Delta_{+} }{\hbar v_F}, \\
	\kappa_n &=& \frac{\Delta_n }{\hbar v_F} \;\; {\rm for\;} n \in \{1,2\}.
\end{eqnarray}
\endnumparts
The localization length of the wave function $\Phi_{\rm mbs, 1}$ is given by
\begin{eqnarray}
\xi_{\rm loc} = \frac{1}{ {\rm Min} (\kappa_{\rm c}, \kappa_{1}, \kappa_{2})},
\label{Eq:xi_loc}
\end{eqnarray}
with ${\rm Min}(\cdots)$ denoting the minimal value.
The second Majorana solution is related to the first one by $\Phi_{\rm mbs, 2} = {\mathcal T} \Phi_{\rm mbs, 1}$  and $\Phi_{\rm mbs, 1} = - {\mathcal T} \Phi_{\rm mbs, 2}$ with the time-reversal operator $\mathcal{T}$ defined above. 
The two solutions $\Phi_{\rm mbs, 1}$ and $\Phi_{\rm mbs, 2}$ localized at $r=0$ form a Kramers pair; another Kramers pair of MBS is formed at $r=L$, with the wave functions given by $\Phi_{\rm mbs, 1}$ and $\Phi_{\rm mbs, 2}$ upon replacing $r \to (L-r)$. 
We remark that, in contrast to MBS appearing in, for example, proximitized Rashba nanowires, the two MBS of each Kramers pair are not spatially separated. Nonetheless, as guaranteed by the Kramers theorem, their wave functions are orthogonal and the MBS do not hybridize unless the time-reversal symmetry is broken.
From the wave functions we see that each MBS has nonlocal and composite nature, with a finite weight on both channels.
We depict the MBS positions in a HOTI nanowire with a square cross section in~\Fref{Fig:HOTI_MKP} and with a hexagonal cross section in~\Fref{Fig:PhaseDiagram_HOTI}. 
Since the zero-energy MBS are located at the corners of the overall three-dimensional crystal structure, they are often named Majorana corner modes or Majorana corner states in the literature on setups utilizing proximitized HOTI; we will review additional proposals for their realization in section~\ref{SubSec:TSU-other}.

Whereas the simplified model \eref{Eq:H_sp} might be further enriched by various single-particle perturbations, such as (co)tunneling processes within (between) channels and spin-orbit coupling~\cite{Reeg:2017,Schrade:2017,Klinovaja:2014,Klinovaja:2015,Hsu:2018}, they do not lead to any gap closure and, therefore, cannot lead to additional topological phase transitions. In addition, since these additional perturbations are less RG relevant than the nonlocal pairing, their strengths are suppressed by interactions more significantly. In conclusion, the MBS are robust against these perturbations.

In summary, a Kramers pair of MBS can be stabilized at each end of the double helical channels when the nonlocal pairing is stronger than the local ones, described by the criterion \eref{Eq:criterion}. As long as the time-reversal symmetry is preserved, the pair does not hybridize. While in noninteracting systems, the local pairing typically dominates~\cite{Reeg:2017}, electron-electron interactions can lead to the opposite~\cite{Thakurathi:2018,Thakurathi:2020}, as we demonstrate below.

\subsection{Interacting double helical channels \label{SubSec:TSU-RG}}
In order to investigate the electron-electron interaction effects on the pairings $\Delta_{n}$ and $\Delta_{\rm c}$, we employ the bosonization description and perform the renormalization-group (RG) analysis. While the analysis presented in \cite{Hsu:2018} focused on HOTI, it also applies to the 2DTI edge channels.

Generalizing the boson fields \eref{Eq:bosonization} to a double-channel system, we introduce $\theta_n$ and $\phi_n$ with the channel index $n \in \{1,2\}$ and get two copies of  $H_{\rm hel}$ in \eref{Eq:hTLL},  
\begin{eqnarray}
H_{\rm hel}^{\rm d} = \sum_{n} \int \frac{\hbar dr}{2\pi} \, \left[ u_n K_n \big( \partial_{r} \theta_n \big)^2 + \frac{u_n}{K_n} \big( \partial_{r} \phi_n \big)^2 \right],
\nonumber \\
\label{Eq:H_dhel}
\end{eqnarray}
with the interaction parameter $K_{n}$ for the channel $n$ and the corresponding velocity $u_{n} = v_F/K_n$. 
The boson fields of a given channel satisfy the commutation relation in \eref{Eq:Commutation} and the fields of different channels commute. 
The local and nonlocal pairing terms are given by 
\numparts
\begin{eqnarray}
H_{\rm loc} &=& \sum_{n} \frac{\Delta_{n}}{\pi a} \int dr \,  \cos (2 \theta_n), 
\label{Eq:H_loc_boson} \\
H_{\rm cap} &=& \frac{2}{ \pi a} \int dr \, \Delta_{\rm c}(r) \cos (\theta_1 + \theta_2 ) \cos (\phi_1 - \phi_2 ).
\label{Eq:H_c_boson}
\end{eqnarray}
\endnumparts
The Hamiltonian $H_{\rm hel}^{\rm d} + H_{\rm loc} + H_{\rm cap}$ describes two interacting parahelical channels in both HOTI and 2DTI settings. 
 
\begin{figure}
	\includegraphics[width=0.52\linewidth]{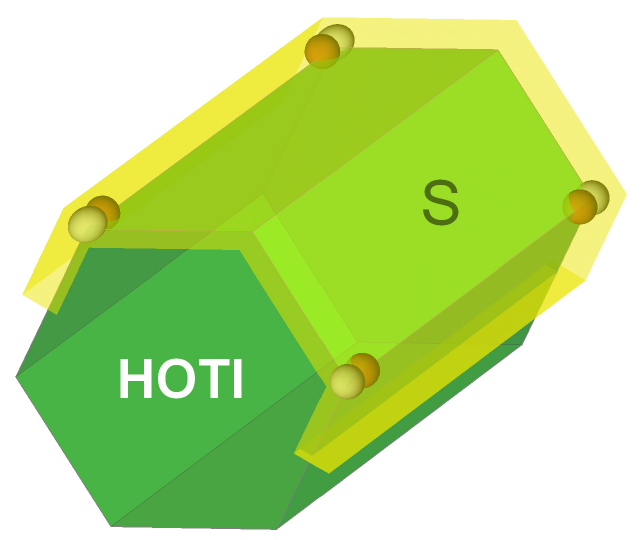} 
	\includegraphics[width=0.45\linewidth]{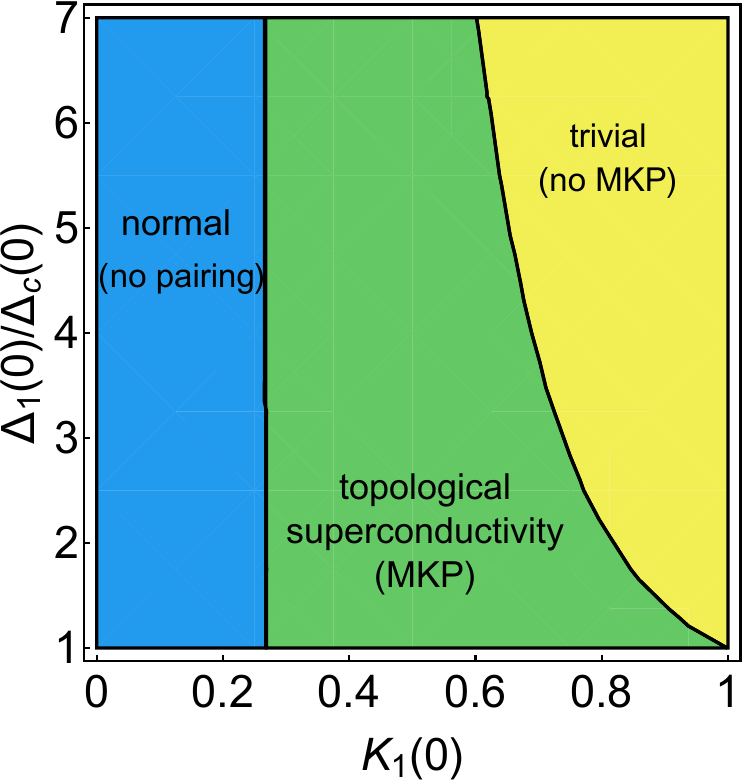} 
	\caption{(Left) MBS in a HOTI nanowire; similar to \Fref{Fig:HOTI_MKP} but for a wire with hexagonal cross section.
		(Right) Phase diagram obtained from the effective-Hamiltonian model. Sufficiently strong interactions lead to Majorana Kramers pairs (denoted as MKP).
		For simplicity, we assume identical initial parameters for the two channels $\Delta_1(0) = \Delta_2(0)$ and $K_1(0) = K_2(0)$.  
			} 
 \label{Fig:PhaseDiagram_HOTI}
	\end{figure}

Since $H_{\rm cap}$ contains the $\phi_n$ field while $H_{\rm loc}$ contains their conjugate field $\theta_n$, the two pairing processes compete with each other. 
As a result, the relative strength of the local and nonlocal pairings varies with the electron-electron interaction strength. Quantitatively, the renormalization of the pairings  is captured by the RG flow equations,
\numparts
\begin{eqnarray}
\frac{d \tilde{\Delta}_n }{d l} &=& \left(2 - \frac{1}{K_n} \right) \tilde{\Delta}_n , \label{Eq:RG_cps1} \\
\frac{d K_n }{d l} &=& \tilde{\Delta}_n^2  + \frac{1}{2} \left( 1 - K_n^2 \right) \tilde{\Delta}_{\rm c}^2 , \\
\frac{d \tilde{\Delta}_{\rm c} }{d l} &=& \left[2 - \frac{1}{4} \sum_{n} \Big( K_n+ \frac{1}{K_n} \Big) \right] \tilde{\Delta}_{\rm c} , \label{Eq:RG_cps2}
\end{eqnarray}
with the dimensionless coupling constants 
\begin{eqnarray}
\tilde{\Delta}_n = \frac{\Delta_n a}{\hbar u_n }, \;\;
\tilde{\Delta}_{\rm c} =  \frac{\Delta_{\rm c}  a }{\hbar \sqrt{u_1 u_2} }. 
\end{eqnarray}
\endnumparts
From the RG flow equations, one can obtain the renormalized values of  $\Delta_n$ and $\Delta_{\rm c}$ at the end of the flow, and examine the existence of MBS Kramers pairs using the topological criterion in \eref{Eq:criterion}.
The results depend on the initial values of the interaction parameters and the gap ratio, as summarized in the phase diagram in~\Fref{Fig:PhaseDiagram_HOTI}.

Crucially, as a result of the distinct scaling dimensions of the cosine terms in \eref{Eq:H_loc_boson} and \eref{Eq:H_c_boson}, the interactions suppress the local pairing gap more significantly than the nonlocal one. Therefore, the repulsive interactions favor the nonlocal pairing. For sufficiently strong interactions, the relative magnitudes of $\tilde{\Delta}_{1,2}$ and $\tilde{\Delta}_{\rm c}$ can be reversed.
It can be shown that, for a given set of initial parameters, the end point of the RG flow is adiabatically connected to the noninteracting point ($K=1$, with renormalized pairings) without closing the system gap~\cite{Hsu:2018}. Since at the noninteracting point the model can be refermionized, one can employ the criterion \eref{Eq:criterion} to justify the presence of MBS. 
Concluding, sufficiently strong interactions can drive the system into a topological superconducting phase with Kramers pairs of MBS either at the corners of the 2DTI layers or at the ends of the HOTI nanowire. For very strong interactions, both $\tilde{\Delta}_{1,2}$ and $\tilde{\Delta}_{\rm c}$ are suppressed, as the corresponding operators are irrelevant in the RG sense. In this limit, both pairing gaps vanish, making the helical channels nonsuperconducting. Interestingly, by comparing \Fref{Fig:PhaseDiagram_HOTI} to the phase diagram for double Rashba wires~\cite{Thakurathi:2018}, one finds that the required interaction strength for the helical channels is weaker than the nonhelical channels, making it easier to achieve topological superconductivity in the former. 

The above analysis is based on the effective Hamiltonian given in \eref{Eq:H_dhel}, \eref{Eq:H_loc_boson} and \eref{Eq:H_c_boson}, where the initial pairing gap values are set by hand. A more realistic model would reflect the actual physical separation $d$ between the helical channels on the scale of the coherence length $\xi_{s}$ of the parent superconductor. To this end, one can use a microscopic model describing the tunnel coupling between the helical states and a proximity $s$-wave superconductor from~\cite{Hsu:2018}. The proximity superconductor is described by
\begin{eqnarray}
H_{\rm sc} &=& \sum_{{\bf K}, \sigma = \uparrow, \downarrow} 
 \frac{ \hbar^2 |{\bf K}|^2  -  \hbar^2 K_{F {\rm s}}^2   }{2 m_e}
 \psi_{{\rm s}, \sigma}^{\dagger}({\bf K}) \psi_{{\rm s}, \sigma}({\bf K}) \nonumber\\
&& + \Delta_{\rm s} \sum_{{\bf K}} \psi_{{\rm s}, \uparrow}({\bf K}) \psi_{{\rm s}, \downarrow}({\bf -K}) + {\rm H.c.}, 
\label{Eq:H_sc}
\end{eqnarray}
with the electron mass $m_e$, the parent pairing gap $\Delta_{\rm s}$, the momentum $\hbar {\bf K}$ and the Fermi wave vector $K_{F {\rm s}}$ of the superconductor. Also, the operator $\psi_{{\rm s}, \sigma}^\dagger$ creates a fermion with spin $\sigma$ in the superconductor.
The tunneling between the superconductor and the helical modes is grasped by
\begin{eqnarray}
H_{\rm tun} &=& \sum_{n=1,2} \int dr d{\bf X} \; \Big\{ t^{\prime}_{n} ({\bf X}, r) \Big[R_n^{\dagger}(r) \psi_{{\rm s}, \downarrow}({\bf X})  \nonumber \\
&& \hspace{0.85in} + L_{n}^{\dagger} (r) \psi_{{\rm s},\uparrow} ({\bf X}) \Big] + {\rm H.c.} \Big\}, 
\label{Eq:H_tun}
\end{eqnarray}
where we introduce the three-dimensional coordinate ${\bf X} \equiv (X,Y,Z)$ of the bulk superconductor and the tunnel amplitude $t^{\prime}_{n}$,
\begin{eqnarray}
t^{\prime}_{n} ({\bf X}, r) &\equiv& t_{n} \delta (Z-r) \delta (X- d_n) \delta (Y),
\end{eqnarray}
with $d_{1} = d/2$ and $d_{2} = -d/2$.  
Here we work in the weak-pairing regime by assuming a weak tunnel coupling $t_{n}$, which does not strongly perturb the helical modes. We remark that a strong tunnel coupling might drastically affect the one-dimensional channels themselves. In the nonhelical case, it can lead to metallization in proximitized Rashba wires~\cite{Reeg:2018b} or formation of a quantum dot in the nanowire~\cite{Awoga:2019}; for the helical case, the hinge channel of a HOTI can also be altered by strong pairing~\cite{Queiroz:2019}.  

\begin{figure}
	\includegraphics[width=0.49\linewidth]{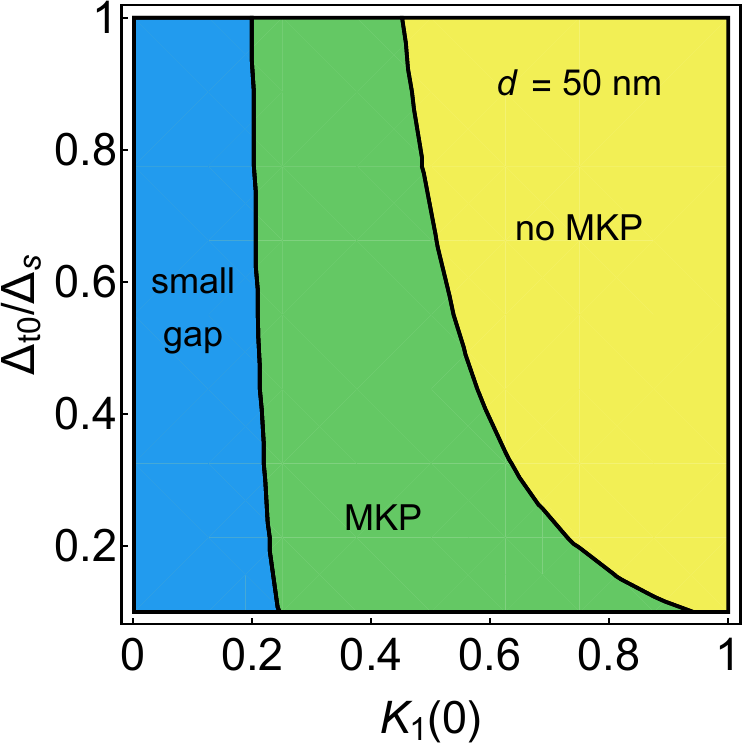} 
	\hspace{0.0\linewidth}
	\includegraphics[width=0.49\linewidth]{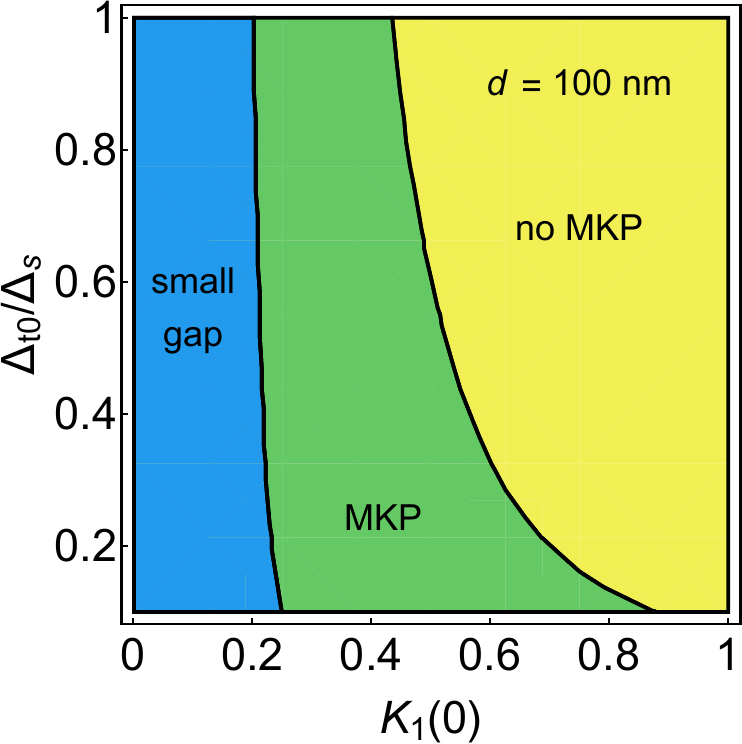}   
		\caption{Phase diagrams obtained from a microscopic model for different interchannel separations $d$.  
		For simplicity, we assume identical parameters for the two channels with the material parameters for Bi HOTI~\cite{Wada:2011,Murani:2017,Schindler:2018,Hsu:2018}: the velocity $u_n= 10^5$~m/s, cutoff $a(0)=5~$nm, channel length $L=1~\mu$m, Fermi wave vector $K_{F {\rm s}}=10^{10}$~m$^{-1}$, and pairing gap $\Delta_{\rm s} = 0.35~$meV corresponding to the coherence length $\xi_{\rm s} = 1.9~\mu$m.
	} 
 \label{Fig:PhaseDiagram_HOTI2}
	\end{figure}

Omitting the details given in~\cite{Hsu:2018}, in~\Fref{Fig:PhaseDiagram_HOTI2} we present the resulting phase diagrams obtained from \eref{Eq:H_dhel}, \eref{Eq:H_sc} and \eref{Eq:H_tun} for two different interchannel distances $d= 50~$nm and 100~nm with the material parameters for Bi HOTI.
The tunnel coupling enters the vertical axis through the parameter 
\begin{eqnarray}
\Delta_{\rm t0} \equiv    \frac{   t_n^2    }{  \Delta_{\rm b} \xi_{\rm s}^2 }   \frac{m_e v_{F {\rm s}}^2}{2\pi \Delta_{\rm s} } K_{0} \left( \frac{ \Delta_{\rm s} }{ \Delta_{\rm b} } \right) 
\end{eqnarray}
with the superconductor Fermi velocity $v_{F {\rm s}}=\hbar k_{F {\rm s}}/m_e$ and the modified Bessel function of the second kind $K_{0}$.
The phase diagrams are consistent with the one obtained from the effective Hamiltonian in~\Fref{Fig:PhaseDiagram_HOTI}. 
Here, the blue region is the parameter regime where the renormalized couplings $\tilde{\Delta}_{\rm c}$ and $\tilde{\Delta}_{1}$ are both less than 0.1, corresponding to the gapless regime in~\Fref{Fig:PhaseDiagram_HOTI}. 
In addition to verifying the result from the effective Hamiltonian, the more elaborate calculation allows one to conclude that, for the interchannel separation $d$ of $O(100 ~{\rm nm})$ and superconducting coherence length $\xi_s$ of $ O(\mu {\rm m})$, one can find a fairly wide regime with Kramers pairs of MBS.
This finding indicates that materials with a micrometer-long superconducting coherence length, such as aluminum, serve as a suitable proximity superconductor.

Remarkably, using a similar setup one can achieve Kramers pairs of parafermions at the system corners~\cite{Klinovaja:2014}. The setup can be achieved by replacing the two 2DTI layers with two-dimensional fractional topological insulators (also known as fractional quantum spin Hall insulator) supporting fractional edge states~\cite{Levin:2009,Maciejko:2015,Stern:2016}. We conclude this subsection by pointing out that since the discussed setups with double helical channels do not invoke magnetic fields, the topological protection of both the helical channels and the zero modes remains intact.

\subsection{Topological superconductivity in other helical systems \label{SubSec:TSU-other}} 
Let us now shortly review additional proposals for realization of topological matter involving helical channels. As already discussed below \eref{Eq:criterion}, in the above double-channel setting, one can exploit that the ``band inversion'' in \eref{Eq:Delta_pbc} takes place even without nonlocal pairing. To this end, one can use a 2DTI bilayer in proximity to two superconductors. Indeed, it was demonstrated that Kramers pairs of Majorana corner states can be stabilized when a phase of $\pi$ between the two superconductors is introduced~\cite{Laubscher:2020}.  

A similar $\pi$ junction can be formed in the following way. Consider a 2DTI layer tunnel coupled to a superconductor through either a layer of randomly oriented magnetic impurities or quantum dots occupied by randomly oriented spins~\cite{Schrade:2015}. The two scenario might be realized, respectively, using magnetic impurities introduced in section~\ref{SubSec:Ensemble} and charge puddles naturally formed in 2DTI introduced in section~\ref{SubSec:ghl}. Changing parameters of such a system, one can invert its effective gap. Therefore, using two sets of such layered structures, one can realize a $\pi$ junction with Kramers pairs of MBS. 
Alternatively, by placing a 2DTI layer on top of an $s$-wave superconductor, one can induce Majorana hinge modes and MBS by applying an external magnetic field in the plane of the heterostructure~\cite{Wu:2020}. Furthermore, parafermion bound states can be formed in a constriction made of a quantum spin Hall insulator in proximity to two $s$-wave superconductors~\cite{Fleckenstein:2019}.

In addition to MBS or parafermions, exotic bound states carrying fractional charge and spin can be realized using 2DTI constrictions~\cite{Klinovaja:2015}. Here, domain walls separate topologically distinct regions with competing gap-opening mechanisms via tunneling, magnetic field, and charge density wave modulation. 
Alternatively, magnetic barriers can be employed to stabilize fractional charges in helical edges~\cite{Fleckenstein:2016}.
Moreover, in proximity to a ferromagnetic insulator and an $s$-wave superconductor, helical liquids can realize unconventional superconductivity with even-parity, spin-triplet and odd-frequency pairing amplitude~\cite{Crepin:2015}.  

Instead of $s$-wave superconductors, one can realize higher-order topological superconductivity by bringing the helical channels in proximity to unconventional superconducting compounds. For even-parity pairing, examples include cuprates with $d_{x^2-y^2}$-wave pairing and iron-based superconductors with $s_{\pm}$-wave pairing. Exploiting the high critical temperature of the parent superconductor and quantum spin Hall state in, for instance, WTe$_2$ monolayer, this setup provides a high-temperature platform for Kramers pairs of MBS at the corners of the two-dimensional structure~\cite{Yan:2018,Wang:2018}. Upon confinement, this setting can form flat bands of Majorana corner modes with a harmonic trapping potential~\cite{Kheirkhah:2020}. 
Alternatively, Majorana corner modes can be induced using a two-dimensional magnetic topological insulator proximitized by either $d_{x^2-y^2}$-wave or $s_{\pm}$-wave superconductors~\cite{Liu:2018}. 
Yet differently, in a heterostructure consisting of a superconducting FeTe$_{1-x}$Se$_{x}$ monolayer and a bicollinear antiferromagnetic FeTe monolayer~\cite{Zhang:2019a} Majorana corner modes appear through the competition of the pairing gap and the ferromagnetic gap in two perpendicular edges. 

Superconductors with odd-parity pairing~\cite{Yan:2019} can be used, too. Here, the early ideas about the first-order topological superconductivity with odd-parity pairing~\cite{Sato:2009b,Sato:2010b,Fu:2010} have been generalized to higher-order topology~\cite{Ahn:2020}. For instance, a second-order topological superconducting phase arises when the $(p_{x} + i p_{y})$-wave pairing is induced in a Dirac semimetal~\cite{Wang:2018b}; a two-dimensional $(p_x + i \sigma p_y)$-wave superconductor is transformed into a second-order topological superconductor hosting Majorana corner modes upon applying an in-plane magnetic field~\cite{Zhu:2018}.
Finally, superconductors with pairing symmetry of mixed parity, such as $(p+id)$-wave pairing, were explored in~\cite{Wang:2018b}.

Instead of 2DTI or HOTI discussed above, one can make use of layered materials with strong Rashba SOI to create helical channels along the boundaries and subsequently use them to realize MBS or parafermion corner states. 
\cite{Volpez:2019} considered a Josephson junction bilayer setup made of either a three-dimensional topological insulator thin film or two tunnel-coupled layers with opposite Rashba SOI, with each layer proximitized by an $s$-wave superconductor. 
When the phase difference between the two pairing amplitudes is $\pi$, the bilayer has topological superconductivity with a Kramers pair of gapless helical states propagating along the edge.
When the interlayer tunneling dominates over the pairing, further applying a weak in-plane Zeeman field drives the system into a second-order topological superconducting states, hosting MBS at diagonally opposite corners.  
Finally, by employing a two-dimensional heterostructure consisting of graphene and transition metal dichalcoginde layers with proximity-induced SOI, one can realize 2DTI phase with helical edge states~\cite{Laubscher:2020b}.  With proximity superconductivity and additionally applied in-plane magnetic field,
Majorana and parafermion corner states can be created in a rectangular sample~\cite{Laubscher:2019,Laubscher:2020b}.

\section{Conclusion and outlook \label{Sec:Conclusion}}
The past decade has witnessed a surge of investigations on helical liquids in topological systems based on semiconductor materials. In this review, we cover several aspects of them, including their realization in various materials, the topological protection for their presence, their properties and characterization, charge transport in their nonsuperconducting phase and their capability of realizing topological superconductivity.
In particular, we present systematic discussions on various mechanisms for resistance sources, including several sets of mutually contradicting references, and summarize the predicted features that can be used to examine the mechanisms in experiments. 
We also review proposals on various settings hosting topological bound states in helical channels with proximity-induced superconductivity.

There are other aspects beyond the scope of this review. 
There have been numerous studies on the ac/dc Josephson effects in the 2DTI edge channels, both experimental~\cite{Hart:2014,Pribiag:2015,Wiedenmann:2016,Deacon:2017,Bocquillon:2017} and theoretical~\cite{Fu:2009,Crepin:2014,HaidekkerGalambos:2020,Novik:2020,Zhang:2020a}; we refer to~\cite{Bocquillon:2018} for a book chapter on Josephson effects in 2DTI-based junctions. Relevant for the time-reversal-invariant settings discussed in section~\ref{Sec:TSU}, it was proposed that Majorana Kramers pairs can be detected through the parity-controlled $2\pi$ Josephson effect~\cite{Schrade:2018}, which might be further extended to perform measurement-based topological quantum computing~\cite{Schrade:2018b}. Concerning the HOTI themselves, the Josephson effect is also useful for identifying the hinge channels, where usual charge transport probes are limited by their insufficient spatial resolution. By fabricating Josephson junctions via  higher-order topological materials, one can deduce the current path from the interference patterns measured from critical current as a function of magnetic flux, thus confirming the presence of hinge channels~\cite{Murani:2017,Schindler:2018,Choi:2020}. 

On the subject of quantum computing, while we focus on the realization of MBS, there are proposals on how to detect or manipulate them in helical systems in order to achieve topological quantum computation~\cite{Liu:2014,Schrade:2018,Schrade:2018b,Pahomi:2020,Plekhanov:2021}, a topic covered by review articles~\cite{Sato:2017,Haim:2019}. In addition to various MBS setups reviewed above, there exist proposals using nanowires or nanoribbons made of three-dimensional topological insulators~\cite{Cook:2011,Cook:2012,Manousakis:2017,Legg:2021} or atomic/optical systems with higher-order topology~\cite{Plekhanov:2019,Luo:2019,Nag:2019,Zeng:2019,Bomantara:2020,Bomantara:2020b}. Finally, other than the zero-energy modes, topological superconductors hosting propagating Majorana modes with linear dispersion and their difference from MBS were reviewed in~\cite{Sato:2017}. Quantum computation exploiting these propagating Majorana modes, as an alternative to the MBS, was reviewed in~\cite{Beenakker:2020}.

There remain puzzles. As discussed above, the intensive studies on charge transport phenomena of 2DTI not only revealed the presence of edge states and their peculiar features, but also led to more puzzles--some of which question even the topological nature of the observed edge states. Meanwhile, the emerging field of van der Waals heterostructures~\cite{Geim:2013} is promising new topological materials~\cite{Qian:2014}. Beyond the first-order topology, HOTI are currently in the spotlight, with increasing numbers of materials or systems having been discovered or under investigation~\cite{Wang:2021}. Concerning MBS, given that their existence in the most intensively investigated setups based on Rashba nanowires remains questionable~\cite{error1,error2}, alternative strategies for their realizations, such as setups using helical channels reviewed here, is called for.

\ack
We thank R.-R. Du for useful remarks on the InAs/GaSb and strained InAs/(Ga,In)Sb heterostructures. 
We acknowledge financial supports from the JSPS Kakenhi Grant No.~19H05610, the Swiss National Science Foundation (Switzerland), the NCCR QSIT and the European Unions Horizon 2020 research and innovation program (ERC Starting Grant, Grant agreement No.~757725). 

\appendix

\section{Derivation of the RG flow equations~\label{App:RG}}
In this section we sketch the derivation of the RG flow equations given in the main text. To be concrete, we take the random-spin-induced backscattering $S_{{\rm rs}}$ in \eref{Eq:S_rs} for illustration.
Following \cite{Giamarchi:1988,Giamarchi:2003}, we start with the correlation function,
\begin{eqnarray}
 \left< e^{i [\phi({\bf r_{1}}) - \phi({\bf r_{2}})] } \right\rangle_{\rm hel + rs} & \equiv& \mathit{Z}_{\rm hel + rs}^{-1} \int D\phi \; e^{- (S_{\rm hel} + S_{\rm rs) } /\hbar} \nonumber \\
 && \hspace{20pt}\times  e^{i [\phi({\bf r_{1}}) - \phi({\bf r_{2}})] },
\end{eqnarray}
where we define the space-time coordinate ${\bf r_{j}} \equiv (r_{j}, y_{j}) = (r_{j}, u \tau_{j})$ for two points $j\in \{1,2\}$ in spacetime and the partition function,
\begin{eqnarray}
\mathit{Z}_{\rm hel + rs} \equiv \int D\phi \; e^{-(S_{{\rm hel}}+ S_{\rm rs} )/\hbar}.
\end{eqnarray} 
The hTLL part $S_{{\rm hel}}$ of the action reads
\begin{eqnarray}
\frac{S_{\rm hel}}{\hbar} 
&=& \int    \frac{  dr d\tau }{2\pi u K  }  \left\{  u^2 \big[ \partial_r \phi (r, \tau) \big]^2 + \big[ \partial_\tau \phi (r, \tau) \big]^2  \right\}, \nonumber \\
\label{Eq:S_hTLL}
\end{eqnarray}
derived from \eref{Eq:hTLL}.

To proceed, we expand the correlation function in orders of the dimensionless coupling $\tilde{D}_{\rm rs} $, which is the overall scale of the backscattering term $S_{{\rm rs}}$.
To the zeroth order, we get
\begin{eqnarray}
\left\langle e^{i [\phi({\bf r_{1}}) - \phi({\bf r_{2}})]} \right\rangle_{\rm hel} &=& 
e^{- K F({\bf r_{1}} - {\bf r_{2}}) / 2},
\end{eqnarray}
with the function,
\begin{eqnarray}
 F({\bf r_{1}} - {\bf r_{2}})  & \equiv & \frac{1}{2} \ln \left[ \frac{(r_{1}-r_{2})^2 + u^2 (\tau_{1} - \tau_{2})^2 }{a^2} \right]  \nonumber \\
 && \hspace{0pt} + \frac{ \tilde{\lambda} }{K} \cos (2 \Theta_{{\bf r_{1}} - {\bf r_{2}}} ), \label{Eq:F(r)} 
\end{eqnarray}
where $\Theta_{{\bf r}}$ is the angle between the vector ${\bf r}$ and the spatial coordinate axis.
The RG flow generates an anisotropic contribution between the spatial and the temporal coordinates, which we express as the $ \tilde{\lambda} $ term. 

The first-order term in $\tilde{D}_{\rm rs} $ is given by 
\begin{eqnarray}
&& \frac{ \tilde{D}_{\rm rs} }{ 8\pi a^3 } \int_{|y-y'|>a} dr dy dy' \; \nonumber \\
&&   \times 
\left\{ \left< e^{i [\phi({\bf r_{1}}) - \phi({\bf r_{2}})]}  
 \cos \left[ 2\phi (r,\tau) - 2\phi (r,\tau')\right] \right>_{\rm hel} \right. \nonumber\\
&&   \left. - \left< e^{i [\phi({\bf r_{1}}) - \phi({\bf r_{2}})]}  \right>_{\rm hel}
\left< \cos \left[ 2\phi (r,\tau) - 2\phi (r,\tau')\right] \right>_{\rm hel} \right\}, \nonumber \\
\end{eqnarray}
which can be written in terms of 
\begin{equation}
e^{-K_{\rm eff } F_{\rm eff }({\bf r_{1}} - {\bf r_{2}}) /2},
\end{equation}
with $F_{\rm eff} ({\bf r_{1}} - {\bf r_{2}})$ in the form of \eref{Eq:F(r)} and the following effective parameters,
\begin{eqnarray}
K_{\rm{eff}} &=& K - \frac{K^2 \tilde{D}_{\rm rs} }{2} \int_{a}^{\infty} \frac{dz}{a} \left( \frac{z}{a}  \right)^{2-2K}  , \\
 \tilde{\lambda}_{\rm{eff}} &=&   \tilde{\lambda}  + \frac{K^2 \tilde{D}_{\rm rs} }{4} \int_{a}^{\infty} \frac{dz}{a} \left( \frac{z}{a}  \right)^{2-2K}.
\end{eqnarray}
Since the correlation function should not change with the cutoff, we keep the effective parameters fixed while changing the cutoff $a \rightarrow a e^{d l} = a + da $.
This procedure gives the RG flow equations,
\begin{eqnarray}
\frac{d \tilde{D}_{\rm rs} (l) }{d l } &=& \left[3-2K(l)\right] \tilde{D}_{\rm rs} (l), \\
\frac{d K (l) }{d l } &=& - \frac{K^2 (l) }{2} \tilde{D}_{\rm rs} (l), \\
\frac{d   \tilde{\lambda}  (l)}{d l } &=& \frac{K^2 (l) }{4} \tilde{D}_{\rm rs} (l),
\end{eqnarray}
where the last line can be transformed into the renormalization of $u$ through  
\begin{eqnarray}
 \frac{du(l)}{dl} &=& -\frac{2 u(l)}{K(l)} \frac{d  \tilde{\lambda} (l)}{d l }.
\end{eqnarray}
The RG flow equations for $u$, $K$ and $\tilde{D}_{\rm rs}$ are given in \eref{Eq:RG_rs1}--\eref{Eq:RG_rs2} in the main text. From the equation for $\tilde{D}_{\rm rs}$, we see that the coupling grows with an increasing cutoff when $K < 3/2$, indicating that the perturbation in \eref{Eq:S_rs} is RG relevant for $K < 3/2$. 

A similar procedure can be employed for the spiral-order-assisted backscattering $S_{\rm sa}$ in \eref{Eq:S_sa}. The generalization to a double-channel system can be utilized to derive the RG flow equations \eref{Eq:RG_cps1}--\eref{Eq:RG_cps2}. 
Due to the magnon-energy dependent factor in \eref{Eq:S_mag}, the calculation for the magnon-induced backscattering $S_{\rm m}$ is more involved. We, therefore, refer the interested readers to Appendix~E of \cite{Hsu:2018b}. \\

\end{document}